\documentclass{aims}

\usepackage{txfonts}
\def\typeofarticle{Research article}
\def\currentvolume{23}
\def\currentissue{5}
\def\currentyear{2026}

\def\ppages{xxx--xxx.}
\def\DOI{}
\def\Received{}
\def\Revised{}
\def\Accepted{}
\def\Published{}

\numberwithin{equation}{section}

\usepackage{tikz}
\usetikzlibrary{arrows.meta,bending}
\usepackage{enumitem}
\usepackage{cite}
\emergencystretch=\maxdimen
\makeatletter
\renewcommand{\@biblabel}[1]{#1\hfill \hspace{-0.2cm}}
\makeatother
\begin{document}

\title{Comparing Bayesian and frequentist inference in
	biological models: A comparative analysis of
	accuracy, uncertainty, and identifiability}

\author{%
	Mohammed A.~Y. Mohammed\affil{1}, Hamed Karami\affil{1,2} and Gerardo Chowell\affil{2,3,}$^*$
}
% \shortauthors is used in copyright information in the end of the paper
\shortauthors{the Author(s)}
\address{%
	\addr{\affilnum{1}}{Department of Mathematics and Statistics, Georgia State University, Atlanta, GA 30303, USA}
	\addr{\affilnum{2}}{Department of Population Health Sciences, School of Public Health, Georgia State University, Atlanta, GA 30303, USA}
	\addr{\affilnum{3}}{Department of Applied Mathematics, Kyung Hee University, Yongin 17104, Korea}
	 %corresponding author
	$^*${\textbf{Correspondence:} Email: gchowell@gsu.edu. }
}

\begin{abstract}
Mathematical models are widely used to study ecological and epidemiological systems, but the accuracy of the resulting inferences and forecasts depends on the estimation framework. This study compares Bayesian and frequentist approaches across three biological models using four datasets: The Lotka–Volterra predator–prey system, a generalized logistic model (GLM) applied to lung injury and mpox data, and the susceptible-exposed-infected (reported and unreported)-recovered (SEIUR) epidemic model for COVID-19. To ensure a fair comparison, both approaches were implemented with a normal error structure.	
We first examined the structural identifiability to determine which parameters can be recovered in principle. We then evaluated practical identifiability and fitting performance using four metrics: Mean absolute error, mean squared error,  prediction interval coverage, and the weighted interval score. For the Lotka--Volterra model, we studied three types of observations: Prey only, predator only, and both species together.	
The frequentist workflow, implemented through QuantDiffForecast, uses nonlinear least squares and parametric bootstraping to quantify uncertainty. The Bayesian workflow, implemented through BayesianFitForecast, uses Hamiltonian Monte Carlo sampling to obtain posterior distributions and diagnostic measures.	
Our results show that frequentist inference performs well when the system comprises fully observed data, as in the GLM or in the Lotka--Volterra model when both species are observed. Bayesian inference performs better when uncertainty is high and observations are limited or indirect, as seen in the SEIUR epidemic model. The identifiability analysis helps explain these differences by showing how observability shapes the reliability of parameter recovery and uncertainty quantification.
\end{abstract}


\maketitle

\section{Introduction}

Mathematical models based on ordinary differential equations (ODEs) are essential tools for understanding and forecasting dynamics in ecology and epidemiology \cite{modeling_overview, mohammed2024trophiccascade, alsammani2025cholera}. Model-based forecasts have been instrumental in managing major public health crises over the past decade. During the COVID-19 pandemic, forecasts guided resource allocation and social distancing policies \cite{dixon2022comparison, cheng2023real, lutz2019applying, shearer2020infectious, chowell2022ensemble, chowell2022sub, karami2016comparative}.  

The US center for Disease Control and Prrevention (CDC’s) FluSight Challenge used models to optimize influenza vaccine's distribution and public health messaging \cite{reich2019collaborative, mcgowan2019collaborative, biggerstaff2014estimates}. During the West African and Democratic Republic of the Congo (DRC) Ebola outbreaks, models predicted the transmission patterns and evaluated the intervention's effectiveness \cite{chowell2017perspectives, funk2019assessing, meltzer2014estimating, chretien2015mathematical, chowell2014catastrophic, roosa2020multimodel}. More recently, forecasting models for mpox predicted its spread and assessed containment measures \cite{bleichrodt2023mpox_rt, bleichrodt2023mpox_ensemble, charniga2024nowcasting, chowell2024growthpredict}.  

In ecology, predator–prey models help predict population dynamics and inform conservation strategies \cite{maclulich1937hare}. These applications demonstrate that reliable forecasts depend on accurate parameter estimation \cite{hyndman2021forecasting, chowell2024quantdiffforecast, karami2025bayesianfitforecast}. However, the ability to recover meaningful parameters from data critically depends on identifiability, that is, whether the model's parameters can be uniquely determined from the available observations \cite{raue2009profilelikelihood, chis2011structural, atanasov2025honeybee}.  

Identifiability can fail when the data are sparse, the models are over-parameterized, or the parameters are strongly correlated, leading to misleading inferences and unreliable forecasts. Recent work has shown that factors such as vaccination behavior, immunity, and data completeness shape both an epidemic's trajectories and the reliability of parameter estimates \cite{alsammani2025vaccination}. Understanding how identifiability interacts with estimation frameworks is essential to ensure that model-based forecasts are interpretable and reproducible. Structural or practical identifiability limitations can cause two inference frameworks to produce different results even under identical models and data.  

Two estimation paradigms dominate the field: Bayesian and Frequentist methods. Frequentist methods typically calibrate ODE models by optimizing a likelihood function or minimizing an objective function such as the sum of squared differences between the observed and predicted values \cite{gneiting2008probabilistic, mwambi2011forceofinfection, chowell2017primer, chowell2020ensembles}. These methods use algorithms such as gradient descent or the Levenberg–Marquardt algorithm, assume specific distributions for measurement errors (e.g., Gaussian, Poisson, or normal), and quantify uncertainty through bootstrapping techniques \cite{banks2014uncertainty, pruitt2024likelihood, transtrum2012optimal, huang2024nonlinear, bates1988nonlinear, cao2012penalized, ramsay2017dynamic, seber2003nonlinear}. Frequentist methods are computationally efficient and often perform well when the data are abundant and of high quality \cite{nocedal2006optimization, chowell2024quantdiffforecast}. The QuantDiffForecast (QDF) toolbox implements this workflow for fitting models and generating predictions with quantified uncertainty \cite{chowell2024quantdiffforecast}.  

Bayesian methods, in contrast, apply Bayes’ theorem to combine the prior distributions of parameters with the likelihood of the observed data, producing posterior distributions that explicitly incorporate uncertainty \cite{greenland2009bias, mckinley2014bayesian, kypraios2017abc, girolami2008diffeq, grinsztajn2021workflow, bouman2024timevarying, gelman2020workflow, belasso2023bayesian}. These methods typically use Markov chain Monte Carlo (MCMC) algorithms to approximate posterior distributions, providing comprehensive measures of parameter uncertainty and  PIs \cite{martin2020computingbayes, dunson2001advantages, harel2018imputation, grinsztajn2021workflow, annis2017stan, kelter2020survival}. Bayesian methods can better navigate complex parameter spaces, avoid local optima, and handle incomplete or noisy data \cite{gelman1996pk, huang2006hiv, huang2020bayesianode}. Tools such as Stan facilitate Bayesian estimation and forecasting, allowing rigorous uncertainty quantification and model validation \cite{gelman2013bayesiandata, vehtari2021rhat, gelman1992rubin, monnahan2017hmc, burkner2017brms}.  

The BayesianFitForecast (BFF) toolbox implements this workflow with diagnostics such as the Gelman–Rubin $\hat{R}$ statistic \cite{karami2025bayesianfitforecast}. Despite the extensive use of both paradigms, prior comparisons often vary models, likelihoods, or preprocessing between methods, making it difficult to attribute observed differences to the estimation framework itself \cite{gneiting2014probabilistic}. To address this gap, we conduct a controlled comparison of Bayesian and frequentist inference under standardized conditions: The same models, the same normal error structure, and harmonized data preprocessing. We purposely use simple, low-dimensional standard models so we can make a clear, fair comparison and use differential algebra tools for structural identifiability.

We analyze three systems and four datasets representing different levels of complexity and observability: The LV predator–prey model (Hudson Bay lynx–hare data), a generalized logistic model (GLM) for lung injury and the 2022 United States mpox outbreak, and an susceptible-exposed-infected (reported and unreported)-recovered (SEIUR) epidemic model for the first COVID-19 wave in Spain \cite{richards1959growth, curran2022hivmpx, arenas2020covidmodel}. For the LV model, we analyze three observation scenarios (prey only, predator only, and both simultaneously) to assess how partial observability affects parameter recovery.  

We integrate a structural identifiability analysis to determine which parameters can theoretically be recovered from the data, separating fundamental data limitations from algorithmic limitations \cite{walter1997identification, stigter2023identifiability, dankwa2022structural}. We then evaluate practical identifiability and forecasting performance using four metrics: Mean absolute error (MAE), mean squared error (MSE), 95\% coverage, and the weighted interval score (WIS) \cite{bracher2021evaluating, gneiting2007scoring, brooks1998convergence, carpenter2018lotkavolterra, elton1924periodic}.  

Our objectives are to (i) compare prediction accuracy across Bayesian and frequentist inference, (ii) assess uncertainty calibration and diagnostics, and (iii) relate observed performance to structural versus practical identifiability under full versus partial observation. This provides practical guidance on when each paradigm is preferable, depending on data richness, observability, and the need of uncertainty quantification \cite{karami2025bayesianfitforecast}.  

The remainder of this paper is organized as follows. Section~\ref{sec:Data} describes the four datasets analyzed in this study. Section~\ref{sec:models} presents the mathematical models used. Section~\ref{sec:methods} details the Bayesian and frequentist inference methodologies. Section~\ref{sec:SI} presents the structural identifiability analysis. Section~\ref{sec:results} reports the empirical results. The Discussion synthesizes our findings, and the Conclusion summarizes the main contributions and implications for biological modeling practice.

\section{Data}
\label{sec:Data}
\begin{figure}[H]
	\begin{center}
		\includegraphics[scale=0.8]{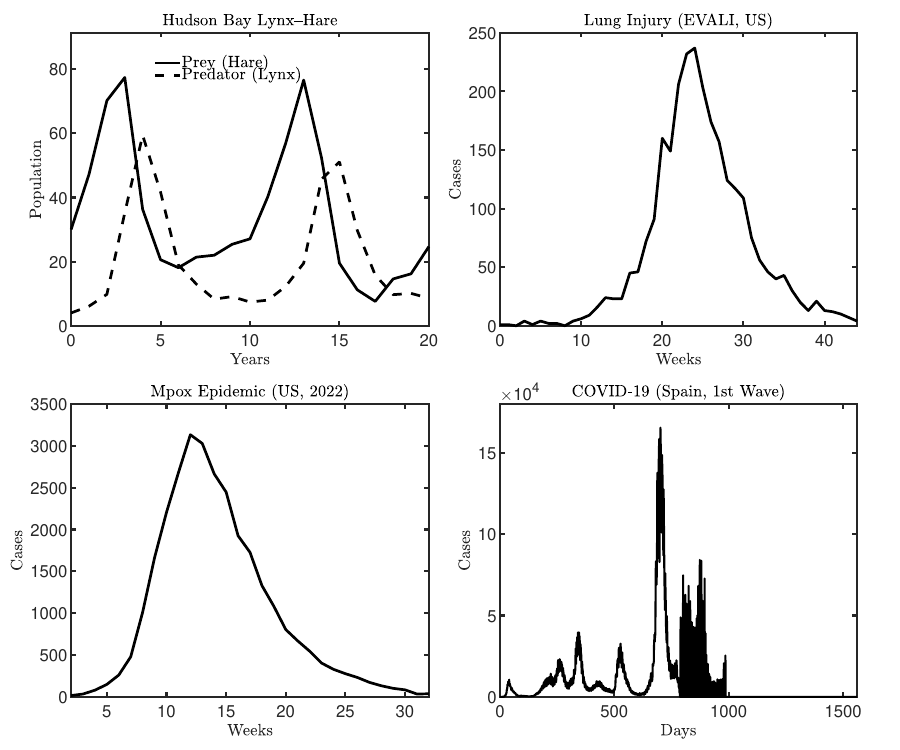}\vspace{0.5cm} 
		\caption{Time series of ecological and epidemic population dynamics. The Hudson Bay Lynx--hare dataset shows annual prey and predator abundance (1900--1920), while the remaining panels depict weekly or daily reported cases from major disease outbreaks: Lung injury (EVALI, US, 2019), mpox (US, 2022), and COVID-19 (Spain, 2020).}
		\label{fig:ecological_epidemic_panel}
	\end{center}
\end{figure}

We analyzed four datasets spanning ecological and epidemiological systems, each selected to represent different levels of data richness, temporal resolution, and observability (Table~\ref{tab:data_summary}; Figure~\ref{fig:ecological_epidemic_panel}). All datasets consist of time-series observations at discrete time points $t_n$, where $n = 1, \ldots, N$ denotes the observation index. The datasets include both population counts (for the ecological system) and incident case counts (for epidemic systems). Data preprocessing was standardized across all datasets to ensure comparability between inference methods; the specific preprocessing steps are summarized at the end of this section.

\subsection{Hudson Bay lynx--hare data}

The Hudson Bay lynx--hare dataset consists of annual counts of Canadian Lynx (\textit{lynx canadensis}) and snowshoe hare (\textit{Lepus americanus}) populations from 1900 to 1920, derived from Hudson's Bay Company pelt return records \cite{carpenter2018lotkavolterra, elton1924periodic}
. These records serve as a proxy for population abundance, with pelt counts reflecting the relative population size. The dataset contains $N = 21$ yearly observations for both prey (hare) and predator (lynx) populations. This dataset is particularly valuable for assessing parameter identifiability in predator--prey dynamics because it provides simultaneous observations of both interacting species over multiple population cycles. The data have been widely used as a benchmark for testing ecological models and parameter estimation methods. No smoothing, interpolation, or imputation was applied to the original data.

\subsection{Lung injury data (EVALI)}

This dataset comprises weekly incident case counts of electronic cigarette or vaping product use-associated lung injury (EVALI) reported in the United States during 2019 \cite{krishnasamy2020evali,lozier2019evaliupdate}
. The outbreak was first identified in mid-June 2019, with cases rapidly escalating through the summer and fall before declining by November 2019. We analyzed $N \approx 21$ weeks of data spanning mid-June to early November 2019, as reported through the Centers for CDC Morbidity and Mortality Weekly Report (MMWR) surveillance system. Case definitions followed CDC criteria for confirmed and probable EVALI cases. The data were aggregated by epidemiological week (MMWR week) with no backfill corrections applied beyond those included in the official CDC reports. This dataset represents a short-duration outbreak with a clear peak, making it suitable for testing phenomenological growth models under well-defined epidemic dynamics.

\subsection{Mpox data (United States, 2022)}

We analyzed weekly incident confirmed or probable mpox cases in the United States during the~2022--2023 outbreak as reported in \cite{cdc2022mpox}. The outbreak began in May 2022 and peaked in August~2022 before declining through early 2023. Data were compiled from CDC national surveillance reports aggregated by MMWR week. Case definitions followed CDC guidelines for confirmed (laboratory-confirmed orthopoxvirus with the specimen typed as monkeypox virus) and probable cases (epidemiologically linked to a confirmed case without laboratory confirmation). The temporal window analyzed corresponds to the public reporting period documented in CDC surveillance dashboards (see the references for access dates). This dataset provides an example of an emerging infectious disease outbreak with intensive surveillance and public health response, allowing evaluation of the model's performance during rapidly evolving epidemic conditions.

\subsection{COVID-19 data (Spain, first wave)}

This dataset consists of daily incident laboratory-confirmed COVID-19 cases reported during the first epidemic wave in Spain from February through to May 2020. Data were compiled from official reports by the Spanish Ministry of Health (Ministerio de Sanidad), with case dates based on the report date rather than the symptom onset date due to data availability constraints during the early pandemic response. The first wave in Spain exhibited rapid exponential growth followed by a peak in late March 2020 and a subsequent decline following the implementation of strict non-pharmaceutical interventions, including a national lockdown. We restricted analysis to the first wave period to avoid complications from changing surveillance protocols, intervention policies, and emergence of new variants that occurred in later waves. Known reporting artifacts explicitly flagged in official data releases were excluded. This dataset represents a large-scale epidemic with significant under-reporting and latent compartments (exposed and unreported infectious individuals), making it ideal for assessing inference methods under partial observability and model complexity.

\begin{table}[H]
	\centering
	\caption{Summary of datasets analyzed in this study, including data sources, temporal coverage, reporting frequency, and number of observations for each system.}\vspace{0.2cm} 
	\small
	\begin{tabular}{llll}
		\hline
		Dataset & Source and Period & Frequency & Observations (N) \\ \hline
		Hudson Bay Lynx--hare & Hudson's Bay Company pelt returns (1900--1920) & Annual & 21 \\
		Lung injury (EVALI, US) & CDC MMWR reports (mid-June--November 2019) & Weekly & $\approx$ 21 \\ 
		Mpox epidemic (US, 2022) & CDC surveillance (May 2022--early 2023) & Weekly & Variable \\ 
		COVID-19 (Spain, 1first wave) & Spanish Ministry of Health (February-May 2020) & Daily & Variable \\ \hline
	\end{tabular}
	\label{tab:data_summary}
\end{table}

\subsection{Preprocessing summary:}
All datasets were processed using a standardized protocol to ensure comparability across inference methods. The preprocessing steps were as follows.

\begin{enumerate}
	\item[\rm{1)}] \textbf{Temporal aggregation:} The Hudson Bay lynx--hare data were used as published without aggregation. Lung injury and mpox data were aggregated at the weekly level (MMWR epidemiological week) as reported in the official CDC surveillance. COVID-19 data for Spain were analyzed at the daily level as reported by the Spanish Ministry of Health.

	\item[\rm{2)}] \textbf{Data type:} The Hudson Bay dataset represents population abundance (pelt counts), while all epidemic datasets (lung injury, mpox, COVID-19) represent incident case counts per reporting period. For the GLM, cumulative case counts were derived by summing incident counts, and model predictions were compared with incident data via the time derivative of cumulative cases.
	
	\item[\rm{3)}] \textbf{Missing values:} No missing values were present in the final temporal windows used for model fitting. No imputation was performed.
	
	\item[\rm{4)}] \textbf{Smoothing:} No smoothing was applied to the data used for parameter estimation or forecasting. Figures may display smoothed overlays for visualization purposes only; these smoothed curves were not used in model fitting.
	
	\item[\rm{5)}] \textbf{Temporal windows:} Modeling windows for calibration and forecasting were selected on the basis of epidemic phase and data availability. Specific windows for each dataset are described in the Results section.
	
	\item[\rm{6)}] \textbf{Data quality control:} For COVID-19 data, known reporting artifacts explicitly flagged in official Spanish Ministry of Health releases were excluded. For all other datasets, the data were used as published in official surveillance reports without additional corrections.
\end{enumerate}

\section{Models}
\label{sec:models}
In this study, we used three compartmental biological and epidemiological models to comprehensively assess the parameter identifiability of our two approaches: (i) the LV model, describing predator-prey interactions and population feedback mechanisms; (ii) the GLM, capturing flexible epidemic growth through nonlinear case dynamics; and (iii) the SEIUR model, explicitly tracking susceptible, exposed, infectious (reported and unreported), and recovered populations to account for both observed and hidden transmission. These models were selected to represent a progression in model complexity and data observability: from a fully observed ecological system LV, to a parsimonious single-equation epidemic model GLM, and to a multi-compartment latent-state model with partial observability (SEIUR). This progression enables a systematic comparison of how Bayesian and frequentist inference across increasing levels of model structure and latent uncertainty. We chose these standard, low-dimensional models on purpose so the comparison is clear and fair and so the structural identifiability can be analyzed directly.

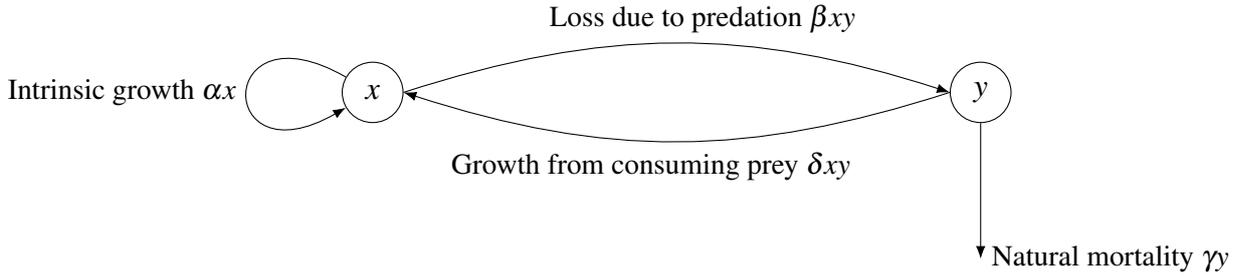
\begin{figure}[H]
	\centering
	\begin{tikzpicture}[>=Latex, node distance=5cm]
		
		\tikzstyle{comp}=[circle, draw, minimum size=8mm, inner sep=0pt]
		\tikzstyle{lab}=[font=\small, align=center]
		
		\node[comp] (x) at (0,0) {$x$};
		\node[comp] (y) at (8,0) {$y$};
		
		\path (x) edge[->, loop left, in=210, out=150, min distance=20mm]
		node[lab, left] {Intrinsic growth $\alpha x$} (x);
		
		\draw[->] (x.east) to[bend left=18] node[lab, above, pos=.55] {Loss due to predation $\beta xy$} (y.west);
		\draw[->] (y.west) to[bend left=18] node[lab, below, pos=.55] {Growth from consuming prey $\delta xy$} (x.east);
		
		\draw[->] (y.south) -- ++(0,-1.8) node[lab, right] {Natural mortality $\gamma y$};
		
	\end{tikzpicture}\vspace{0.5cm} 
	\caption{ LV predator--prey diagram. Circles represent the prey $x$ and predator $y$ populations. The self-loop on $x$ indicates intrinsic growth at a rate $\alpha$. Curved arrows between $x$ and $y$ represent interactions: Prey loss due to predation and predator growth from consuming prey. The downward arrow on $y$ represents natural mortality.}
	\label{fig:lotka_volterra}
\end{figure}

\paragraph{Lotka-Volterra model.} The LV equations (Lotka 1925; Volterra 1926, 1927) consist of ODEs describing the population dynamics of two interacting species: One predator and one prey. Let $x$ denote the prey population and $y$ the predator population at time $t$. Volterra modeled the temporal dynamics of these populations as follows:

\begin{equation} 
	\label{eq:LV}
	\frac{dx}{dt} = \alpha x - \beta xy, \quad
	\frac{dy}{dt} = -\gamma y + \delta xy,
\end{equation}
with the initial conditions
\begin{equation}
	\label{eq:ICLV}
	x(0) = x_0, \quad y(0) = y_0,
\end{equation}
where $\alpha$ is the intrinsic growth rate of the prey population, $\beta$ is the predation rate coefficient, $\gamma$ is the mortality rate of the predator population, and $\delta$ is the predator’s growth rate per prey consumed. The initial conditions $x_0$ and $y_0$ specify the starting population sizes of the prey and predator, respectively. The LV model serves as a classical benchmark in population ecology, offering a well-understood test case for examining parameter identifiability and uncertainty propagation under full versus partial observability. Figure~\ref{fig:lotka_volterra} illustrates the interaction structure of the LV predator--prey model.

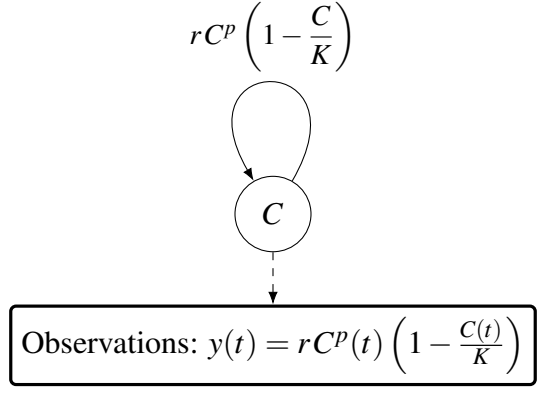
\begin{figure}[H]
	\centering
	\begin{tikzpicture}[>=Latex]
		
		\tikzstyle{comp}=[circle, draw, minimum size=10mm, inner sep=0pt]
		\tikzstyle{lab}=[font=\small, align=center]
		
		\node[comp] (C) at (0,0) {$C$};
		
		\path (C) edge[->, loop above, in=120, out=60, min distance=20mm]
		node[lab, above] {$r \, C^{p} \left(1 - \dfrac{C}{K}\right)$} (C);
		
		\node[draw, very thick, rounded corners=2pt, fill=white,
		minimum width=10mm, minimum height=7mm,
		anchor=north] (OBS) at (0,-1.2) {Observations: $y(t) = r \, C^{p}(t) \left(1 - \frac{C(t)}{K}\right)$};
		
		\draw[dashed, ->] (C.south) -- (OBS.north);
		
	\end{tikzpicture}\vspace{0.5cm} 
	\caption{Diagram of the GLM. The circle represents the cumulative cases $C(t)$. The self-loop indicates growth governed by the generalized logistic equation. The dashed arrow indicates the observed incidence.}
	\label{fig:GLM}
\end{figure}

\paragraph{Generalized logistic model.} 
The GLM is a flexible extension of the logistic function used to model S-shaped epidemic growth curves. It is given by

\begin{equation}
	\label{eq:GLM}
	\frac{dC}{dt} = r \, C^{p}(t) \left(1 - \frac{C(t)}{K}\right),
\end{equation}
with the initial condition
\begin{equation}
	\label{eq:ICGLM}
	C(0) = C_0,
\end{equation}
where $C(t)$ is the cumulative number of cases at time $t$, $r$ is the generalized growth rate, $K$ is the final epidemic size, and $p \in [0,1]$ controls the growth dynamics: $p = 0$ corresponds to constant incidence, $0 < p < 1$ to subexponential growth, and $p = 1$ to exponential growth. Here, $C_0$ is the number of reported cases on day 0. Moreover, $\frac{dC}{dt}$ serves as the observation operator mapping the model states to the reported incidence data. The GLM was chosen because it provides a parsimonious, single-equation description of epidemic trajectories and serves as a tractable case for testing the accuracy and uncertainty of inferences when the full time series is directly observed. This model has also been widely applied in epidemiological settings to capture early and full-phase epidemic dynamics \cite{chowell2016zika}
. The GLM has been widely used as a phenomenological growth model to capture epidemic curves with a rapid increase and saturation \cite{chowell2016zika, pell2018ebola}. Figure~\ref{fig:GLM} illustrates the GLM and its observation process.

\paragraph{SEIUR model} 
The SEIUR model tracks susceptible ($S$), exposed ($E$), reported infectious ($I$), unreported infectious ($U$), recovered ($R$), and cumulative reported cases ($C$). The dynamics are

\begin{equation} 
\label{eq:SEIUR}
\begin{gathered}
\frac{dS}{dt} = -\beta_f(t) \frac{(I + U)S}{N}, \quad
\frac{dE}{dt} = \beta_f(t) \frac{(I + U)S}{N} - \kappa E, \quad
\frac{dI}{dt} = \kappa \rho E - \gamma I, \\
\frac{dU}{dt} = \kappa (1 - \rho) E - \gamma U, \quad
\frac{dR}{dt} = \gamma (I + U), \quad
\frac{dC}{dt} = \kappa \rho E,
\end{gathered}
\end{equation}
\noindent
with the initial conditions
\begin{equation} 
\label{eq:ICSEIUR}
(S(0), E(0), I(0), U(0), R(0), C(0)) = (N - C_0, 0, C_0, 0, 0, C_0),
\end{equation}
where $N$ is the total population, and $\beta_f(t)$ is the time-dependent transmission rate, $\kappa$ is the incubation rate, $\gamma$ is the recovery rate, $\rho$ is the reporting proportion. Here, $C_0$ is the number of reported cases on Day 0. The initial condition assumes that all initially infected individuals are reported and that the remaining population is susceptible.
The time-dependent transmission rate $\beta_f(t)$ is defined as the~ following: 
\[
\beta_f(t) =
\begin{cases}
\beta_0, & \text{if } t < t_{\text{int}}, \\[6pt]
\beta_1 + (\beta_0 - \beta_1)\, e^{-q_1 (t - t_{\text{int}})}, & \text{if } t \ge t_{\text{int}}.
\end{cases}
\]

The time-dependent transmission rate $\beta_f(t)$ is modeled in a piecewise manner to reflect changes in the population behavior and control measures following an intervention at time $t_{\mathrm{int}}$. Before the intervention, transmission is assumed to be constant at a level $\beta_0$, representing relatively uncontrolled epidemic spread. After $t_{\mathrm{int}}$, transmission decreases smoothly toward a lower asymptotic level $\beta_1$ through an exponential decay, capturing the gradual adoption and increasing effectiveness of interventions such as mobility reduction, masking, or behavioral adaptation. The decay rate $q_1$ controls how rapidly these effects take hold, allowing the model to represent both abrupt and slow transitions while remaining parsimonious and identifiable.

\begin{figure}[H]
\centering
\begin{tikzpicture}[>=Latex]

\tikzstyle{comp}=[circle, draw, minimum size=8mm, inner sep=0pt]
\tikzstyle{lab}=[font=\small, align=center]

\node[comp] (S) at (0,0) {$S$};
\node[comp] (E) at (4,0) {$E$};
\node[comp] (I) at (7,2.2) {$I$};
\node[comp] (U) at (7,-2.5) {$U$};
\node[comp] (R) at (15.7,0) {$R$};

\draw[->] (S) -- (E)
node[lab, pos=0.55, above] {$\dfrac{\beta (I+U) S}{N}$};

\draw[->] (E) -- (I)
node[lab, pos=0.56, above, xshift=-3mm] {$\kappa \rho E$};

\draw[->] (E) -- (U)
node[lab, pos=0.56, below, xshift=-7mm] {$\kappa (1-\rho) E$};

\draw[->] (I) -- (R)
node[lab, pos=0.55, above] {$\gamma I$};

\draw[->] (U) -- (R)
node[lab, pos=0.55, below] {$\gamma U$};

\node[draw, very thick, rounded corners=2pt, fill=white,
minimum width=.8mm, minimum height=9mm,
anchor=west] (OBS) at (5.4,0)
{Observations: $y(t)=\kappa \rho E(t)$};

\draw[dashed, ->] (E) -- (OBS.west);

\end{tikzpicture}\vspace{0.5cm} 
\caption{Compartmental diagram of the SEIUR model with under-reporting. Circles represent the epidemiological compartments. Solid arrows indicate transitions between compartments, and the dashed arrow indicates the source of observed cases.}
\label{fig:seiur}
\end{figure}
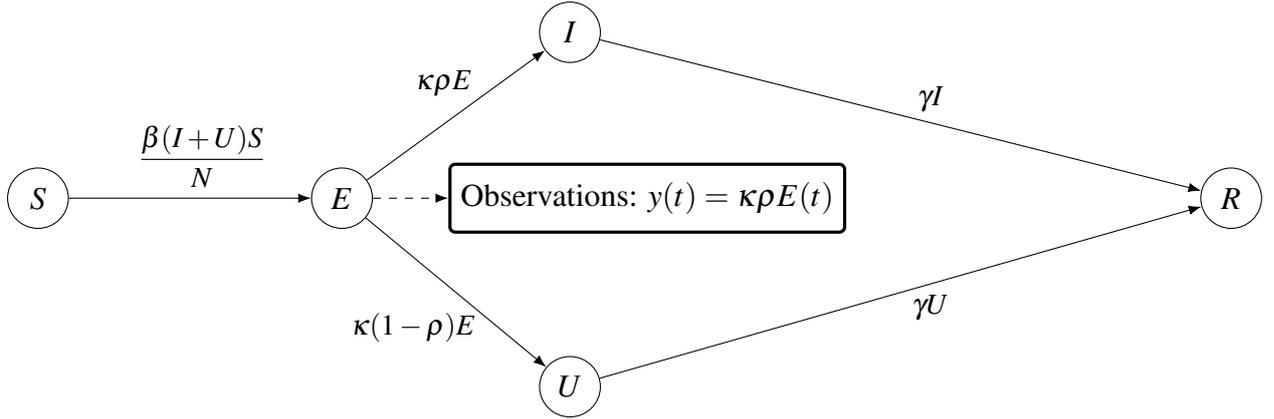

This model represents a more realistic epidemic process with both observed and hidden states, making it ideal for testing how Bayesian and frequentist methods handle latent variables, parameter coupling, and partial identifiability. Figure~\ref{fig:seiur} shows the compartmental structure of the SEIUR model with under-reporting.

\section{Methods}
\label{sec:methods}

This section describes the Bayesian and frequentist estimation frameworks used for parameter inference, uncertainty quantification, and forecasting. Let $Y = (y_{t_1}, \ldots, y_{t_n})$ denote the observed data at discrete time points $t_1, \ldots, t_n$, and let $\boldsymbol{\theta}$ represent the vector of unknown model parameters to be estimated. For the LV model, $\boldsymbol{\theta} = (\alpha, \beta, \gamma, \delta)$. For the GLM, $\boldsymbol{\theta} = (r, p, K)$. For the SEIUR model, $\boldsymbol{\theta} = (\beta_0, \beta_1, q_1, \rho, \kappa, \gamma_1)$. Both estimation approaches assume a normal error structure to ensure the comparability of results.

\subsection{Bayesian inference}

Bayesian inference integrates prior knowledge with observed data to obtain the posterior distribution of the model's parameters \cite{vanDeSchoot2021}. This probabilistic framework is particularly effective when prior information is available or when data are sparse, noisy, or subject to partial observability, as it provides comprehensive uncertainty quantification through posterior distributions \cite{grinsztajn2021bayesian}.

\subsubsection{Bayes' rule}

According to Bayes' theorem, the posterior distribution of the parameters, given the data, is proportional to the product of the prior distribution and the likelihood:
\begin{equation}
	\label{eq:bayes}
	p(\boldsymbol{\theta} \mid Y) \propto p(\boldsymbol{\theta}) \, p(Y \mid \boldsymbol{\theta}),
\end{equation}
where $p(\boldsymbol{\theta})$ is the prior distribution encoding existing knowledge about the parameters before observing the data, $p(Y \mid \boldsymbol{\theta})$ is the likelihood function representing the probability of observing the data given the parameter values, and $p(\boldsymbol{\theta} \mid Y)$ is the posterior distribution of parameters after incorporating the observed data.

\subsubsection{Likelihood function}

We assume that observation errors are independent and identically distributed (i.i.d.) following a normal distribution with constant variance. For the LV model with observed prey and predator populations, the likelihood is:
\begin{equation}
	y_{t_j}^{\text{prey}} \mid \boldsymbol{\theta} \sim \mathcal{N}(x_{\boldsymbol{\theta}}(t_j), \sigma_{\text{prey}}^2), \quad
	y_{t_j}^{\text{predator}} \mid \boldsymbol{\theta} \sim \mathcal{N}(y_{\boldsymbol{\theta}}(t_j), \sigma_{\text{predator}}^2),
\end{equation}
independently for $j = 1, \ldots, n$, where $x_{\boldsymbol{\theta}}(t)$ and $y_{\boldsymbol{\theta}}(t)$ denote the prey and predator population solutions to the LV ODE system given the parameters $\boldsymbol{\theta}$, and $\sigma_{\text{prey}}^2$ and $\sigma_{\text{predator}}^2$ are the observation error variances.

For the GLM, the likelihood is based on incident cases (the time derivative of cumulative cases):
\begin{equation}
	y_{t_j} \mid \boldsymbol{\theta} \sim \mathcal{N}\left( \frac{dC_{\boldsymbol{\theta}}(t_j)}{dt}, \sigma^2 \right), \quad j = 1, \ldots, n,
\end{equation}
where $C_{\boldsymbol{\theta}}(t)$ is the solution to the GLM ODE given the parameters $\boldsymbol{\theta}$.

For the SEIUR model, the likelihood is based on incident reported cases:
\begin{equation}
	y_{t_j} \mid \boldsymbol{\theta} \sim \mathcal{N}\left( \kappa \rho E_{\boldsymbol{\theta}}(t_j), \sigma^2 \right), \quad j = 1, \ldots, n,
\end{equation}
where $E_{\boldsymbol{\theta}}(t)$ is the exposed compartment solution to the SEIUR ODE system given parameters $\boldsymbol{\theta}$, and $\kappa \rho E$ represents the flow of individuals from exposed to reported infectious status.

\subsubsection{Prior distributions}

Prior distributions encode the existing knowledge or beliefs about parameter values before observing the data \cite{greenland2009bias}. For parameters with established biological or epidemiological interpretations, we specify informative priors based on previous studies. When prior knowledge is limited, we use weakly informative or uniform priors over plausible parameter ranges. For structural identifiability analyses with simulated data, the priors were centered on true parameter values with varying degrees of precision to assess the impact of the prior strength on parameter recovery. Tables~\ref{tab:parameter_priors_bounds},~\ref{tab:parameter_ranges_GLM_mpox_BFF}, and \ref{tab:parameter_ranges_QFF} summarizes the prior distributions used for LV model. The normal error variance $\sigma^2$ is assigned a weakly informative inverse-gamma or half-Cauchy prior to allow the data to inform the observation error scale.

\subsubsection{Posterior sampling and convergence diagnostics}

Posterior distributions are approximated using MCMC sampling implemented through the Hamiltonian Monte Carlo (HMC) algorithm in Stan \cite{vehtari2021rhat, burkner2017brms}. For each model, we ran four independent chains with a sufficient number of iterations (typically 2000--4000) after a burn-in period to ensure convergence to the stationary distribution. Convergence was assessed using the Gelman--Rubin diagnostic $\hat{R}$ statistic \cite{burkner2017brms}, which compares between-chain and within-chain variance; values of $\hat{R} < 1.05$ (ideally $\hat{R} \approx 1.01$) indicate successful convergence. Effective sample size (ESS) was monitored to ensure adequate sampling of the posterior distribution. Trace plots and posterior density plots were visually inspected to confirm mixing and convergence.

Posterior summaries are reported as medians with 95\% PIs, which represent the central 95\% of the posterior distribution. For forecasting, we propagate parameter uncertainty by sampling parameters values from the posterior distribution, solving the ODE for each sample, and summarizing the resulting forecast distribution with the median and 95\%  PIs.

\subsubsection{Computational implementation}
Bayesian inference was conducted using the BFF toolbox \cite{karami2025bayesianfitforecast}, an R package designed for fitting and forecasting ODE-based epidemic models. The BFF toolbox provides an automated workflow that generates Stan code based on user-specified model structures, priors, and data inputs, eliminating the need for users to program directly in Stan. The toolbox outputs include posterior parameter distributions, convergence diagnostics ($\hat{R}$, ESS), trace plots, posterior density plots, and forecasting results with quantified uncertainty. Performance metrics including MAE, MSE, WIS, and 95\% PI coverage are automatically computed to evaluate model fit and forecast accuracy.

\subsection{Frequentist inference}

Frequentist estimation treats model parameters as fixed but unknown quantities and estimates them by optimizing an objective function based solely on the observed data, without incorporating prior information \cite{chowell2017primer, chowell2020ensembles}. Uncertainty quantification is performed through bootstrap resampling procedures.

\subsubsection{Parameter estimation}

Under the assumption of normally distributed observation errors with constant variance (consistent with the Bayesian approach), parameter estimates are obtained by minimizing the sum of squared residuals between observed data and model predictions. This corresponds to nonlinear least squares (NLS) estimation:
\begin{equation}
	\label{eq:nls}
	\hat{\boldsymbol{\theta}} = \underset{\boldsymbol{\theta}}{\text{argmin}} \sum_{j=1}^{n} \left( y_{t_j} - \mu_j(\boldsymbol{\theta}) \right)^2,
\end{equation}
where $\mu_j(\boldsymbol{\theta})$ is the model-predicted value at time $t_j$ given parameters $\boldsymbol{\theta}$. For the LV model, $\mu_j(\boldsymbol{\theta})$ consists of predicted prey and predator populations $(x_{\boldsymbol{\theta}}(t_j), y_{\boldsymbol{\theta}}(t_j))$. For the GLM, $\mu_j(\boldsymbol{\theta}) = dC_{\boldsymbol{\theta}}(t_j)/dt$. For the SEIUR model, $\mu_j(\boldsymbol{\theta}) = \kappa \rho E_{\boldsymbol{\theta}}(t_j)$.

In our study, the frequentist inference is performed using MATLAB’s fmincon with the sequential quadratic programming (SQP) algorithm. To mitigate sensitivity to the initial conditions and reduce the risk of convergence to local minima, the optimization is wrapped within a MultiStart framework using a CustomStartPointSet, with an optional RandomStartPointSet.

The observed data variance is estimated as follows:
\begin{equation}
\hat{\sigma}^2 = \frac{1}{n} \sum_{j=1}^{n} \left( y_{t_j} - \mu_j(\hat{\boldsymbol{\theta}}) \right)^2.
\end{equation}

\subsubsection{Uncertainty quantification via parametric bootstraping}

To quantify parameter uncertainty and construct PIs, we use a parametric bootstrap procedure \cite{chowell2017primer, banks2014uncertainty}. This approach generates synthetic datasets by resampling from the fitted model and re-estimates the parameters for each synthetic dataset. The distribution of bootstrap parameter estimates characterizes the sampling variability. The steps are as follows.

\begin{enumerate}
	\item[\rm{1)}] \textbf{Generate bootstrap samples:} For each bootstrap replicate $b = 1, \ldots, B$ (typically $B = 200$--$500$), generate a synthetic dataset $\{y_{t_1}^b, \ldots, y_{t_n}^b\}$ by sampling from the fitted model:
	\begin{equation}
		y_{t_j}^b \sim \mathcal{N}\left( \mu_j(\hat{\boldsymbol{\theta}}), \hat{\sigma}^2 \right), \quad j = 1, \ldots, n,
	\end{equation}
	where $\hat{\boldsymbol{\theta}}$ and $\hat{\sigma}^2$ are the parameter estimates from the original data.
	
	\item[\rm{2)}] \textbf{Re-estimate parameters:} For each bootstrap sample $\{y_{t_1}^b, \ldots, y_{t_n}^b\}$, solve the NLS optimization problem (Eq~(\ref{eq:nls})) to obtain the bootstrap parameter estimates $\hat{\boldsymbol{\theta}}^b$.
	
	\item[\rm{3)}] \textbf{Construct  PIs:} Use the empirical distribution of $\{\hat{\boldsymbol{\theta}}^b : b = 1, \ldots, B\}$ to construct  PIs. For example, the 95\%  PI for parameter $\theta_k$ is given by the 2.5th and 97.5th percentiles of $\{\hat{\theta}_k^b : b = 1, \ldots, B\}$.
\end{enumerate}

For forecasting, we propagate parameter uncertainty by solving the ODE model for each bootstrap parameter estimate $\hat{\boldsymbol{\theta}}^b$ to obtain the forecast trajectories $\hat{y}^b(t_{n+h})$ for $h$-step-ahead predictions. The 95\% PI for $y(t_{n+h})$ is constructed from the 2.5th and 97.5th percentiles of $\{\hat{y}^b(t_{n+h}) : b = 1, \ldots, B\}$.

\subsubsection{Computational implementation}
Frequentist inference was implemented using the QuantDiffForecast (QDF) MATLAB toolbox \cite{chowell2024quantdiffforecast}, which provides a comprehensive framework for parameter estimation, uncertainty quantification, and forecasting for ODE models. The QDF toolbox supports multiple optimization algorithms, flexible error structures (normal, Poisson, negative binomial), user-defined ODE systems, and automated bootstrap-based uncertainty quantification. The toolbox outputs parameter estimates with  PIs, fitted model trajectories, forecast distributions, and performance metrics (MAE, MSE, WIS,~95\% PI coverage).

\subsection{Performance metrics}

To evaluate and compare the performance of Bayesian and frequentist inference methods, we computed four complementary metrics that assess both their point forecast accuracy and uncertainty calibration \cite{gneiting2007scoring, bracher2021evaluating}. Let $t_i$ for $i = 1, \ldots, N$ denote the observation times, $y_{t_i}$ the observed data, $f(t_i, \hat{\boldsymbol{\theta}})$ the model prediction at time $t_i$ using estimated parameters $\hat{\boldsymbol{\theta}}$, and $N$ the number of observations in the evaluation period (either calibration or forecasting). The metrics are defined as follows.

\subsubsection{Mean absolute error}

MAE measures the average absolute deviation between model predictions and observed data:
\begin{equation}
	\label{eq:mae}
	\text{MAE} = \frac{1}{N} \sum_{i=1}^{N} \left| f(t_i, \hat{\boldsymbol{\theta}}) - y_{t_i} \right|.
\end{equation}
MAE provides a direct measure of forecast accuracy on the original data scale. Lower MAE values indicate better point forecast performance. MAE is less sensitive to outliers compared with MSE.

\subsubsection{Mean squared error}

MSE measures the average squared deviation between the model's predictions and the observed~data
\begin{equation}
	\label{eq:mse}
	\text{MSE} = \frac{1}{N} \sum_{i=1}^{N} \left( f(t_i, \hat{\boldsymbol{\theta}}) - y_{t_i} \right)^2.
\end{equation}
MSE penalizes larger errors more heavily than MAE due to the squaring operation, making it more sensitive to outliers. Lower MSE values indicate a better fit. The square root of MSE (RMSE) is sometimes reported to return the metric to the original data scale.

\subsubsection{Coverage of the 95\%  PI}

The 95\% PI coverage quantifies the proportion of observed data points that fall within the 95\%  PI, providing a measure of uncertainty calibration:
\begin{equation}
	\label{eq:coverage}
	\text{95\% PI Coverage} = \frac{1}{N} \sum_{i=1}^{N} \mathbf{1}(L_{t_i} < y_{t_i} < U_{t_i}),
\end{equation}
where $L_{t_i}$ and $U_{t_i}$ are the lower and upper bounds of the 95\%  PI at time $t_i$, and $\mathbf{1}(\cdot)$ is an indicator function equal to 1 if the condition is true and 0 otherwise. Ideally, the coverage should be close to 95\%. Coverage significantly below 95\% indicates overly narrow intervals (underestimation of uncertainty), while coverage above 95\% suggests overly wide intervals (overestimation of uncertainty).

\subsubsection{Weighted interval score}

WIS is a proper scoring rule that evaluates the quality of the entire predictive distribution by combining sharpness (interval width) and calibration (penalties for observations outside intervals) \cite{gneiting2007scoring, bracher2021evaluating}. The interval score (IS) for a single  PI at level $\alpha$ is:
\begin{equation}
	\label{eq:IS}
	\text{IS}_{\alpha}(F, y) = (u - l) + \frac{2}{\alpha} (l - y) \mathbf{1}(y < l) + \frac{2}{\alpha} (y - u) \mathbf{1}(y > u),
\end{equation}
where $l$ and $u$ represent the $\frac{\alpha}{2}$ and $\left(1 - \frac{\alpha}{2}\right)$ quantiles of the forecast distribution $F$, respectively. The IS consists of the following three components:
\begin{itemize}
	\item \textbf{Sharpness:} $(u - l)$, which is the width of the central $(1 - \alpha) \times 100\%$  PI. Narrower intervals receive lower scores, rewarding precise forecasts.
	\item \textbf{Underprediction penalty:} $\frac{2}{\alpha} (l - y) \mathbf{1}(y < l)$, penalizes observations falling below the lower bound $l$, with the penalty proportional to the distance $(l - y)$.
	\item \textbf{Overprediction penalty:} $\frac{2}{\alpha} (y - u) \mathbf{1}(y > u)$, which penalizes observations exceeding the upper bound $u$, with the penalty proportional to the distance $(y - u)$.
\end{itemize}

To comprehensively evaluate the full predictive distribution, we compute the WIS by averaging the interval scores over multiple interval levels $(1 - \alpha_1) < (1 - \alpha_2) < \cdots < (1 - \alpha_K)$ along with the predictive median $\tilde{y}$ (which can be viewed as an interval at the level $(1 - \alpha_0) \to 0$):
\begin{equation}
	\label{eq:wis}
	\text{WIS}_{\alpha_{0:K}}(F, y) = \frac{1}{K + \frac{1}{2}} \left( w_0 |y - \tilde{y}| + \sum_{k=1}^{K} w_k \, \text{IS}_{\alpha_k}(F, y) \right),
\end{equation}
where $w_k = \frac{\alpha_k}{2}$ for $k = 1, \ldots, K$ and $w_0 = \frac{1}{2}$. The WIS quantifies how close the entire predictive distribution is to the observed data in units on the original data scale. Lower WIS values indicate better forecast performance, balancing sharpness and calibration.

\subsubsection{Interpretation}

Lower values of MAE, MSE, and WIS indicate better model performance. For 95\% PI coverage, values close to 95\% indicate well-calibrated uncertainty estimates. Coverage values substantially above 95\% indicate overconservative uncertainty intervals, whereas values substantially below 95\% indicate underestimation of the uncertainty. Together, these metrics provide a comprehensive assessment of both point forecast accuracy (MAE, MSE), distributional forecast quality (WIS), and uncertainty calibration (95\% PI coverage). We report these metrics separately for the calibration period (in-sample fit) and forecasting period (out-of-sample prediction) to distinguish between model fitting and predictive performance.

\section{Structural identifiability}
\label{sec:SI}

Structural identifiability (SI) analysis determines whether a model's parameters can, in principle, be uniquely recovered from perfect, noise-free observations of the system's outputs, given the model's structure and observation scheme \cite{raue2009profilelikelihood, stigter2023identifiability}. A parameter is \textit{structurally identifiable} if its value can be uniquely determined from the input output relationship defined by the model equations and the set of observable variables. Conversely, a parameter is \textit{structurally unidentifiable} if multiple distinct parameter values produce identical model outputs, making unique parameter recovery impossible regardless of the data quality or quantity of the data.

SI is a prerequisite for meaningful parameter estimation: If a parameter is structurally unidentifiable, no amount of data or sophisticated inference methods can recover its true value \cite{chis2011structural, raue2009profilelikelihood}. SI analysis is particularly important in compartmental models where only a subset of state variables is observed, as partial observability often leads to identifiability deficits \cite{dankwa2022structural}
. Understanding which parameters are identifiable under different observation scenarios guides experimental design, informs prior specification in Bayesian inference, and helps interpret the estimation's results.

\subsection{Methodology}

We conducted symbolic structural identifiability analysis using \textit{StructuralIdentifiability.jl} \cite{stigter2023identifiability}, a Julia package that applies differential algebra methods to determine the identifiability of parameters in ODE models. For each model, we analyzed two scenarios: (1) Known initial conditions, where the initial values of all state variables are assumed to be known exactly, and (2) unknown initial conditions, where the initial conditions are treated as additional unknown parameters to be estimated. The distinction between known and unknown initial conditions is critical, as the identifiability results can differ substantially depending on whether initial conditions must be inferred from the data \cite{walter1997identification}.

For the (LV) model, we examined three observation schemes to assess how partial observability affects identifiability.
\begin{itemize}
	\item \textbf{LV-1 (both prey and predator observed):} Both the $x(t)$ (prey) and $y(t)$ (predator) populations are observed.
	\item \textbf{LV-2 (predator only):} Only the $y(t)$ (predator) is observed; prey population $x(t)$ is unobserved.
	\item \textbf{LV-3 (prey only):} Only the $x(t)$ (prey) is observed; predator population $y(t)$ is unobserved.
\end{itemize}

For the GLM, we analyzed the identifiability when observing incident cases (the time derivative of cumulative cases $C(t)$). For the SEIUR model, we considered the scenario where only incident reported cases (the flow $\kappa \rho E(t)$) are observed, representing typical epidemic surveillance data where only a fraction of infections are detected and reported.

\subsection{Results}

Table~\ref{tab:identifiability_all_models} summarizes the structural identifiability results for all models under different observation schemes and initial condition assumptions.

\begin{table}[H]
	\centering
	\caption{Structural identifiability results for all models under different observation schemes. The table indicates which parameters and state variables are identifiable or unidentifiable when initial conditions (ICs) are known versus when they are unknown. LV-1, LV-2, and LV-3 refer to LV scenarios with both species observed, predator only, and prey only, respectively. ``NULL'' indicates that no states or parameters are unidentifiable.}\vspace{0.2cm} 
	\resizebox{\textwidth}{!}{%
		\begin{tabular}{llllll}
			\hline
			Model & Observations & Identifiable (unknown IC) & Unidentifiable (unknown IC) & Identifiable (known IC) & Unidentifiable (known IC) \\ \hline
			
			LV-1  & $x(t),y(t)$       
			& $\alpha,\beta,\delta,\gamma,x(t),y(t)$ 
			& -- 
			& All states and parameters 
			& -- \\ 
			
			LV-2  & $y(t)$ only       
			& $\alpha,\beta,\gamma,y(t)$             
			& $\delta,x(t)$ 
			& All states and parameters 
			& -- \\ 
			
			LV-3  & $x(t)$ only       
			& $\alpha,\delta,\gamma,x(t)$            
			& $\beta,y(t)$  
			& All states and parameters 
			& -- \\ 
			
			GLM   & $C(t),\, dC/dt$  
			& $C,\, r,\, K,\, p$ 
			& -- 
			& All states and parameters 
			& -- \\
			
			SEIUR & $\kappa \rho E(t)$     
			& $C,\, \kappa,\, \gamma_1$ 
			& $S,\, E,\, I,\, U,\, R,\, N,\, \rho$ 
			& All states and parameters 
			& -- \\ \hline
			
		\end{tabular}%
	}
	\label{tab:identifiability_all_models}
\end{table}

\subsection{Interpretation}

\paragraph{Lotka--Volterra model.}
When both the predator and prey time series are observed (LV-1), all model parameters ($\alpha, \beta, \gamma, \delta$) are structurally identifiable regardless of whether the initial conditions are known or unknown. The coupled dynamics of the two species provide sufficient information to uniquely determine all interaction parameters.

With partial observation, identifiability losses emerge when the initial conditions are unknown. When only the predator is observed (LV-2), the prey's growth-from-consumption parameter $\delta$ and the prey population's trajectory $x(t)$ become unidentifiable. The observed predator dynamics constrain only the product $\delta x(t)$ rather than $\delta$ and $x(t)$ separately, creating a structural non-uniqueness. Similarly, when only the prey is observed (LV-3), the predation rate $\beta$ and the predator population $y(t)$ are unidentifiable because the prey dynamics constrain only the product $\beta y(t)$.

Importantly, when the initial conditions are known, all parameters become identifiable even under partial observation scenarios (LV-2 and LV-3). Knowledge of the initial prey and predator populations breaks the symmetry and allows unique recovery of all parameters. This highlights the value of accurate initial condition estimates in ecological field studies.

\paragraph{Generalized logistic model.}
For the GLM, observation of cumulative cases $C(t)$ and incident cases $dC/dt$ yields the full structural identifiability of all parameters $(r, p, K)$ under both known and unknown initial conditions. The single-equation structure of the GLM, combined with direct observation of both the state variable and its derivative, ensures that all growth parameters are uniquely determined. This robust identifiability makes the GLM a reliable choice for phenomenological epidemic modeling when the full case trajectory is observed.

\paragraph{SEIUR model.}
The SEIUR model exhibits substantial identifiability challenges due to its multi-compartment structure and partial observability. Even when all compartments $(S, E, I, U, R, C)$ are theoretically observable, only the cumulative reported cases $C$, the incubation rate $\kappa$, and the recovery rate $\gamma_1$ remain structurally identifiable when the initial conditions are unknown. The remaining state variables ($S, E, I, U, R$) and critical parameters including the total population size $N$ and the reporting proportion $\rho$ are structurally unidentifiable.

This identifiability deficit arises because the observed incidence data (new reported cases per day) depend on the product $\kappa \rho E(t)$ rather than on $\kappa$, $\rho$, and $E(t)$ individually. Multiple combinations of the reporting proportion $\rho$, the exposed population $E(t)$, and the total population $N$ can produce identical observed case counts. The unobserved unreported infectious compartment $U(t)$ further compounds the identifiability problem, as the split between reported and unreported infections is not directly constrained by the data.

When the initial conditions are known, all SEIUR parameters become identifiable. This underscores the importance of accurate estimates of the initial susceptible and exposed populations (e.g., from seroprevalence surveys or contact tracing data) for reliable parameter inference in epidemic models with under-reporting.

\subsection{Implications for inference}

The structural identifiability analysis provides critical context for interpreting the empirical parameter estimates and forecasting results presented in the Results section. For the LV model, we expect both the Bayesian and frequentist methods to successfully recover all parameters when both species are observed (LV-1), but the parameter estimates may be unreliable or exhibit high uncertainty in the partial observation scenarios (LV-2, LV-3), where the structural identifiability is compromised.

For the GLM applied to lung injury and mpox data, the robust structural identifiability suggests that both inference methods should yield well-constrained parameter estimates and reliable forecasts, provided that the data quality is sufficient.

For the SEIUR model applied to COVID-19 data in Spain, the limited structural identifiability indicates that several model parameters cannot be uniquely determined from case incidence data alone. We expect the parameter estimates for $\rho$, $N$, and the latent compartments to exhibit substantial uncertainty and potential nonuniqueness. In this setting, Bayesian inference may have an advantage by incorporating prior information to regularize the otherwise ill-posed inverse problem, while frequentist methods may struggle without additional constraints or data sources.

These predictions will be assessed empirically in the Results section, where we compare the accuracy of parameter estimation, uncertainty quantification, and forecast performance across models and observation scenarios.

\subsection{Summary of models, domains, data sources, and observation scenarios}
As summarized in Table~\ref{tab:models-summary}, the models, application domains, data sources,
and observation scenarios used in this study are presented.

\begin{table}[H]
	\centering
	\caption{The models, application domains, data sources, and observation scenarios used in this study. The datasets span ecological, clinical, and epidemiological systems.}\vspace{0.2cm} 
	\begin{tabular}{p{3cm}p{3cm}p{5cm}p{4cm}}
		\hline
		Model & Domain & Data source & Scenarios \\ \hline
		LV & Ecology & Hudson--Bay lynx--hare population & predator + prey, prey only, predator only \\ 
		GLM & Clinical & Lung injury progression & Full data \\ 
		GLM & Epidemiology & 2022 US\ mpox outbreak & Full data \\ 
		SEIUR & Epidemiology & COVID-19 incidence in Spain & First-wave data \\ \hline
	\end{tabular}
	\label{tab:models-summary}
\end{table}

\section{Results}
\label{sec:results}

This section reports the parameter estimates, calibration performance metrics, fitting visualization, and Bayesian convergence diagnostics across the models. We organize results by model and, when applicable, by observation scenario.

\subsection{Lotka--Volterra model} 

We fitted the LV model to the Hudson Bay lynx hare data under three observation scenarios: (1) Both prey and predator observed (LV-1), (2) predator only (LV-2), and (3) prey only (LV-3). These scenarios allow us to assess how partial observability affects parameter recovery and forecasting performance in light of the structural identifiability results presented in Section~\ref{sec:SI}.

\subsubsection{LV-1: Joint predator--prey observations}

\begin{figure}[H]
\centering
\includegraphics[width=0.8\linewidth]{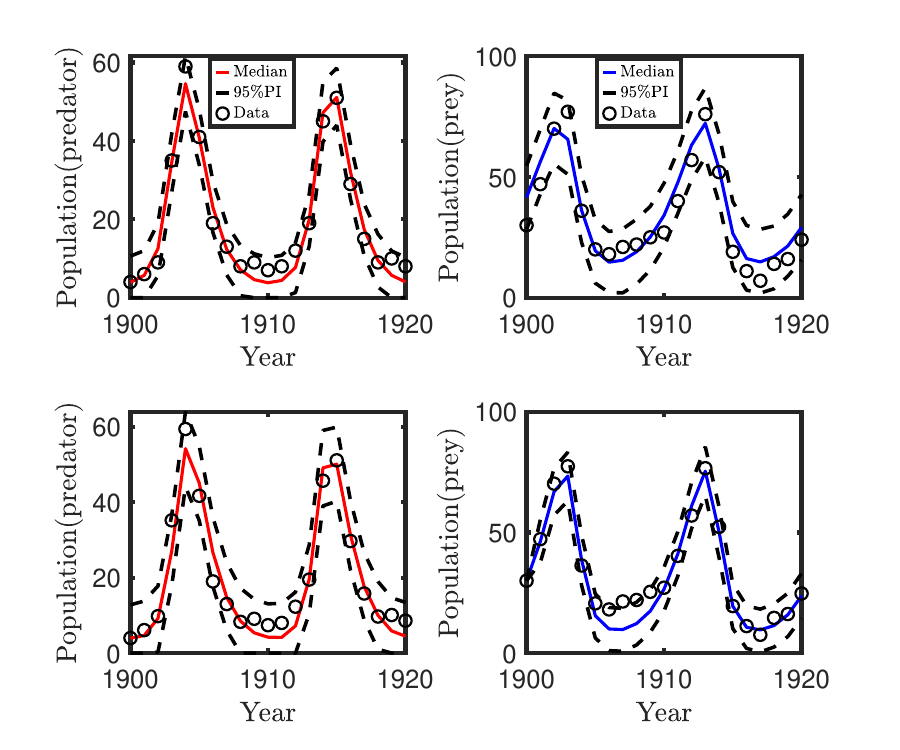}  \vspace{0.5cm} 
\caption{Fitting visualization for the LV model, with the Hudson Bay lynx hare data, observing both the predator and prey, using the QDF and BFF approaches. \textbf{Top row:} BFF. \textbf{Bottom row:} QDF. \textbf{Left column:} Predator. \textbf{Right column:} Prey.}
\label{fig:1}
\end{figure}

As shown in Figure~\ref{fig:1}, with both time series available, both methods recover the characteristic coupled oscillations with well-calibrated uncertainty. However, BFF provides better coverage of the prey data for 1905--1910. the parameter estimates are almost similar for $\beta$ (0.02 vs. 0.025) and $\delta$ (both~0.03) but differ for other parameters, as shown in Table~\ref{tab:parameter__prey__predator}.

The calibration results show a trade-off that depends on the series (Table~\ref{tab:performance___prey___predator}). For the predator series, QDF has lower errors, such as a lower MAE and MSE. For the prey series, BFF shows lower point-error metrics and 100\% PI coverage, which exceeds the nominal 95\% level and therefore suggests slightly overconservative uncertainty intervals, whereas QDF’s 90.48\% coverage is closer to the nominal target but reflects mild undercoverage. Table~\ref{tab:convergence__prey__predator} also shows that the Bayesian chains mixed well, with $\hat{R}$ values close to 1 for all parameters.

\begin{table}[H]
\centering
\caption{Parameter estimates for the LV model, with Hudson Bay lynx hare data, observing both predator and prey, using the QDF and BFF approaches.}\vspace{0.2cm} 
\begin{tabular}{lllll}
\hline
& \textbf{$\alpha$} & \textbf{$\beta$} & \textbf{$\delta$} & \textbf{$\gamma$} \\ \hline
BFF  
& 0.40 (0.36, 0.44) 
& 0.02 (0.02, 0.02) 
& 0.03 (0.03, 0.03) 
& 1.06 (0.95, 1.17) \\ 
QDF 
& 0.55 (0.51, 0.60)  
& 0.025 (0.022, 0.029) 
& 0.03 (0.02, 0.03)  
& 0.84 (0.76, 0.92) \\ \hline
\end{tabular}
\label{tab:parameter__prey__predator}
\end{table}

\begin{table}[H]
\centering
\caption{Performance metrics for the LV model,  with the Hudson Bay lynx hare data, observing both the predator and prey, using the QDF and BFF approaches.}\vspace{0.2cm} \setlength{\tabcolsep}{7mm} 
\begin{tabular}{llllll}
\hline
&                              & \textbf{$MAE$} & \textbf{$MSE$} & \textbf{$WIS$} & \textbf{$95\%PI$} \\ \hline
Predator & \textit{BFF} & 4.95           & 35.93          & 3.07           & 100            \\ \cline{2-6} 
& \textit{QDF}   & 2.87           & 13.56          & 1.88           & 100              \\ \hline
Prey     & \textit{BFF} & 2.24           & 8.04           & 1.46           & 100              \\ \cline{2-6} 
& \textit{QDF}   & 3.29           & 21.96          & 2.27           & 90.48             \\ \hline
\end{tabular}
\label{tab:performance___prey___predator}
\end{table}

\begin{table}[H]
\centering
\caption{Convergence diagnostic for the LV model,  with the Hudson Bay lynx hare data, observing both the predator and prey in the Bayesian approach.}\vspace{0.2cm} \setlength{\tabcolsep}{7mm} 
\begin{tabular}{lllll}
\hline
& $\alpha$ & $\beta$ & $\delta$ & $\gamma$ \\ \hline
$\textbf{$\hat{R} $}$ & 1.01     & 1.01    & 1.01     & 1.01     \\ \hline
\end{tabular}
\label{tab:convergence__prey__predator}
\end{table}

\subsubsection{LV-2: Predator-only observations}
We next fit the LV model using predator observations only. As expected under reduced observability (Section~\ref{sec:SI}), the parameter uncertainty increases and multiple parameter combinations can explain the observed predator cycles (Table~\ref{tab:parameter_predator}). Despite this, point fit performance is similar across methods (Table~\ref{tab:performance_prey_predator}): QDF has a slightly lower MAE (3.8 vs.\ 3.6 for BFF), while BFF has a slightly lower MSE and WIS. PI coverage is 100\% for BFF and 95.24\% for QDF; the latter is closer to the nominal 95\% level, while 100\% coverage indicates slightly overconservative intervals, and the Bayesian chains show $\hat{R}=1$ in this fit (Table~\ref{tab:convergence_prey_predator}).

\begin{table}[H]
\centering
\caption{Parameter estimates for the LV model,  with the Hudson Bay lynx hare data, observing only predator, using the QDF and BFF approaches.}\vspace{0.2cm}
\begin{tabular}{llllll}
\hline
&  & $\alpha$ & $\beta$ & $\delta$ & $\gamma$ \\ \hline
Predator 
& \textit{BFF}
& 1.20 (0.75, 1.90) 
& 0.03 (0.01, 0.06)  
& 0.02 (0.01, 0.03)    
& 0.36 (0.22, 0.62)     \\ \cline{2-6} 
& \textit{QDF} 
& 0.97 (0.64, 1.00)     
& 0.02 (0.02, 0.06)    
& 0.025 (0.02, 0.036)     
& 0.46 (0.40, 0.70)     \\ \hline
\end{tabular}
\label{tab:parameter_predator}
\end{table}

\begin{figure}[H]
\centering
\includegraphics[width=0.8\linewidth]{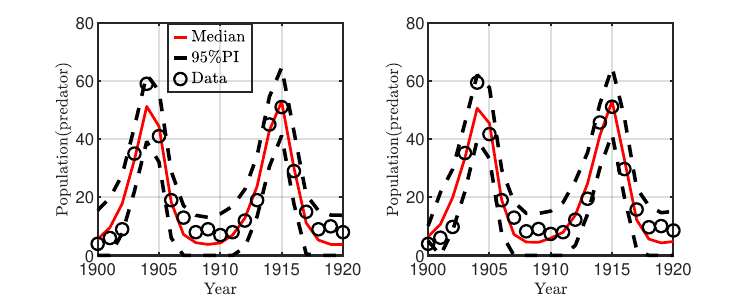}\vspace{0.5cm}
\caption{Fitting visualization for the LV model,  with the Hudson Bay lynx hare data, observing only predator, using QDF and BFF approaches. \textbf{Left:} BFF; \textbf{Right:} QDF.}
\label{fig:only_Predator}
\end{figure}

\noindent
BFF and QDF yield comparable predator-only fits (Figure~\ref{fig:only_Predator}), but the implied prey dynamics are necessarily weakly constrained without prey observations.

% (Narrative interpretation provided above; table retained for completeness.)

\begin{table}[H]
\centering
\caption{Performance metrics for the LV model,  with the Hudson Bay lynx hare data, observing only predator's, using the QDF and BFF approaches.}\vspace{0.2cm}\setlength{\tabcolsep}{7mm}
\begin{tabular}{llllll}
\hline
&                              & \textbf{$MAE$} & \textbf{$MSE$} & \textbf{$WIS$} & \textbf{$95\%PI$} \\ \hline
Predator & \textit{BFF} & 3.6           & 18.07          & 2.22           & 100            \\ \cline{2-6} 
&\textit{QDF}  & 3.8           & 19.7          & 2.33           & 95.24                         \\ \hline
\end{tabular}
\label{tab:performance_prey_predator}
\end{table}

\begin{table}[H]
\centering
\caption{Convergence diagnostic for the LV model,  with the Hudson Bay lynx hare data, observing only predator's in the Bayesian approach.}\setlength{\tabcolsep}{7mm}\vspace{0.2cm}
\begin{tabular}{llllll}
\hline
&         & $\alpha$ & $\beta$ & $\delta$ & $\gamma$ \\ \hline
Predator & 
\textbf{$\hat{R} $} & 1    & 1    & 1     & 1     \\ \hline
\end{tabular}
\label{tab:convergence_prey_predator}
\end{table}

\subsubsection{LV-3: Prey-only observations}
Finally, we fit the LV model using prey observations only. In this scenario, QDF achieves lower point-error metrics (Table~\ref{tab:performance_predator}: MAE 3.3 vs.\ 4.6; MSE 15.83 vs.\ 27.34; WIS 2.1 vs.\ 2.8), while BFF achieves 100\% PI coverage, indicating overconservative intervals relative to the nominal 95\% target, whereas QDF’s 90.48\% suggests mild undercoverage. Parameter estimates remain plausible but are less constrained than in the joint case (Table~\ref{tab:parameter_prey_predator}), and the Bayesian fit again shows $\hat{R}\approx 1.01$ (Table~\ref{tab:convergence_prey}).

\begin{table}[H]
\centering
\caption{Performance metrics for the LV model,  with the Hudson Bay lynx hare data, observing only the prey, using the QDF and BFF approaches.}
\begin{tabular}{llllll}
\hline
&                              & \textbf{$MAE$} & \textbf{$MSE$} & \textbf{$WIS$} & \textbf{$95\%PI$} \\ \hline

Prey     & \textit{BFF} & 4.6           & 27.34           & 2.8           & 100             \\ \cline{2-6} 
& \textit{QDF}   & 3.3           & 15.83          & 2.1           & 90.48             \\ \hline
\end{tabular}

\label{tab:performance_predator}
\end{table}

\begin{figure}[H]
\centering
\includegraphics[width=0.8\linewidth]{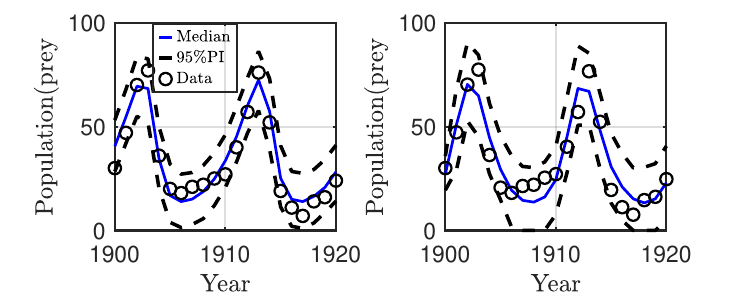}\vspace{0.5cm}
\caption{Fitting visualization for the LV model,  with the Hudson Bay lynx hare data, observing only the prey, using the QDF and BFF approaches. \textbf{Left:} BFF, \textbf{Right:} QDF.}
\label{fig:only_Prey}
\end{figure}

\noindent
Both methods track the main prey fluctuations under prey-only calibration (Figure~\ref{fig:only_Prey}), but implied predator dynamics remain uncertain without direct predator observations.

\begin{table}[H]
\centering
\caption{Parameter estimates for the LV model,  with the Hudson Bay lynx hare data, observing only the prey, using the QDF and BFF approaches.}
\begin{tabular}{llllll}
\hline
&  & $\alpha$ & $\beta$ & $\delta$ & $\gamma$ \\ \hline
Prey    
& \textit{BFF} 
& 0.34 (0.28, 0.52)    
& 0.01 (0.00, 0.05)    
& 0.04 (0.02, 0.05)     
& 1.40 (0.75, 1.90)     \\ \cline{2-6} 
& \textit{QDF} 
& 0.42 (0.34, 0.55)    
& 0.013 (0.005, 0.035)    
& 0.03 (0.02, 0.04)     
& 1.20 (0.78, 1.50)      \\ \hline
\end{tabular}
\label{tab:parameter_prey_predator}
\end{table}
Table~\ref{tab:convergence_prey} displays the Gelman Rubin diagnostic ($\hat{R}$) values for BFF, all of which are 1.01, indicating successful MCMC convergence.
\begin{table}[H]
\centering
\caption{Convergence diagnostic for the LV model, observing only the prey in the Bayesian~approach.}\setlength{\tabcolsep}{7mm}\vspace{0.2cm}
\begin{tabular}{llllll}
\hline
&         & $\alpha$ & $\beta$ & $\delta$ & $\gamma$ \\ \hline
Prey     & 
\textbf{$\hat{R} $} & 1.01     & 1.01    & 1.01     & 1.01     \\ \hline
\end{tabular}
\label{tab:convergence_prey}
\end{table}
\noindent

\begin{figure}[H]
\centering    
\includegraphics[width=1\textwidth]{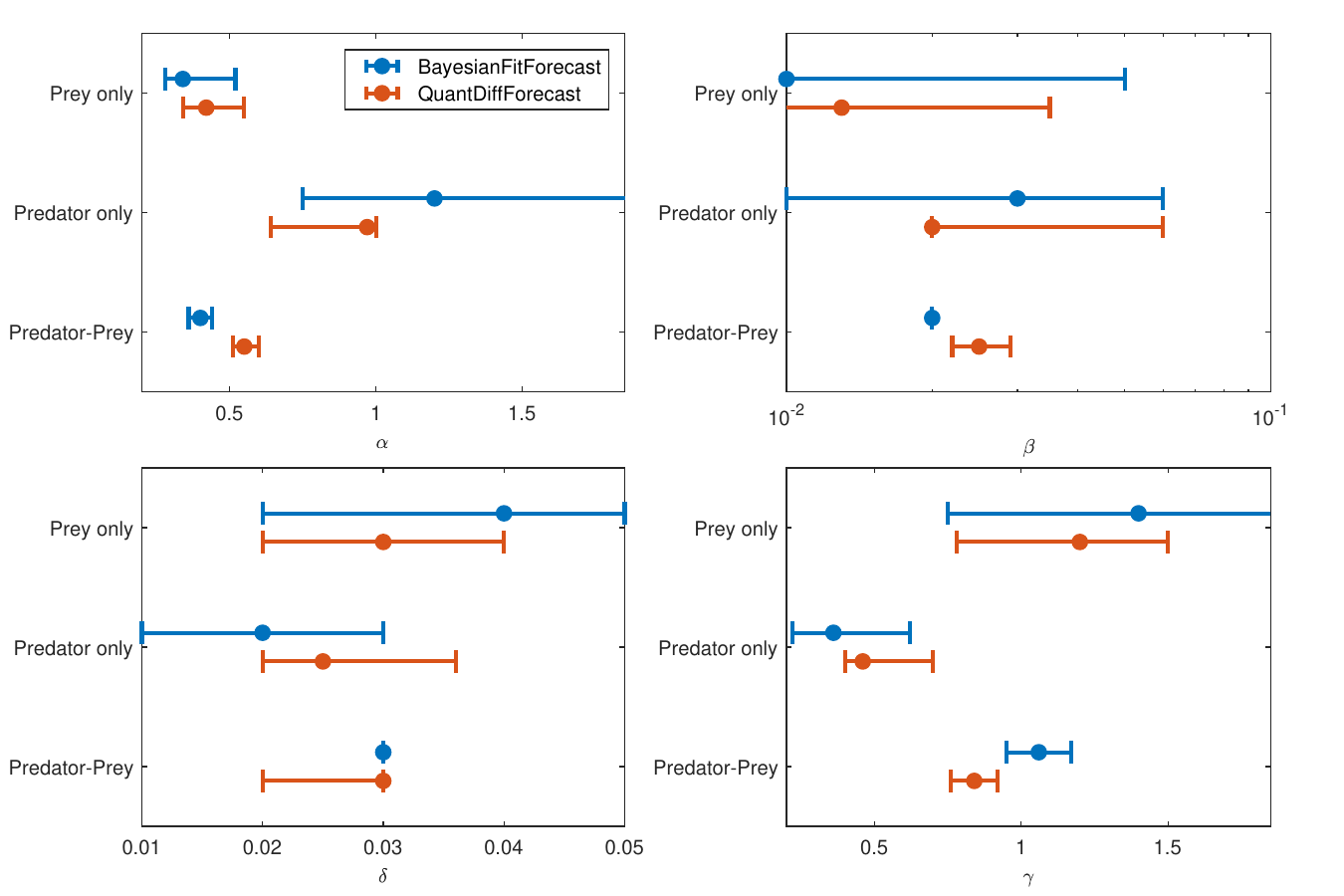}\vspace{0.5cm}
\caption{Parameter comparison for the LV model with the Hudson Bay lynx hare data across observation schemes: joint predator--prey, predator only, and prey only. We use a log scale for the $x$-axis in the $\beta$ panel.}

\label{fig:LV_parameter estimates}
\end{figure}

Across the three observation scenarios, parameter estimates and interval widths vary in a parameter-specific manner (Figure~\ref{fig:LV_parameter estimates}). Joint predator--prey observation primarily affects interaction-related parameters, most notably $\beta$, for which the point estimates differ substantially across inference methods, reflecting stronger parameter coupling and improved identifiability. By contrast, for other the parameters ($\alpha$, $\delta$, and $\gamma$), the  interval widths and estimates overlap considerably across observation scenarios, indicating that joint observation does not uniformly reduce uncertainty relative to single-series calibration.

\begin{figure}[H]
	\centering
	\includegraphics[width=0.8\textwidth]{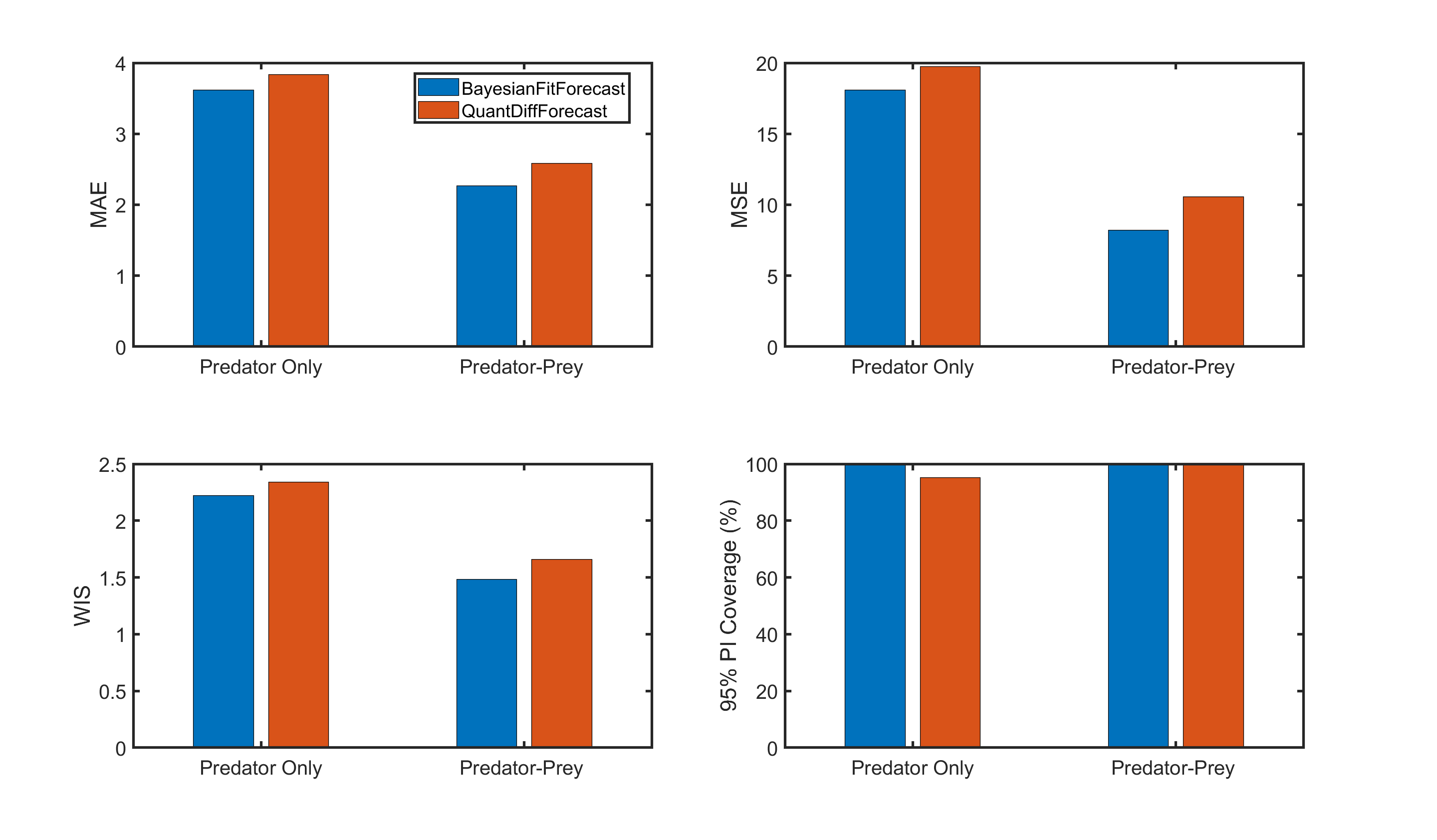}\vspace{0.5cm}
	\caption{Predictive accuracy vs. data availability in the LV. Across the MAE, MSE, WIS, and 95\% PI, joint predator–prey observation improves the performance for both frameworks compared with predator-only data.}
	
	\label{fig:parameters_1}
\end{figure}

Comparing predator-only vs.\ joint data availability, adding prey observations improves the  calibration performance for both methods across MAE, MSE, WIS and PI coverage (Figure~\ref{fig:parameters_1}), underscoring the value of joint observations for dynamic ecological systems.

\begin{figure}[H]
\centering
\includegraphics[width=0.8\textwidth]{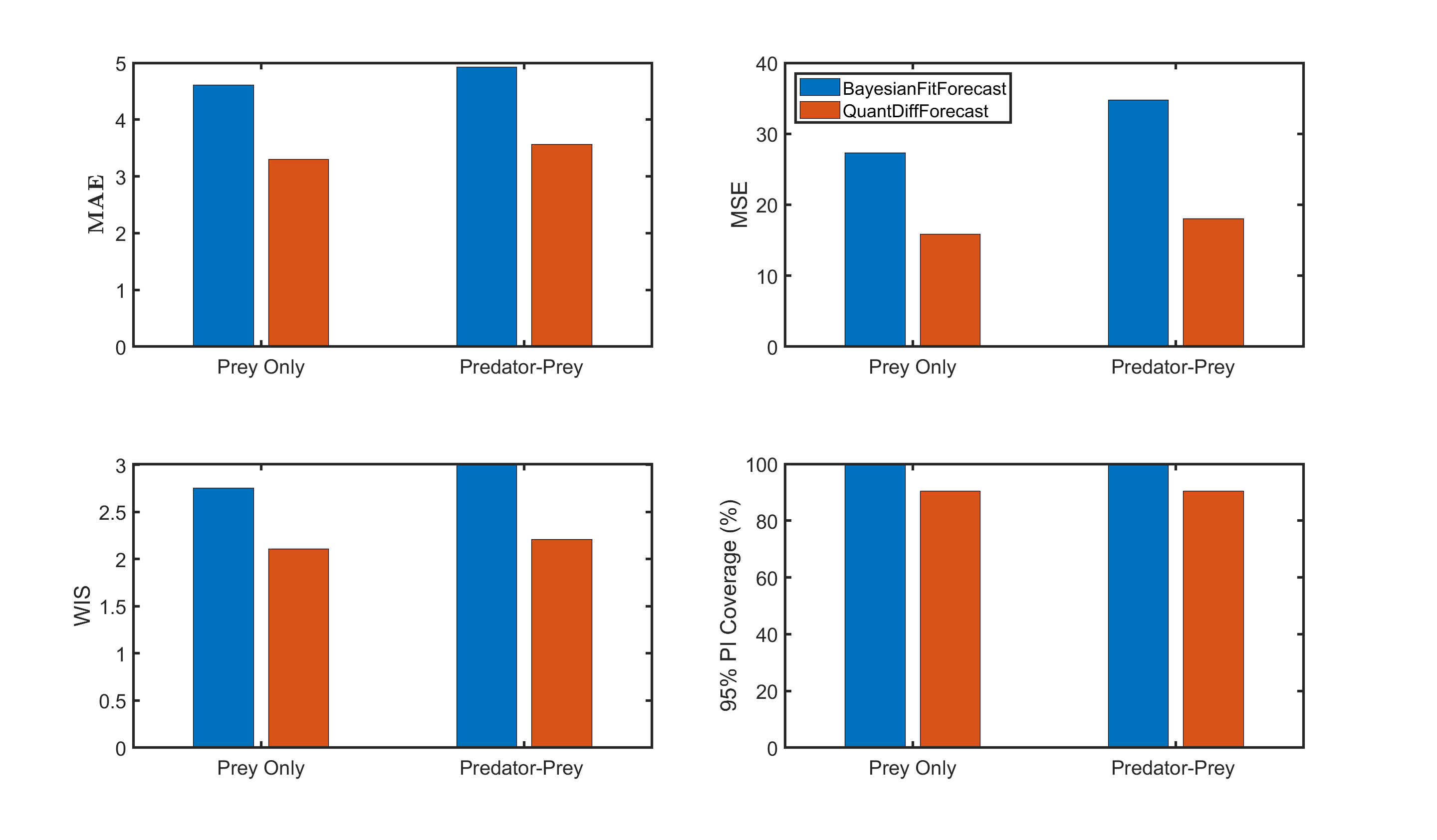}\vspace{0.5cm}
\caption{Adding predator data to the prey-only scenario slightly increases the MAE, MSE, and WIS, while maintaining similar 95\% PI coverage for both BFF and QDF relative to prey-only fits.}
\label{fig:parameters_2}
\end{figure}

Adding predator observations to a prey-only fit slightly increases the MAE, MSE, and WIS, while maintaining similar 95\% PI coverage for both frameworks (Figure~\ref{fig:parameters_2}).

\begin{figure}[H]
	\begin{center}
		\includegraphics[width=0.25\textwidth]{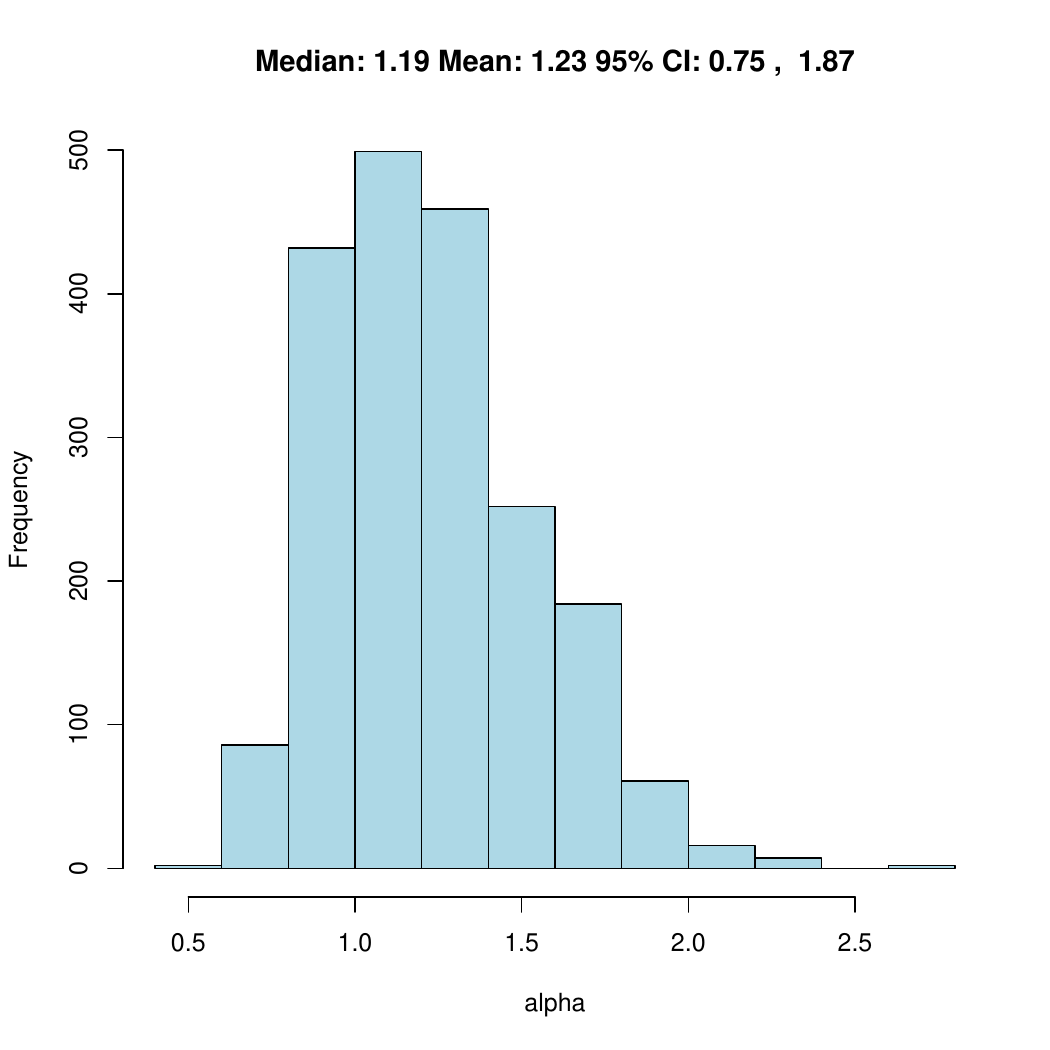}
		\includegraphics[width=0.25\textwidth]{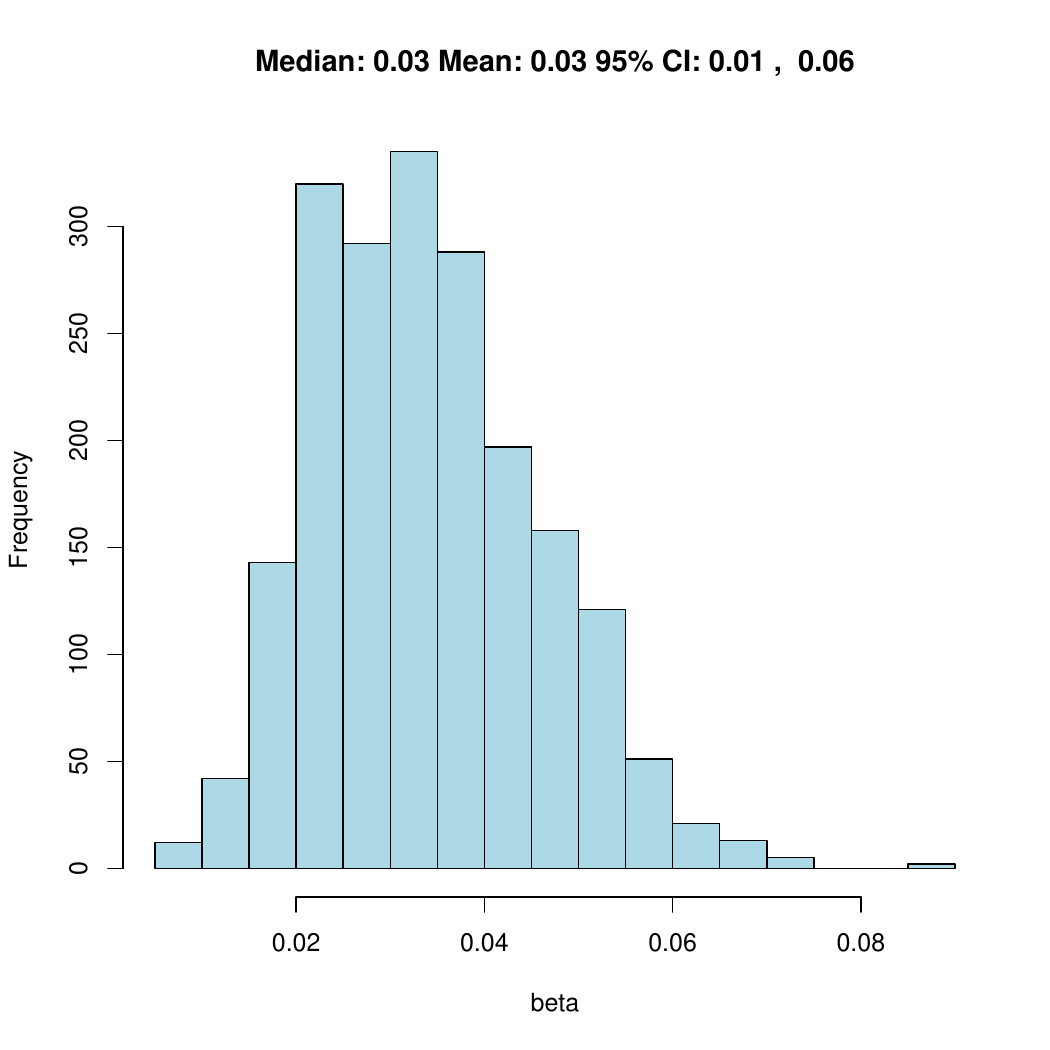}\\
		\includegraphics[width=0.25\textwidth]{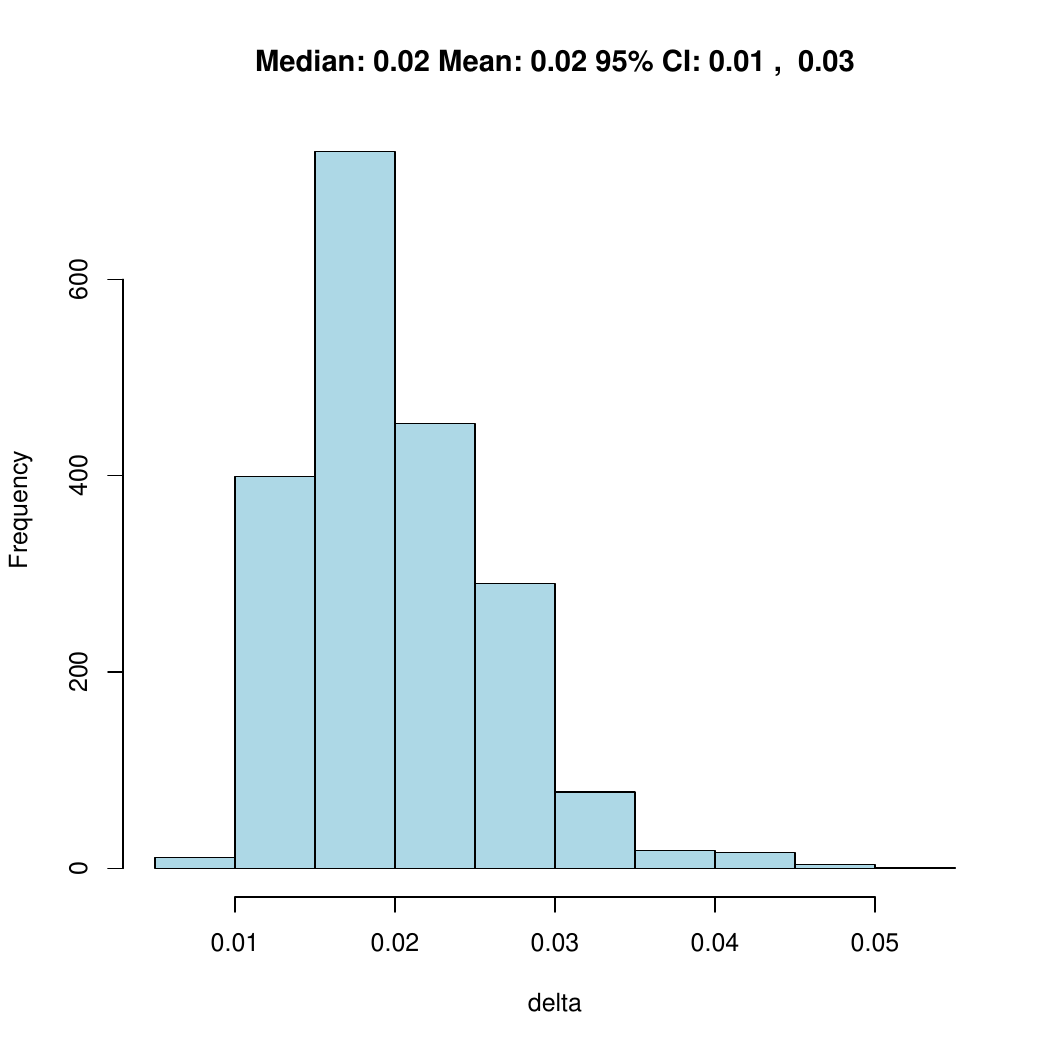}
		\includegraphics[width=0.25\textwidth]{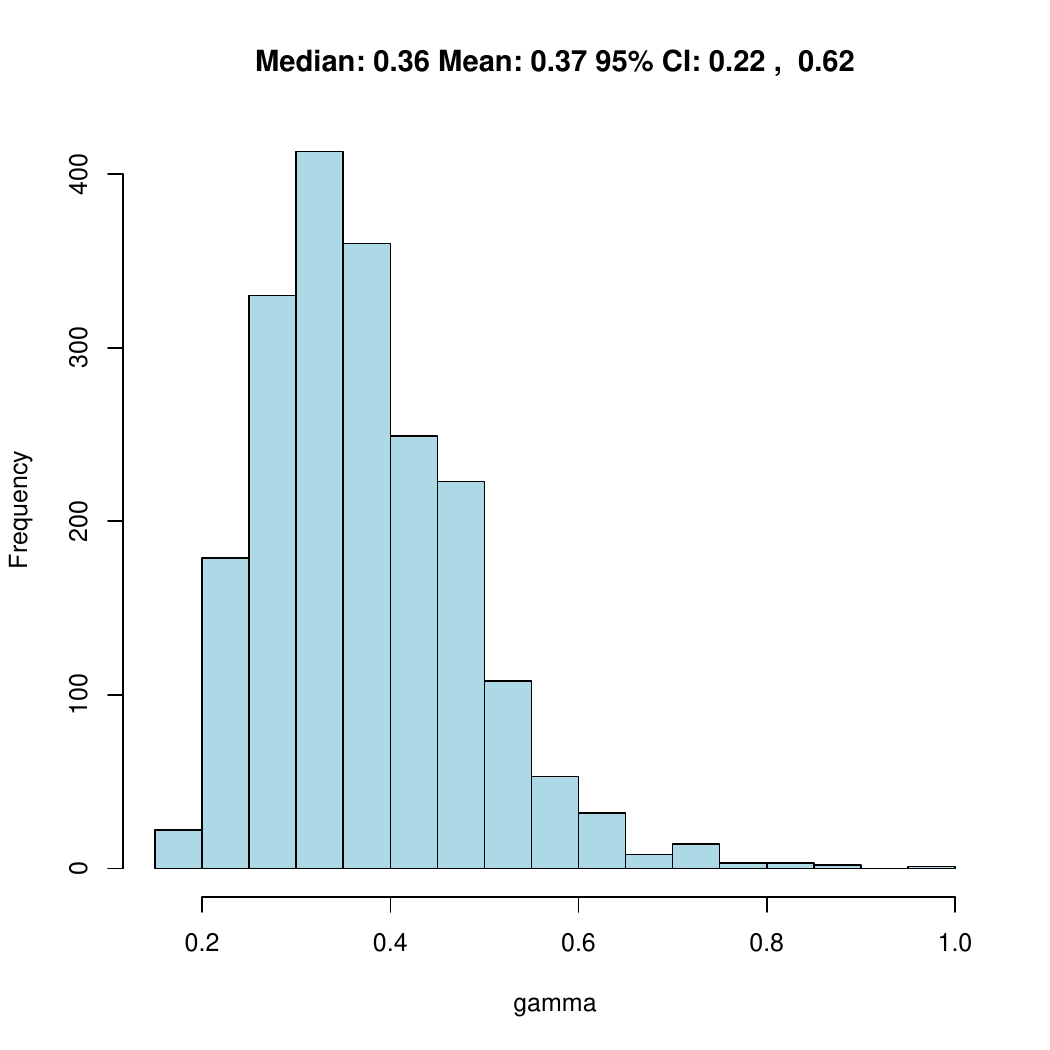}
		\caption{Histograms of the LV parameters $(\alpha, \beta, \delta, \gamma)$.}
		\label{fig:parameters_bff_both}
	\end{center}
\end{figure}

\begin{figure}[H]
	\centering
	\includegraphics[width=0.28\textwidth]{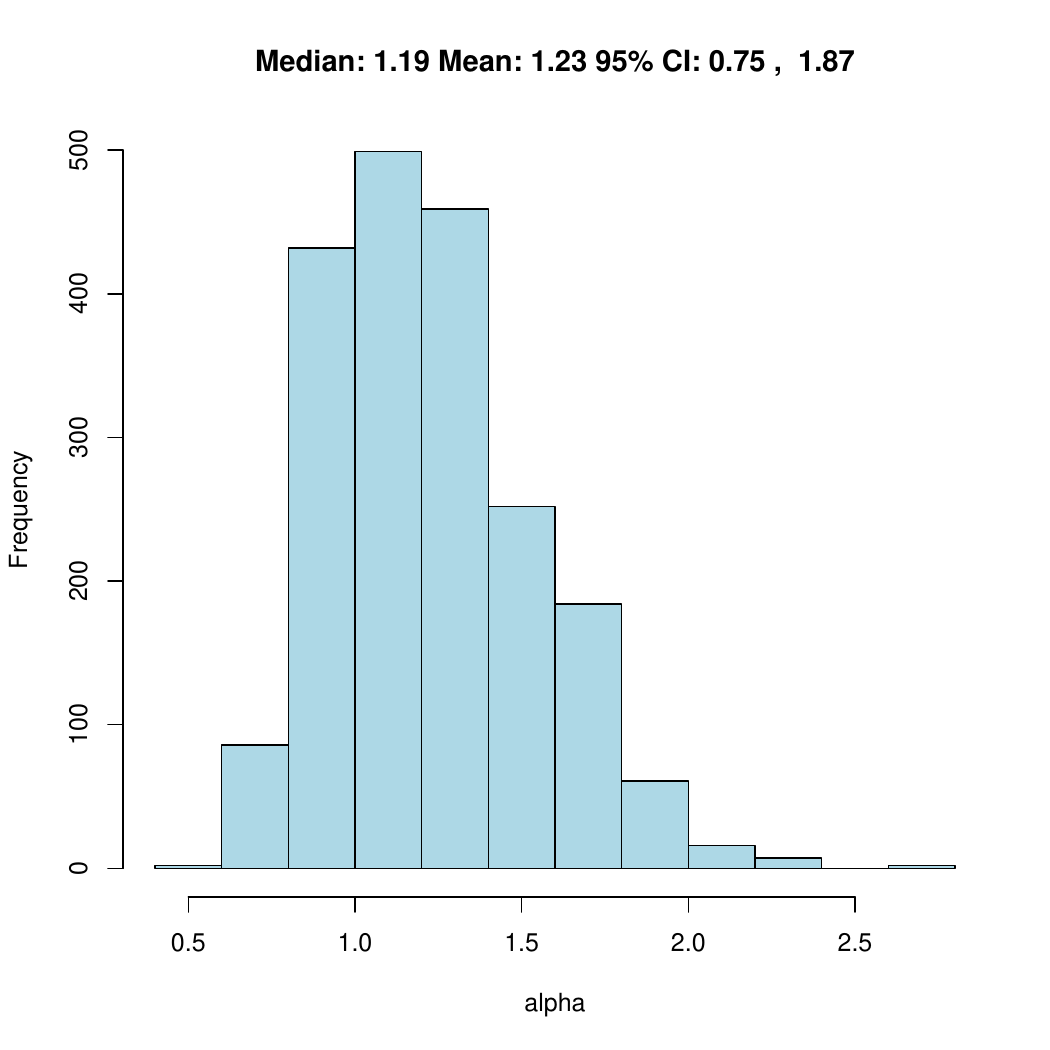}
	\includegraphics[width=0.28\textwidth]{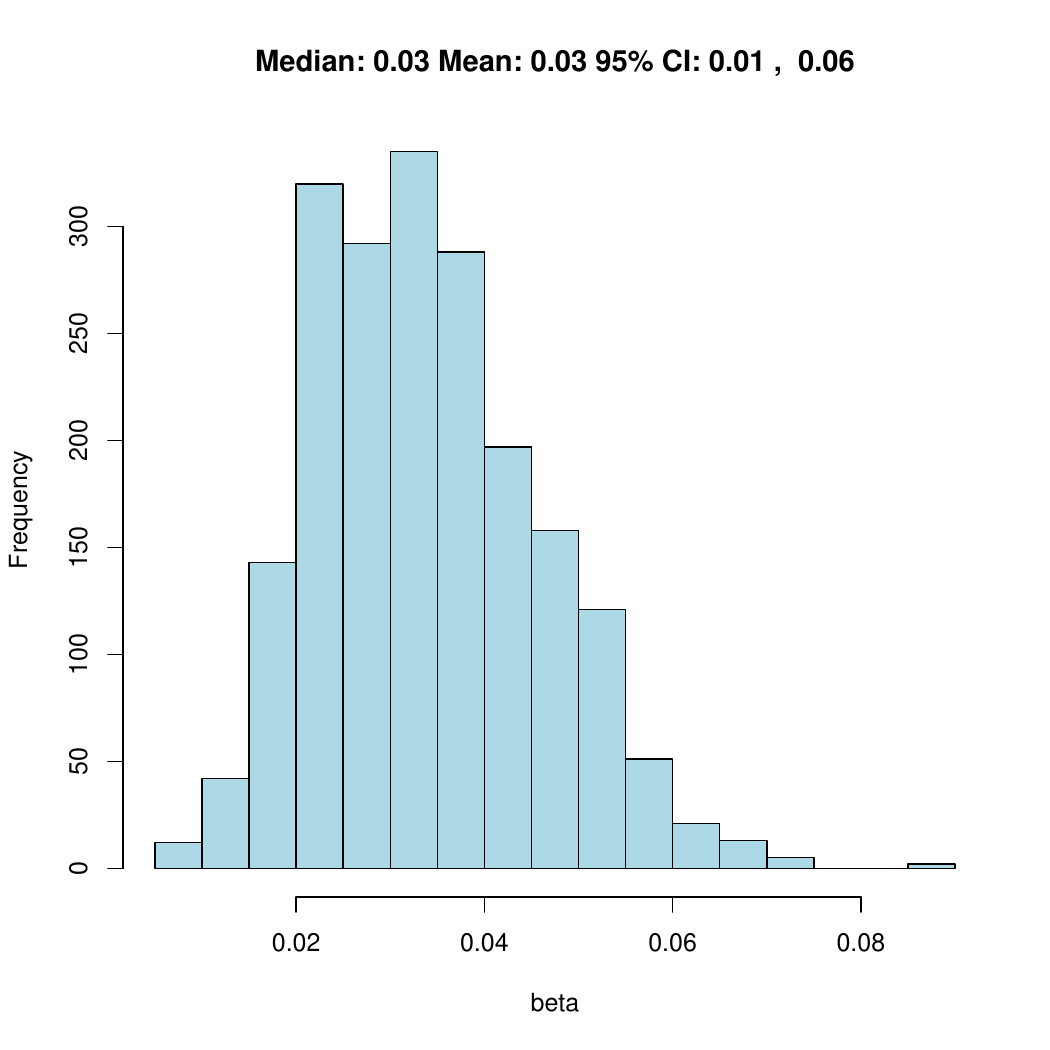}
	\\
	\includegraphics[width=0.28\textwidth]{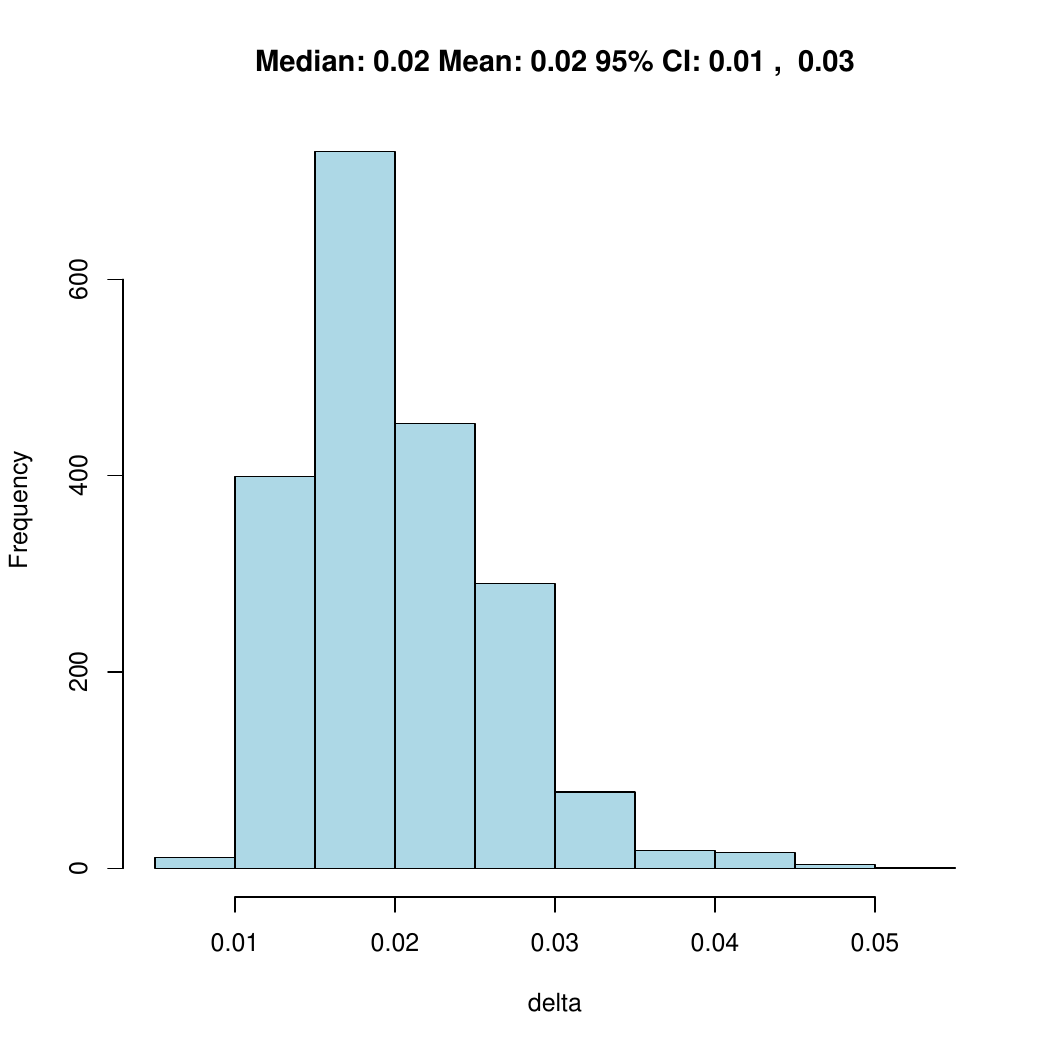}
	\includegraphics[width=0.28\textwidth]{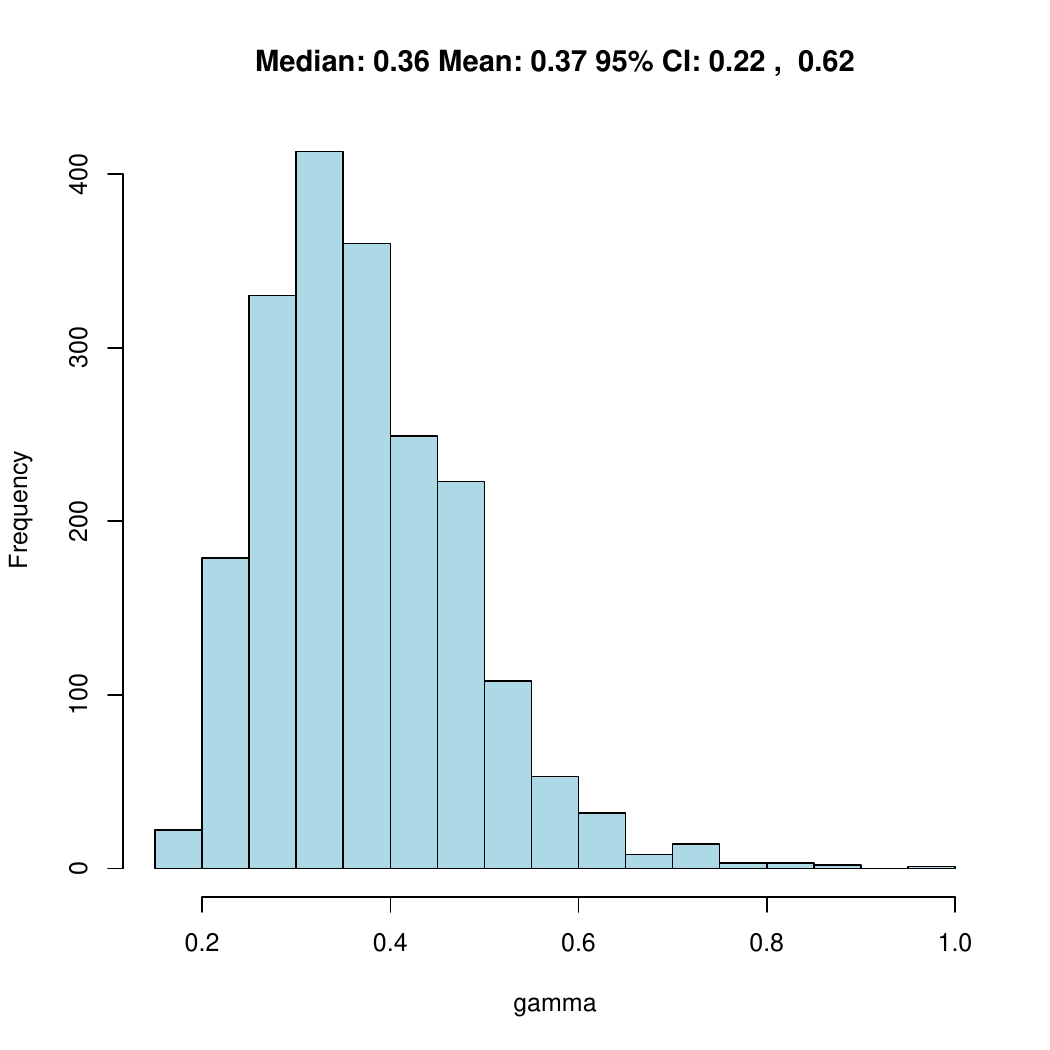}
	\caption{Histograms for predator data only with $(\alpha, \beta, \delta, \gamma)$ parameters.}        \label{fig:parameters_bff_predator}
	\label{fig:parameters}
\end{figure}

\begin{figure}[H]
	\centering
	\includegraphics[width=0.28\textwidth]{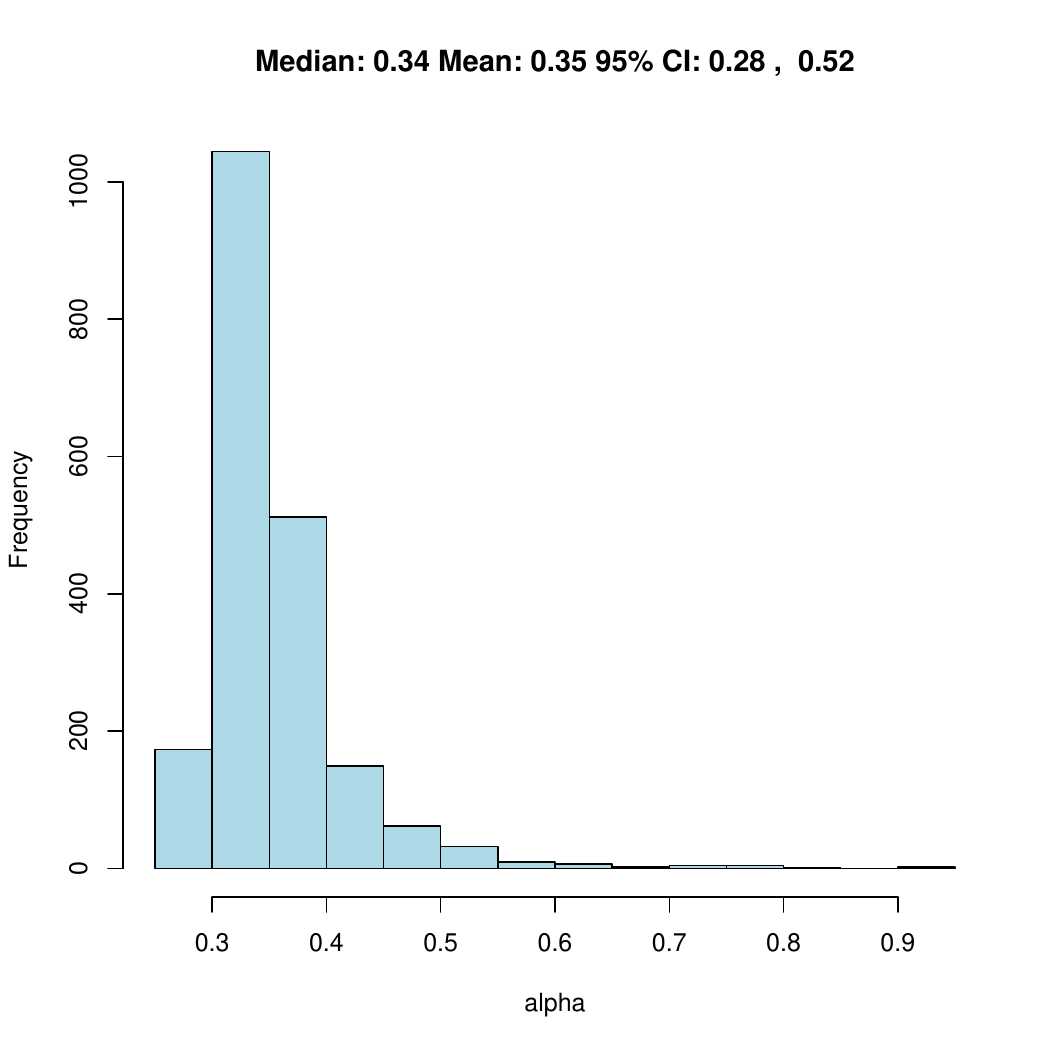}
	\includegraphics[width=0.28\textwidth]{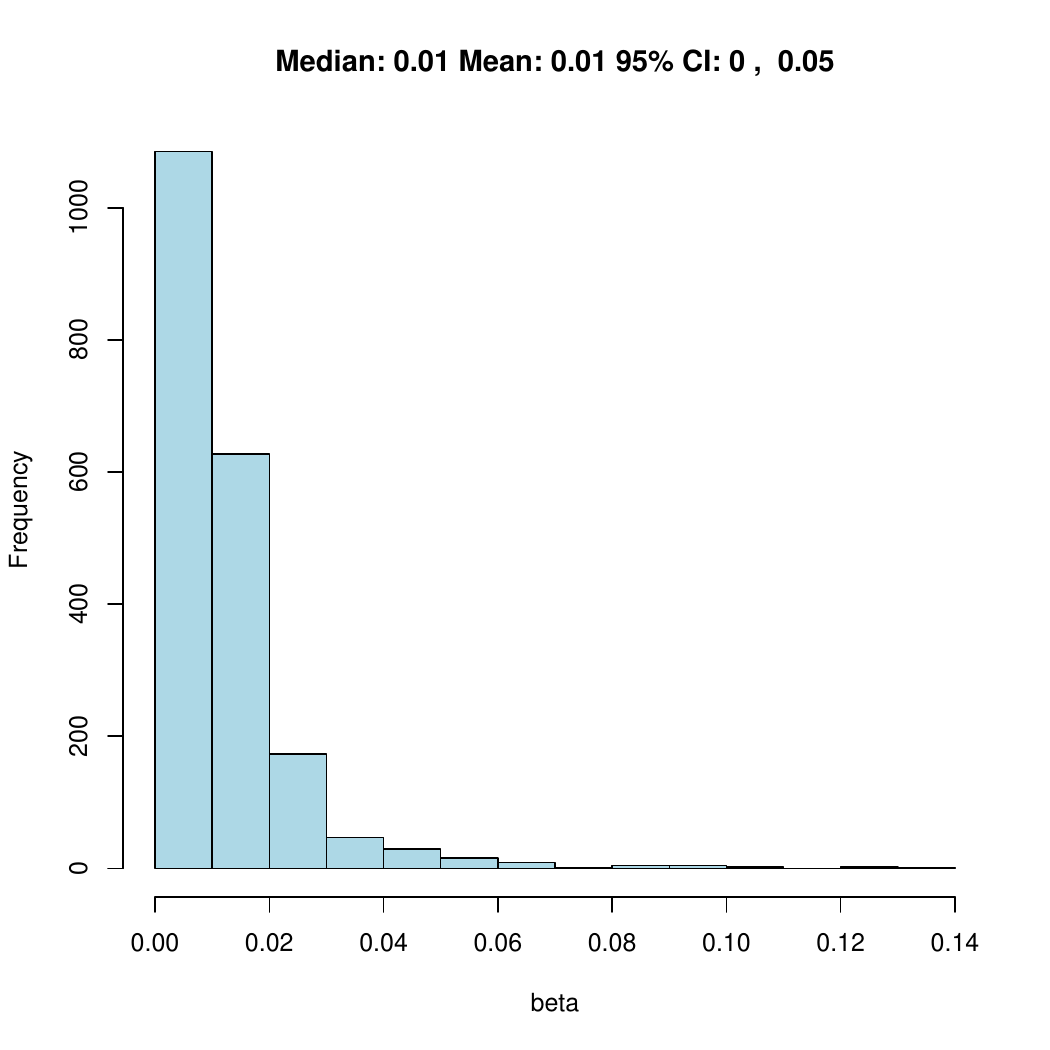}
	\\
	\includegraphics[width=0.28\textwidth]{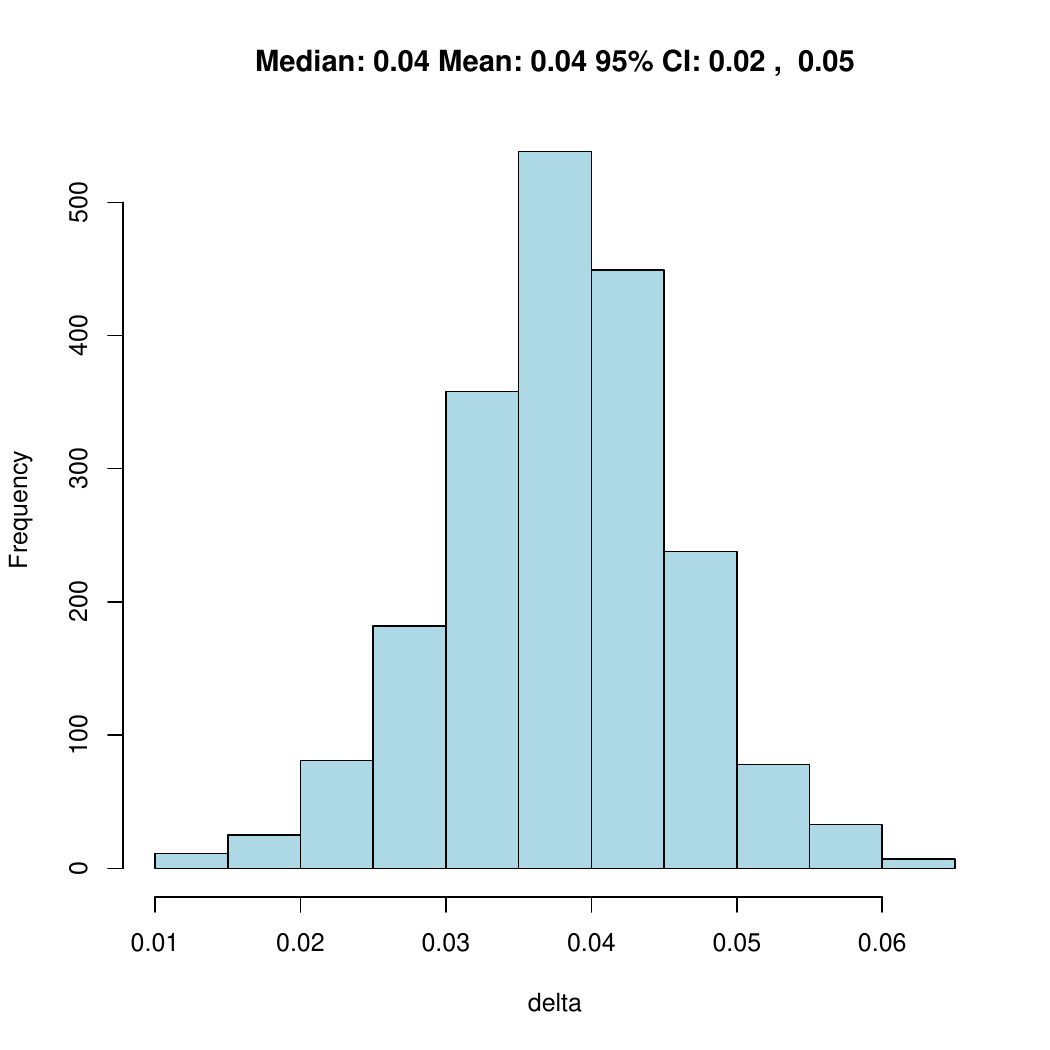}
	\includegraphics[width=0.28\textwidth]{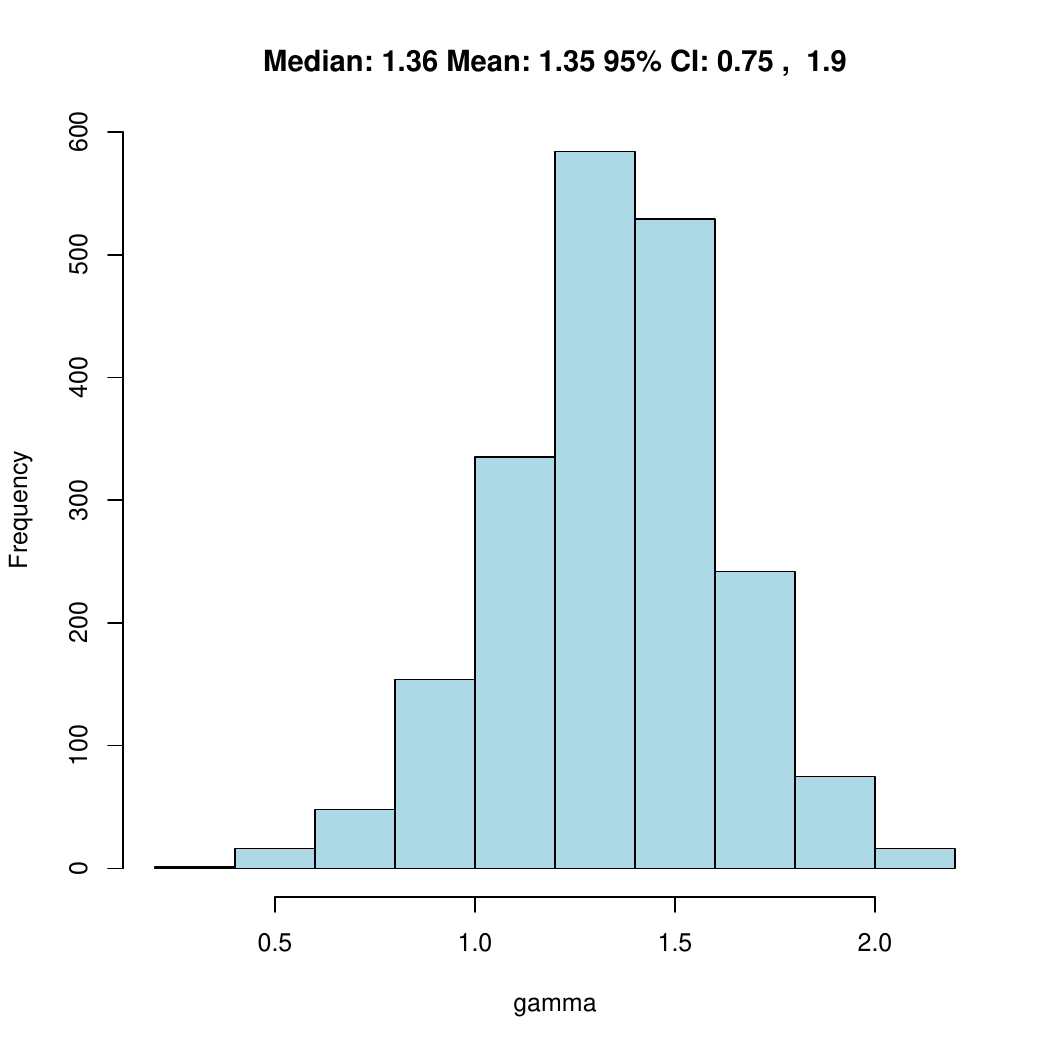}
	\caption{Histograms for prey data only of $(\alpha, \beta, \delta, \gamma)$ parameters}      \label{fig:parameters_bff_prey}
\end{figure}

\begin{figure}[H]
	\centering
	\includegraphics[width=0.18\textwidth]{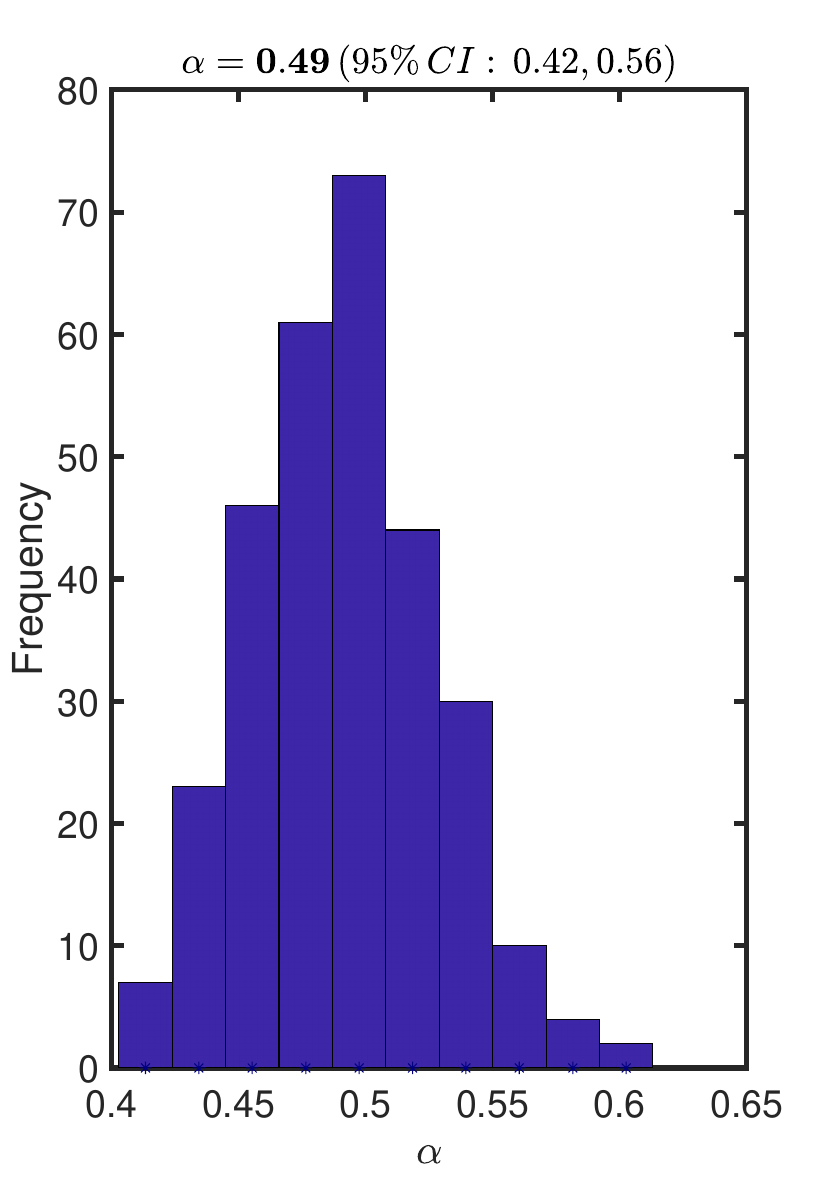}
	\includegraphics[width=0.18\textwidth]{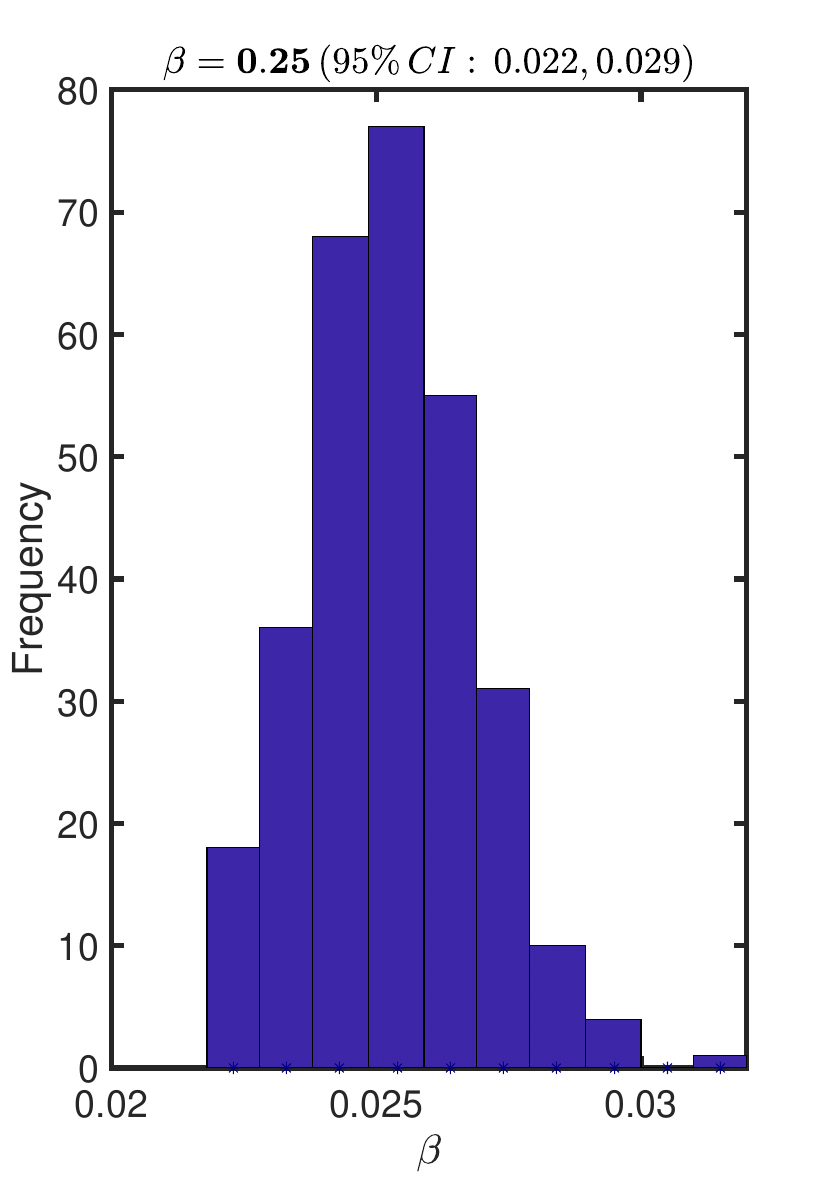}
	\\
	\includegraphics[width=0.18\textwidth]{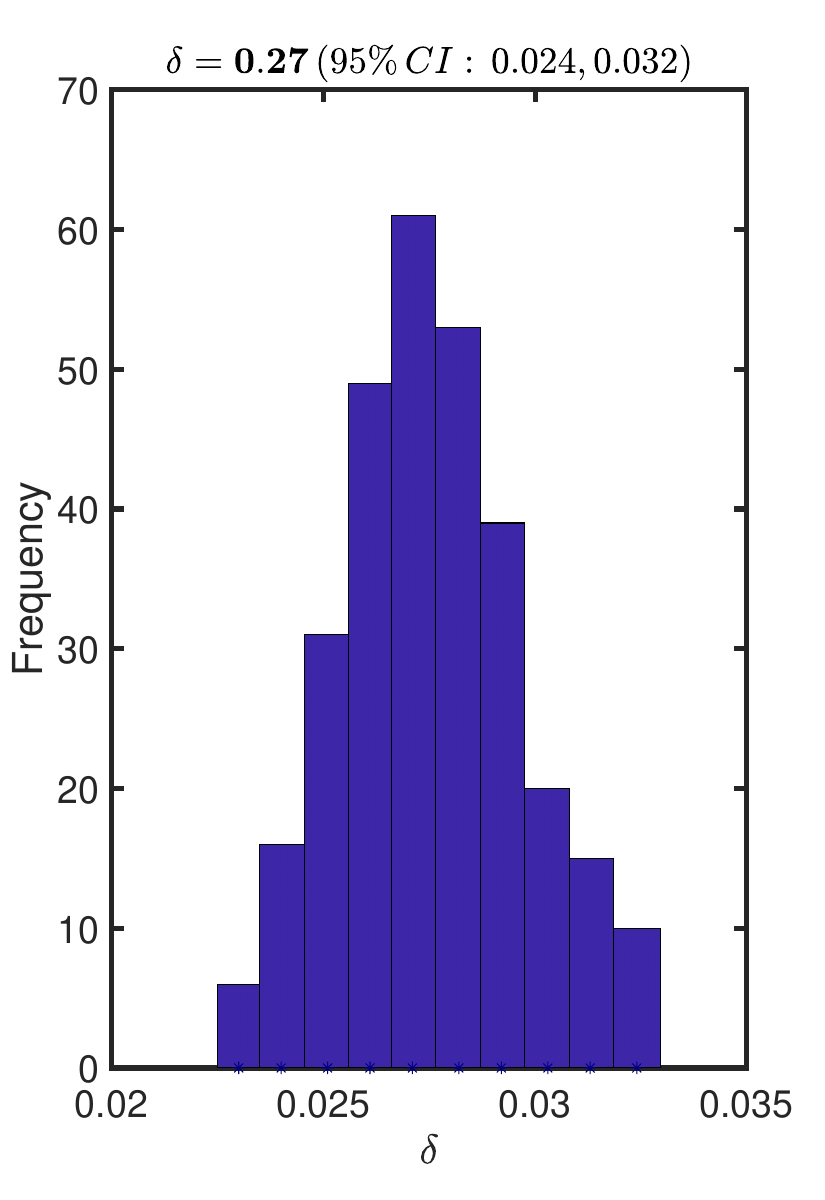}
	\includegraphics[width=0.18\textwidth]{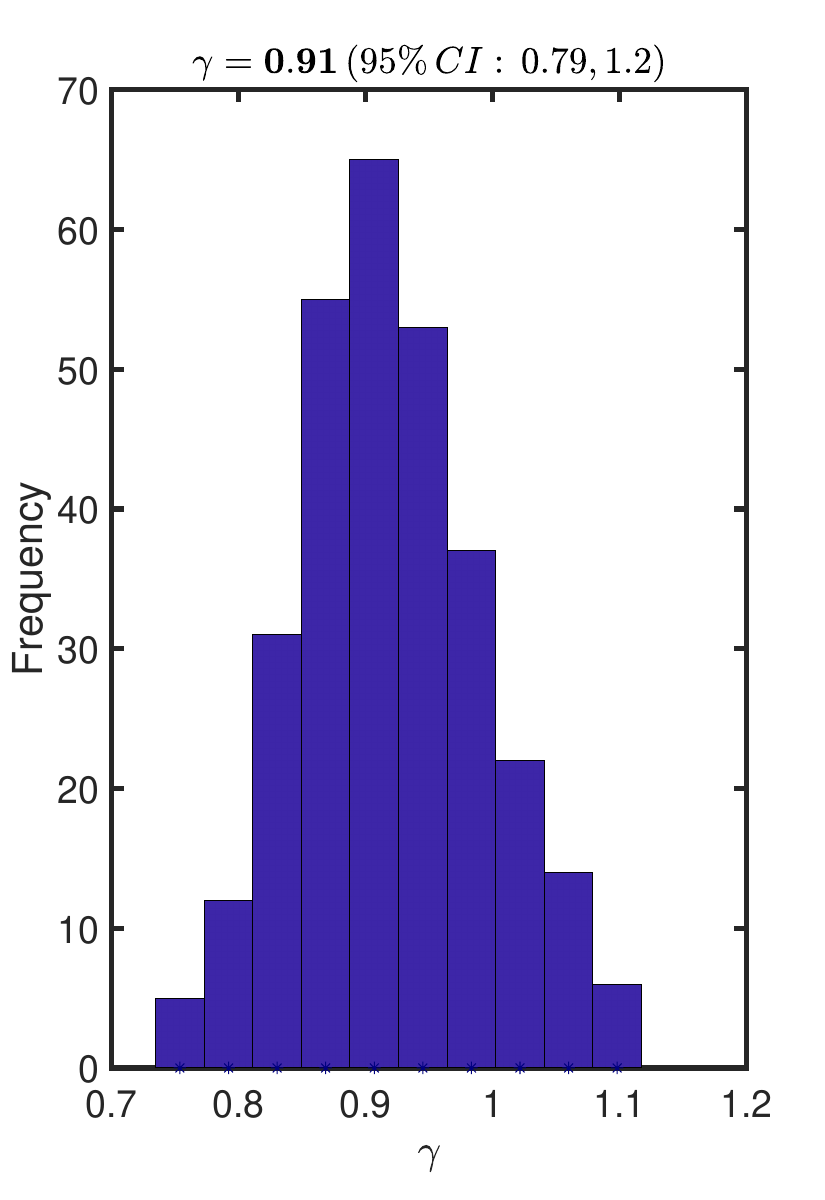}
	\caption{Histograms for predator prey data with the $(\alpha, \beta, \delta, \gamma)$ parameters.}      \label{fig:parameters_qdf_both}
\end{figure}

\begin{figure}[H]
	\centering
	\includegraphics[width=0.18\textwidth]{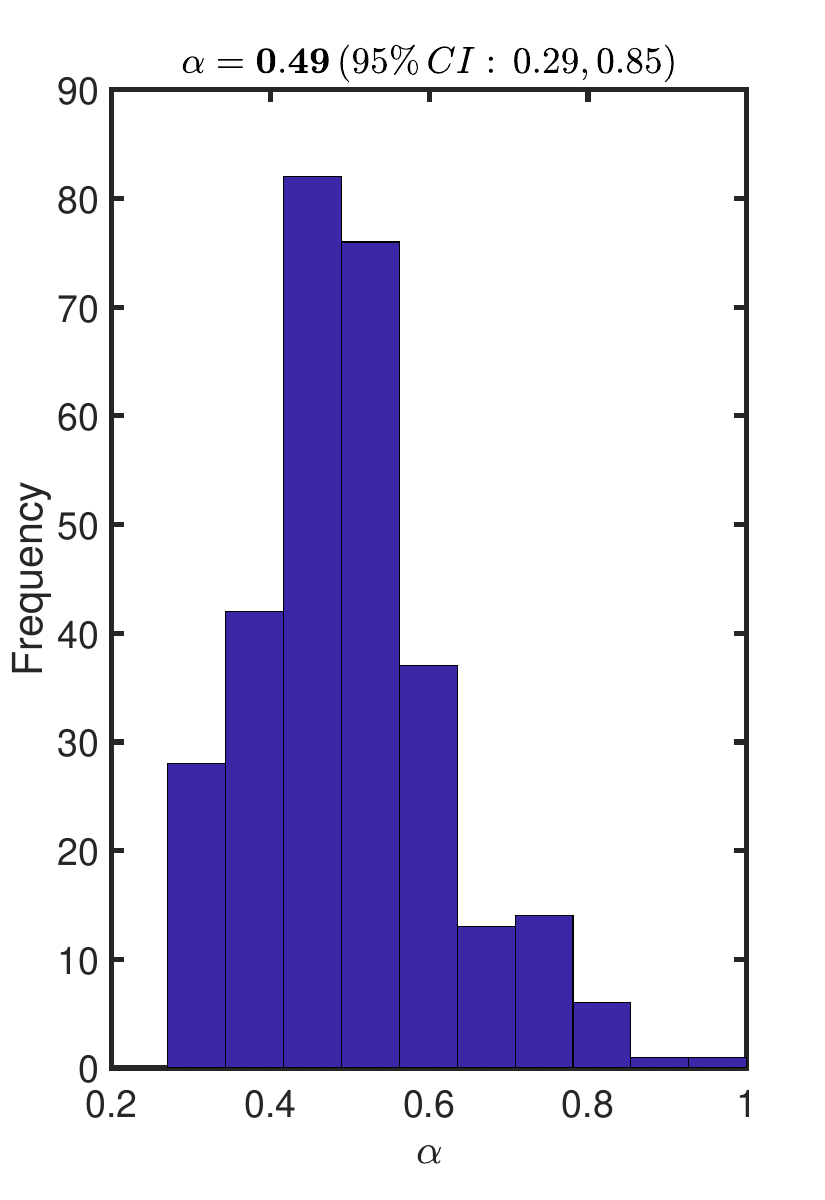}
	\includegraphics[width=0.18\textwidth]{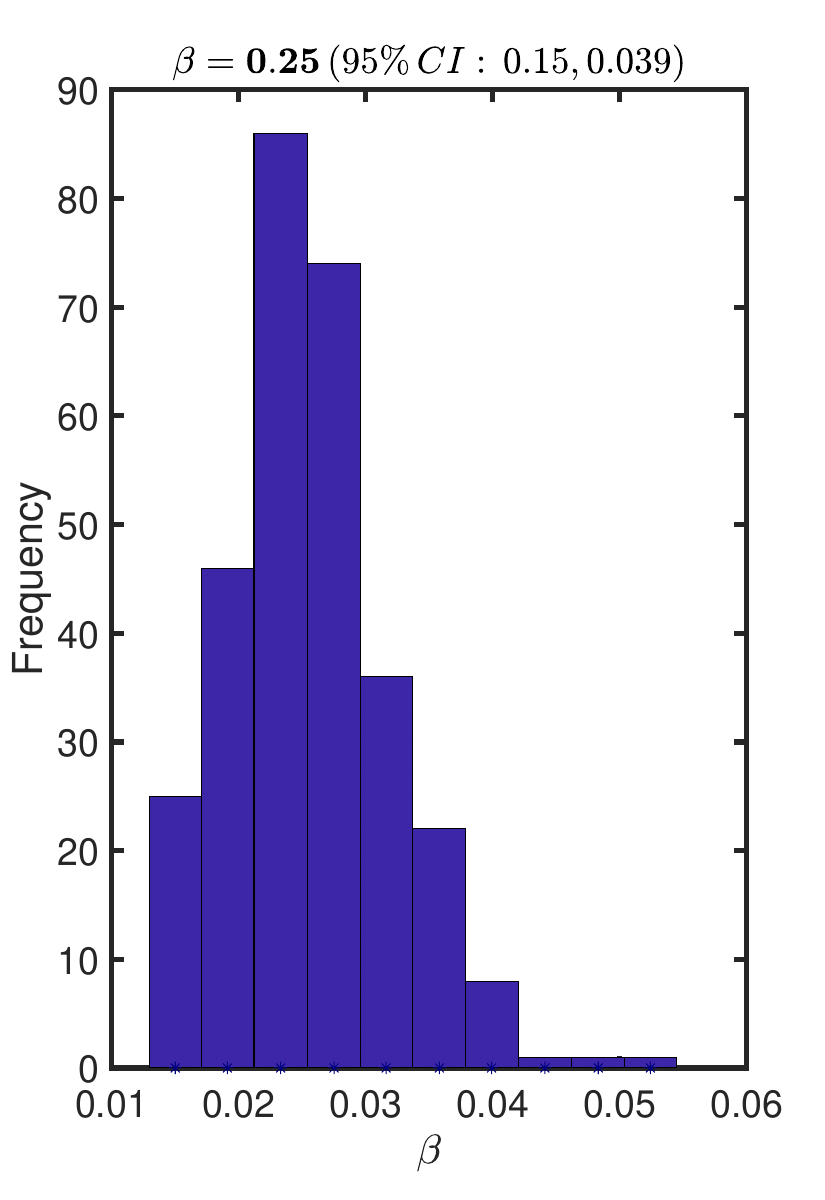}\\
	\includegraphics[width=0.18\textwidth]{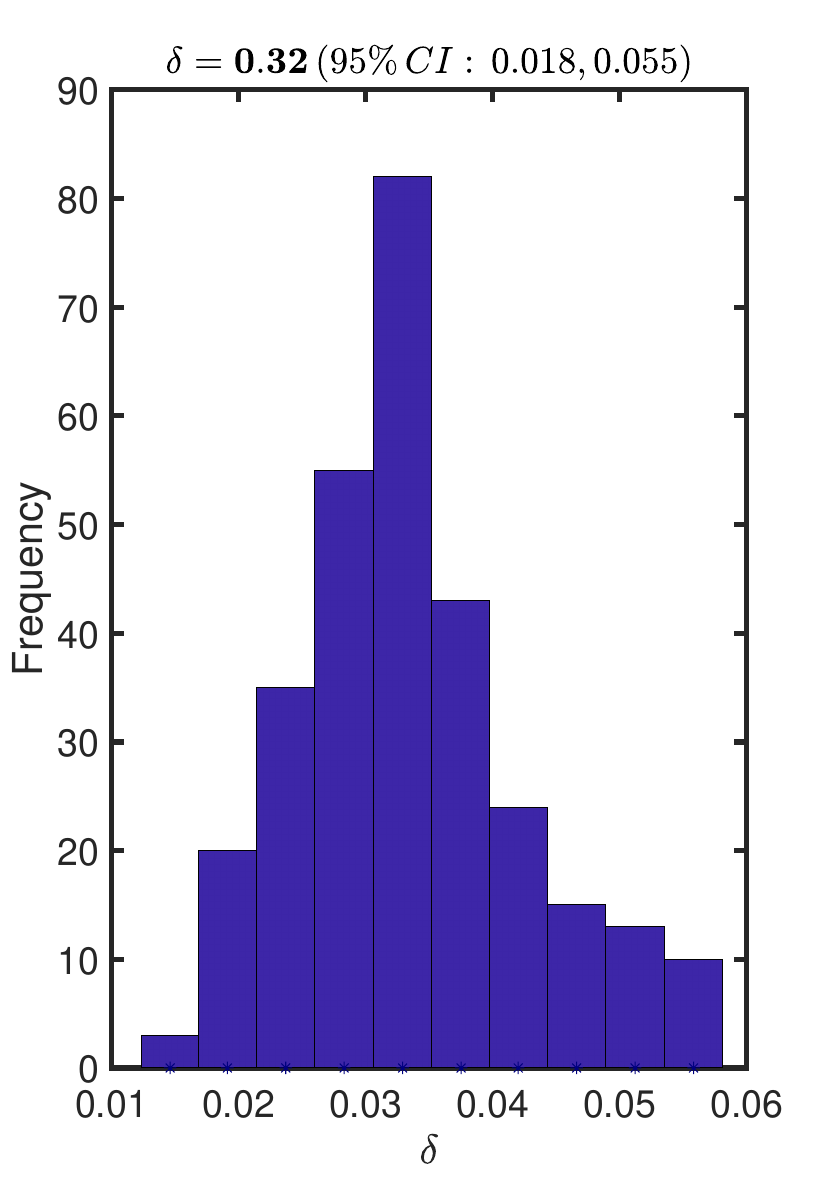}
	\includegraphics[width=0.18\textwidth]{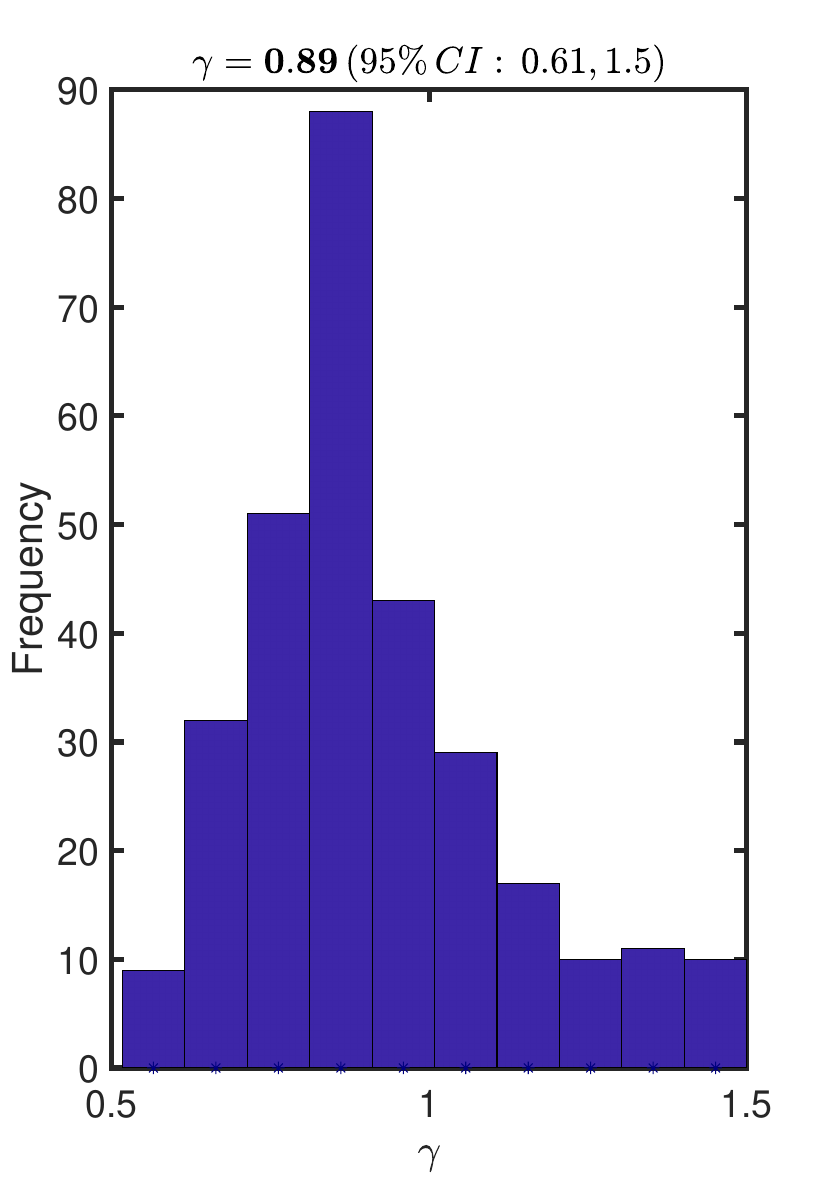}
	\caption{Histograms for predator data only with the  $(\alpha, \beta, \delta, \gamma)$ Parameters.}      \label{fig:parameters_qdf_predator}
\end{figure}

\begin{figure}[H]
	\centering
	\includegraphics[width=0.18\textwidth]{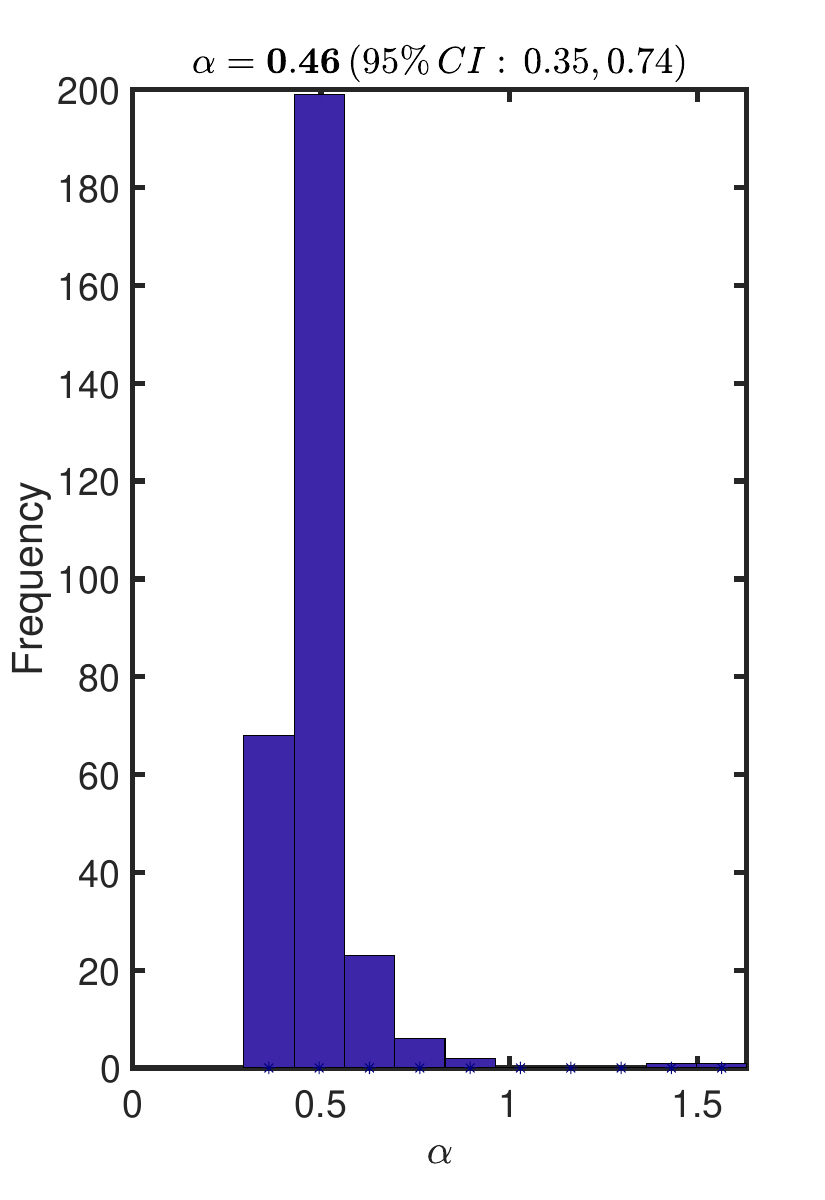}
	\includegraphics[width=0.18\textwidth]{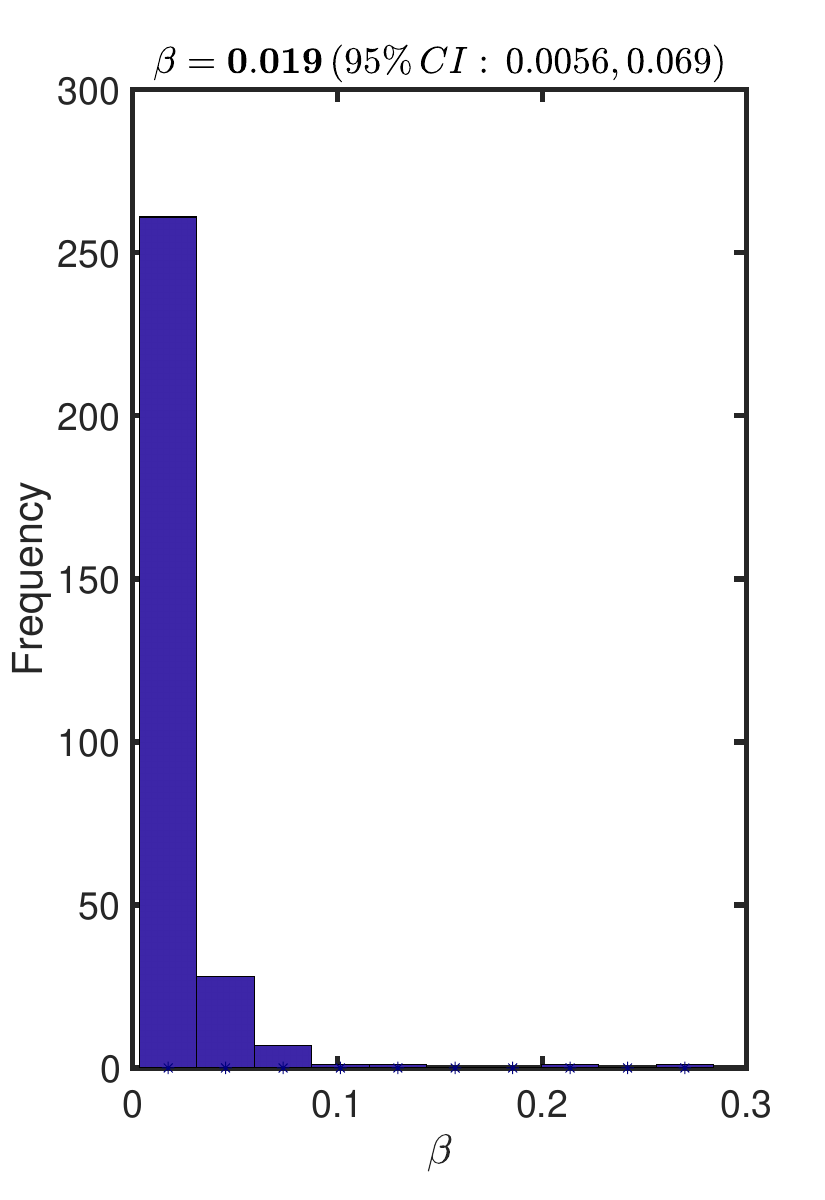}\\
	\includegraphics[width=0.18\textwidth]{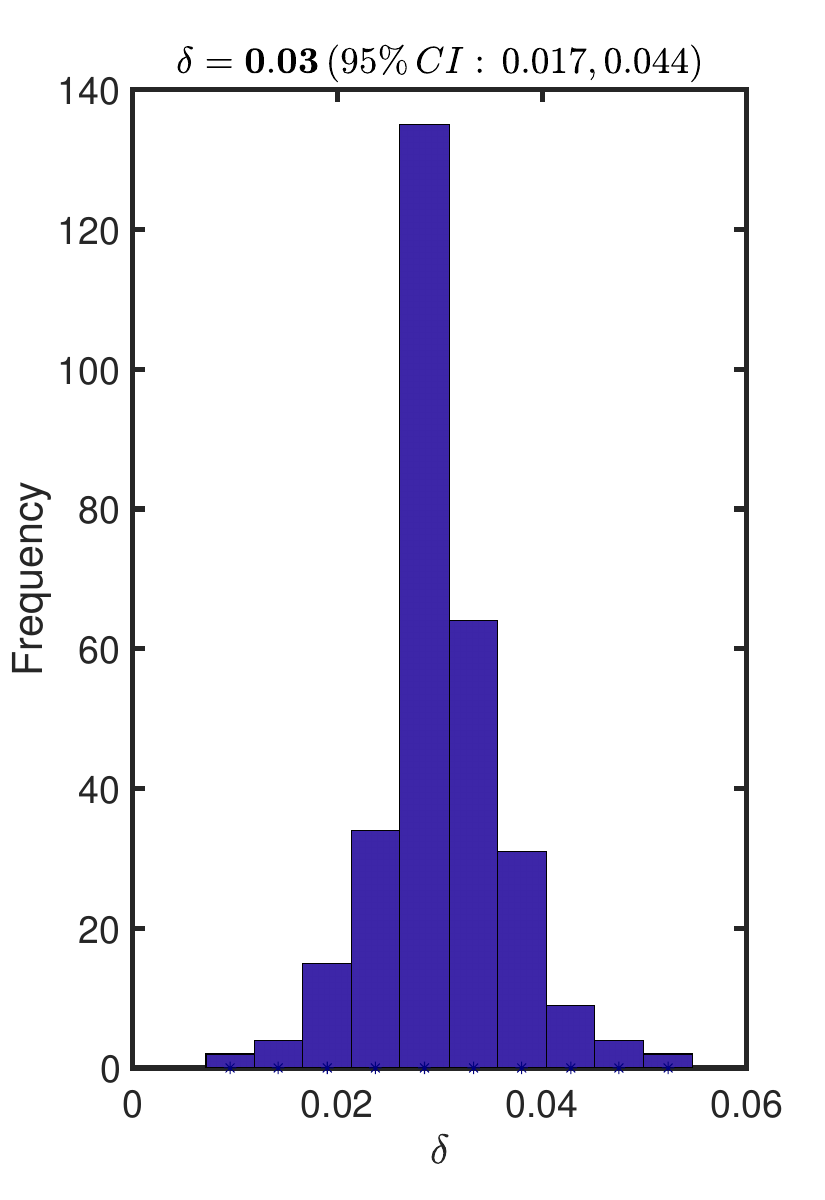}
	\includegraphics[width=0.18\textwidth]{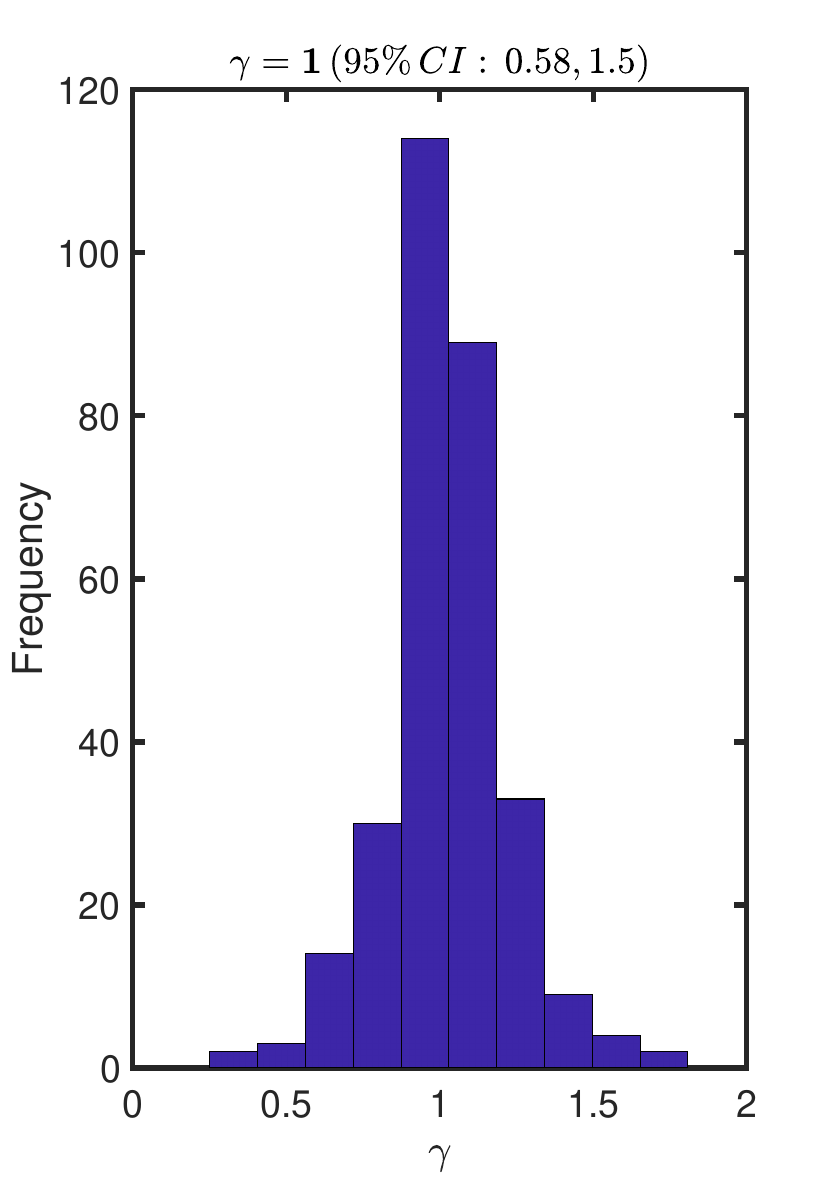}
	\caption{Histograms for prey data only with the $(\alpha, \beta, \delta, \gamma)$ parameters.} \label{fig:parameters_qdf_prey}
\end{figure}

The histogram summaries (Figures~\ref{fig:parameters_bff_both}--\ref{fig:parameters_qdf_prey}) reinforce these patterns: Joint data produce tighter, more concentrated uncertainty summaries, while the predator-only and prey-only settings yield broader distributions, consistent with reduced information under partial observation (Section~\ref{sec:SI}). 

%%%%%%%%%%%%%%%%%%%%%%%%%%%%%%%%%%%%%%%%%%%%%%%%%%
%%%%%%%%%%%%%%%%%%%%%%%%%%%%%%%%%%%%%%%%%%%%%%%%%
%%%%%%%%%%%%%%%%%%%%%%%%%%%%%%%%%%%%%%%%%%%%%%%%%%
%%%%%%%%%%%%%%%%%%%%%%%%%%%%%%%%%%%%%%%%%%%%%%%%%
%%%%%%%%%%%%%%%%%%%%%%%%%%%%%%%%%%%%%%%%%%%%%%%%%%

Tables~\ref{tab:parameter_priors_bounds} and~\ref{tab:parameter_ranges} presents the lower and upper bounds in this analysis.

\begin{table}[H]
	\centering
	\caption{Prior distributions and bounds for the parameters in the LV model. The priors are truncated normal distributions with a mean of 0.5 and a standard deviation of 0.5, constrained to be non-negative (T[0,]).}\vspace{0.2cm}
		\begin{tabular}{llll}
		\hline
		Parameter & Prior distribution & Lower bound (LB) &Upper bound (UB)  \\ \hline
		$\alpha$ & Normal (0.5, 0.5) T[0,] & 0 & NA \\ 
		$\beta$ & Normal (0.5, 0.5) T[0,] & 0 & NA \\ 
		$\delta$ & Normal (0.5, 0.5) T[0,] & 0 & NA \\
		$\gamma$ & Normal (0.5, 0.5) T[0,] & 0 & NA \\ \hline
	\end{tabular}
	
	\label{tab:parameter_priors_bounds}
\end{table}

\begin{table}[H]
	\centering
	\caption{Parameter ranges for the LV model of QDF.}\vspace{0.2cm}\setlength{\tabcolsep}{8mm}
		\begin{tabular}{lll}
		\hline
		Parameter & Lower bound (LB) & Upper bound (UB) \\
		\hline
		$\alpha$     & 0.001            & 1.0              \\
	
		$\beta$      & 0.001            & 0.5              \\
		
		$\delta$     & 0.001            & 0.5              \\
	
		$\gamma$     & 0.001            & 1.5              \\
		
		%\bottomrule
		\hline
	\end{tabular}
	\label{tab:parameter_ranges}
\end{table}

The difference in the parameter ranges reflects a key methodological difference between the two approaches. QDF uses hard parameter bounds because it relies on deterministic optimization, where bounds are needed for numerical stability. In contrast, the BFF is based on Bayesian inference and uses prior distributions rather than strict bounds to regularize the parameters, allowing the values to be penalized probabilistically rather than excluded. Using identical hard bounds for both methods would unnecessarily restrict the Bayesian model and reduce its flexibility. Instead, we ensured a fair comparison by using the same model structure, data, and evaluation metrics, and by setting reasonable priors for Bayesian estimation. Different bounds or prior support may affect the parameter estimates in weakly identifiable settings, but our main comparisons focus on predictive accuracy and uncertainty calibration, which remain directly comparable across methods. The Bayesian priors were chosen to be weakly informative so that inference is primarily driven by the data.

\subsection{GLM model for lung injury}

\begin{figure}[H]
\centering
\includegraphics[width=0.8\linewidth]{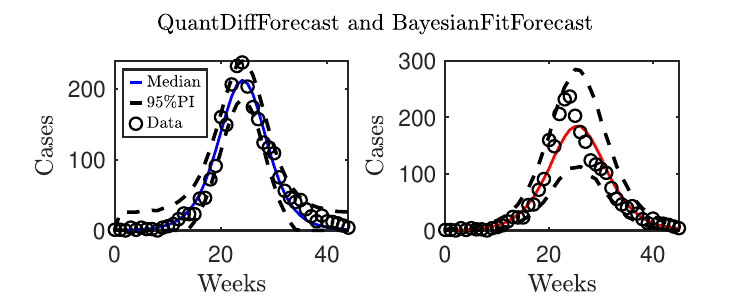}\vspace{0.5cm}
\caption{Fitting visualization for the GLM model,  with lung injury data, using the QDF and BFF approaches. \textbf{Left:} QDF, \textbf{Right:} BFF.}
\label{fig:glm-lung-fit}
\end{figure}

The GLM provides a well-identified baseline example in which the data support stable parameter recovery under both frameworks (Section~\ref{sec:SI}). Both approaches capture the rise and decline in the lung injury time series, with QDF providing a better fit (Figure~\ref{fig:glm-lung-fit}).
\noindent
The inferred growth parameters are close across methods (Table~\ref{tab:glm-lung-params}), with $r$ near 0.32--0.33 and $p$ near 1.0, indicating consistent recovery of the growth shape. The main discrepancy is in the estimated carrying capacity $k$, where BFF yields a higher central estimate than QDF.

In calibration performance, QDF achieves lower point-error metrics (MAE 9.42 vs.\ 11.85; MSE~152.80 vs.\ 412.29; WIS 5.75 vs.\ 7.15) while both methods achieve well calibrated PI coverage near the nominal 95\% level (97.83\% for BFF; 100\% for QDF), noting that 100\% reflects slightly overconservative intervals (Table~\ref{tab:glm-lung-metrics}). This pattern is consistent with an identifiable, low-dimensional setting where optimization-based estimation can yield very sharp point fits without compromising interval calibration.

The Bayesian fit converged well for this dataset ($\hat{R}=1$ for all GLM parameters; Table~\ref{tab:glm-lung-rhat}), supporting the reliability of the reported posterior summaries.

\begin{table}[H]
\centering
\caption{Parameter estimates for the GLM model,  with lung-injury data, using QDF and BFF approaches.}\vspace{0.2cm}\setlength{\tabcolsep}{7mm}
\begin{tabular}{lllll}
\hline
& $r$ & $p$ & $k$  \\ \hline
BFF & 0.33 (0.3, 0.36)  & 0.98 (0.96, 1)    & 2734.03 (2732.03, 2735.82)     \\ 
QDF   & 0.32 (0.30, 0.35)  & 1 (0.99, 1.02) & 2549.5 (2438.9, 2662.9)    \\ \hline
\end{tabular}
\label{tab:glm-lung-params}
\end{table}

\begin{table}[H]
\centering
\caption{Parameter estimates for the GLM model,  with lung injury data, using the QDF and BFF approaches.}\vspace{0.2cm}\setlength{\tabcolsep}{7mm}

\begin{tabular}{lllll}
\hline
& MAE & MSE & WIS &95$\% $PI \\ \hline
BFF & 11.85  & 412.2943    & 7.15  & 97.83   \\
QDF   & 9.42  & 152.80 & 5.75  & 100 \\ \hline
\end{tabular}

\label{tab:glm-lung-metrics}
\end{table}

\begin{table}[H]
\centering
\caption{Convergence diagnostics for the GLM model with lung injury data, using the QDF and BFF approaches.}\setlength{\tabcolsep}{12mm}\vspace{0.2cm} 
\begin{tabular}{lllll}
\hline
& $r$ & $p$ & $k$  \\ \hline
$\textbf{$\hat{R} $}$ & 1     & 1    & 1        \\ \hline
\end{tabular}

\label{tab:glm-lung-rhat}
\end{table}

\begin{figure}[H]
\centering
\includegraphics[width=0.8\textwidth]{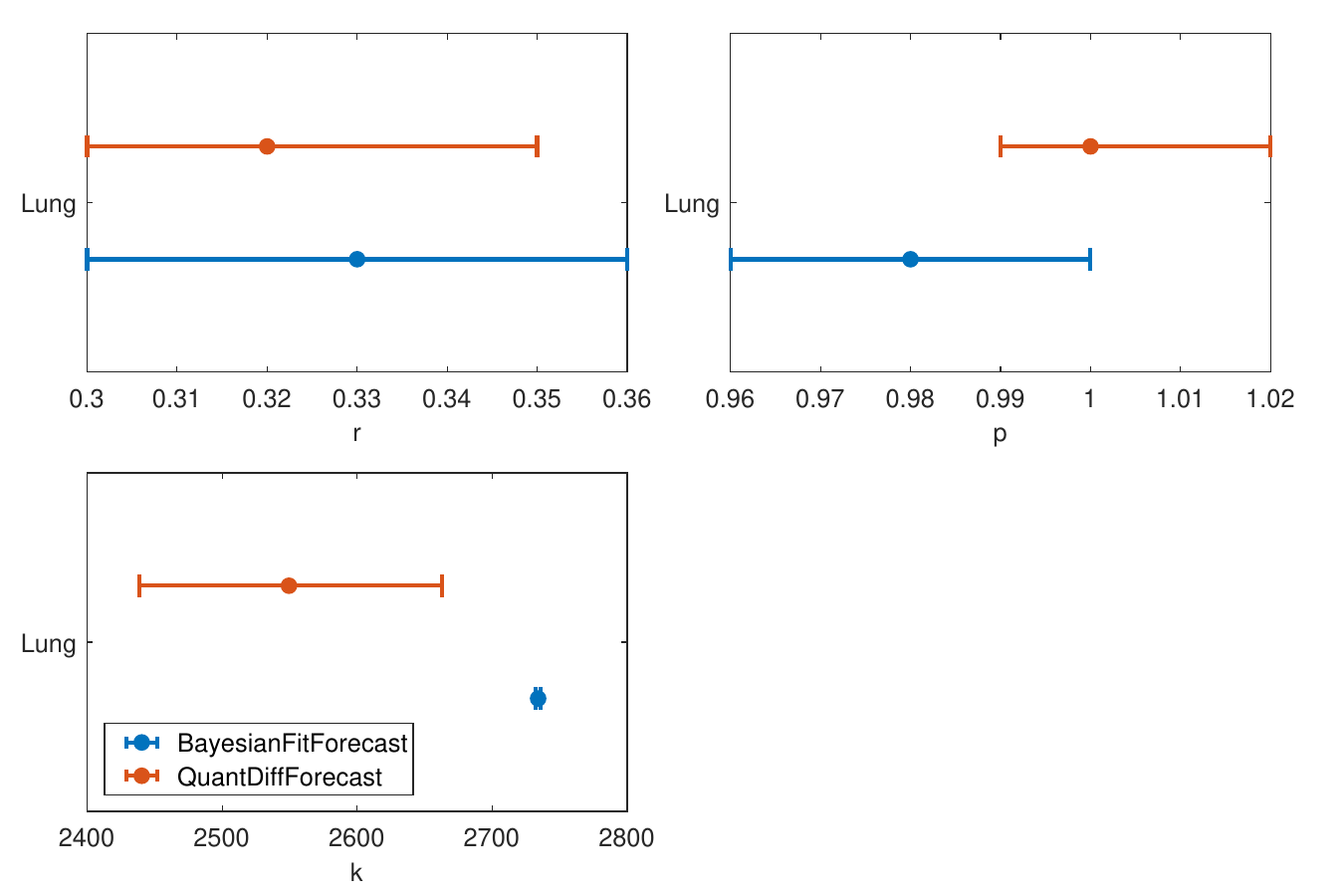}
\caption{Parameter comparison for the GLM model, with lung injury data, using the QDF and BFF approaches.}\vspace{0.5cm} 
\label{fig:glm-lung-paramcmp}
\end{figure}

The cross-method parameter comparison reinforces that $(r,p)$ are tightly constrained in both frameworks, with only modest differences in their uncertainty widths (Figure~\ref{fig:glm-lung-paramcmp}).

\begin{figure}[H]
% \centering
\centering
\includegraphics[width=0.8\textwidth]{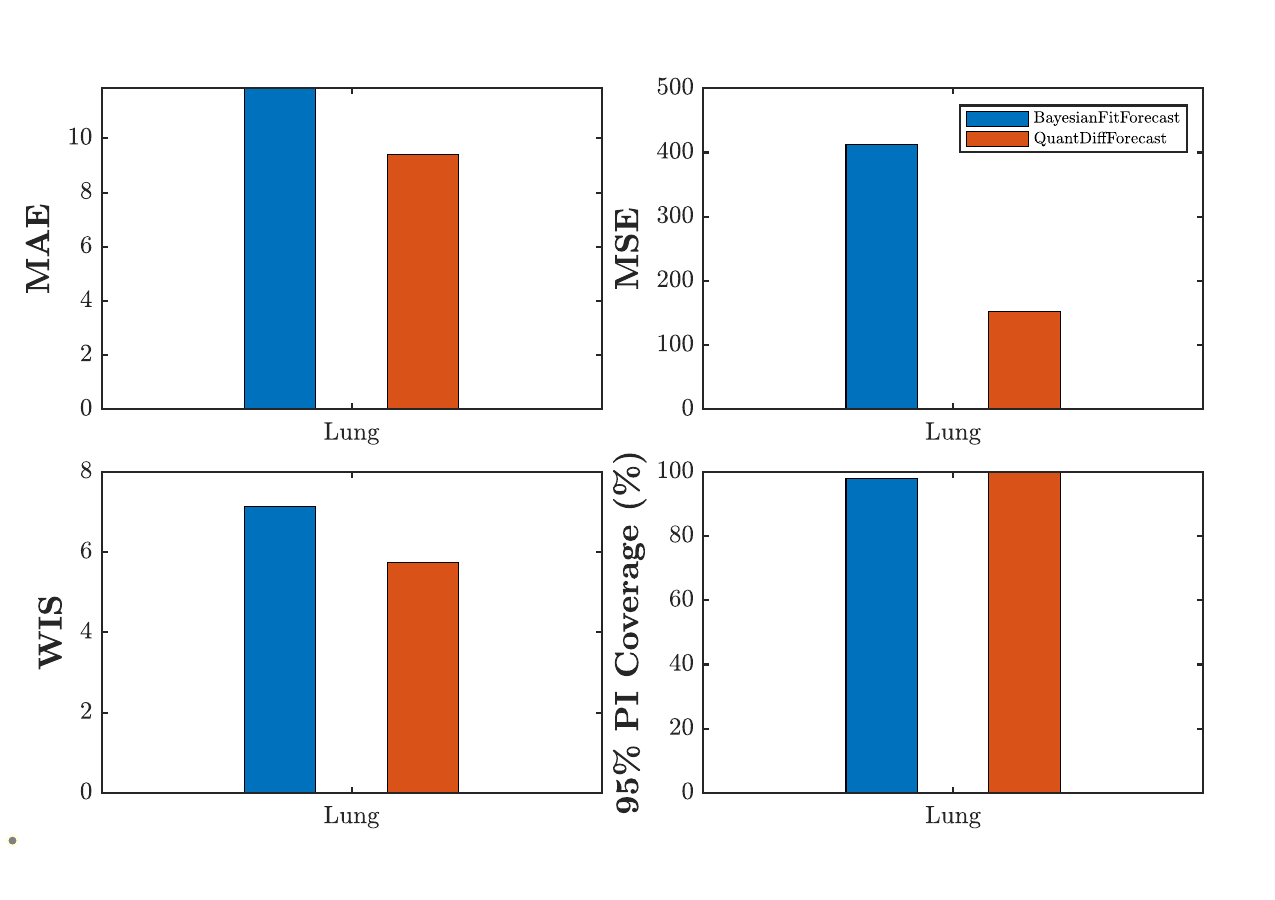}\vspace{0.5cm} 
\caption{Performance metrics comparison for the GLM model,  with lung injury data, using the QDF and BFF approaches.}
\label{fig:glm-lung-perf}
\end{figure}

Across the summary metrics, the same ranking is apparent: QDF yields lower MAE, MSE, WIS while the PI coverage remains comparable (Figure~\ref{fig:glm-lung-perf}).

\label{sec:glm_hists}

The parameter bounds used for applying the BFF and QDF for this model are provided in Tables~\ref{tab:parameter_ranges_QFF} and~\ref{tab:parameter_rangesQDF}.

\label{sec:glm_params_bff}

\begin{table}[H]
\centering
\caption{Parameter bounds for the GLM model with lung injury data, using the BFF~approach.}
\begin{tabular}{llll}
\hline
Parameter & Prior distribution & Lower bound (LB) & Upper bound (UB) \\
\hline
$r$ & Normal $(0.5, 0.5)$ T$[0,\ ]$ & 0.0 & NA \\

$p$ & Normal $(0.5, 0.5)$ T$[0,\ ]$ & 0.0 & NA \\

$k$ & Normal $(2632, 10)$ T$[0,\ ]$ & 0.0 & NA \\
\hline
\end{tabular}
\label{tab:parameter_ranges_QFF}
\end{table}

\begin{table}[H]
\centering
\caption{Parameter bounds for the GLM model with lung injury data, using the QDF~approach.}\vspace{0.2cm} 
\begin{tabular}{lll}
\hline
Parameter & Lower bound (LB) & Upper bound (UB) \\
\hline
$r$     & 0.01            & 5.0              \\

$p$      & 0.0            & 2.0              \\

$k$     & 0.01            & 1,000,000.0                         \\
\hline
\end{tabular}
\label{tab:parameter_rangesQDF}
\end{table}

\begin{figure}[H]
	\centering
	\includegraphics[width=0.2\textwidth]{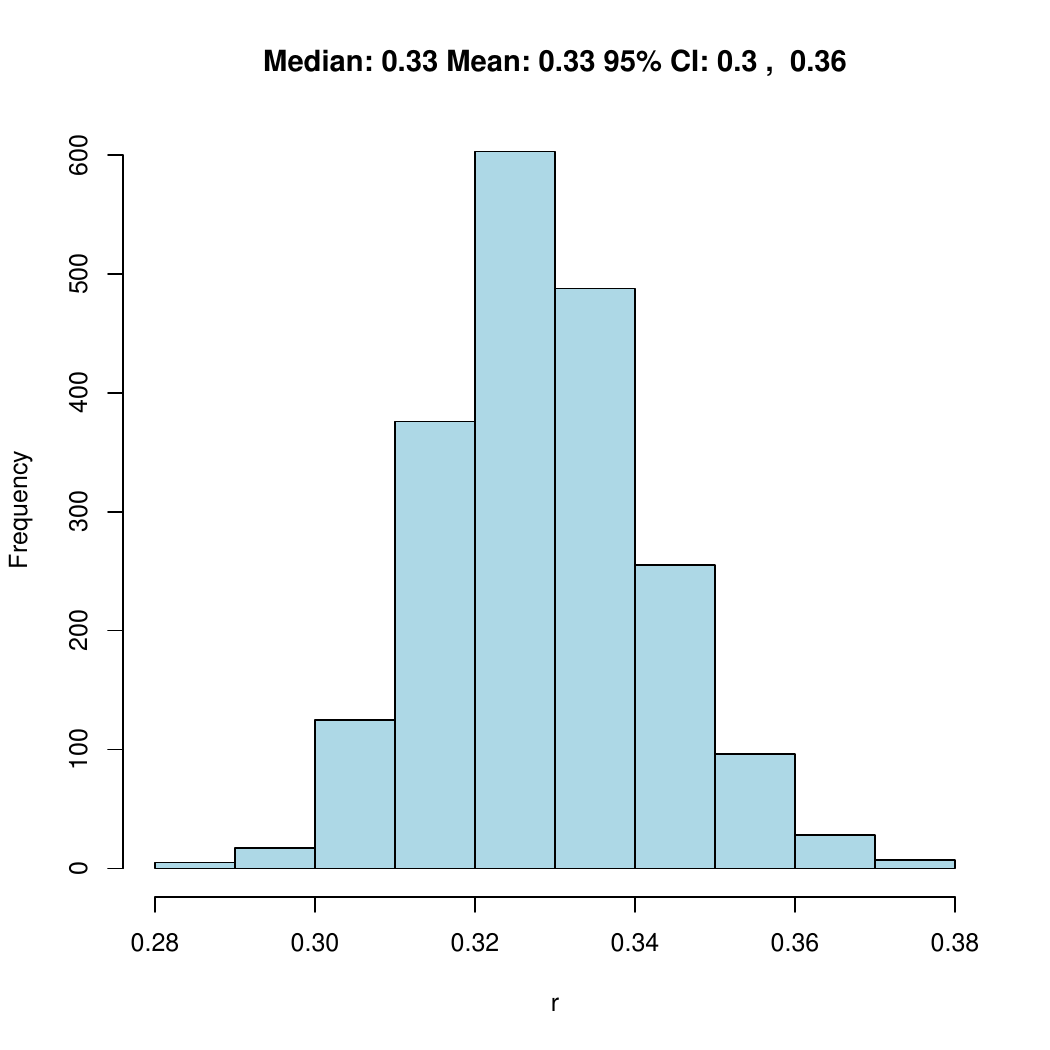}
	\includegraphics[width=0.2\textwidth]{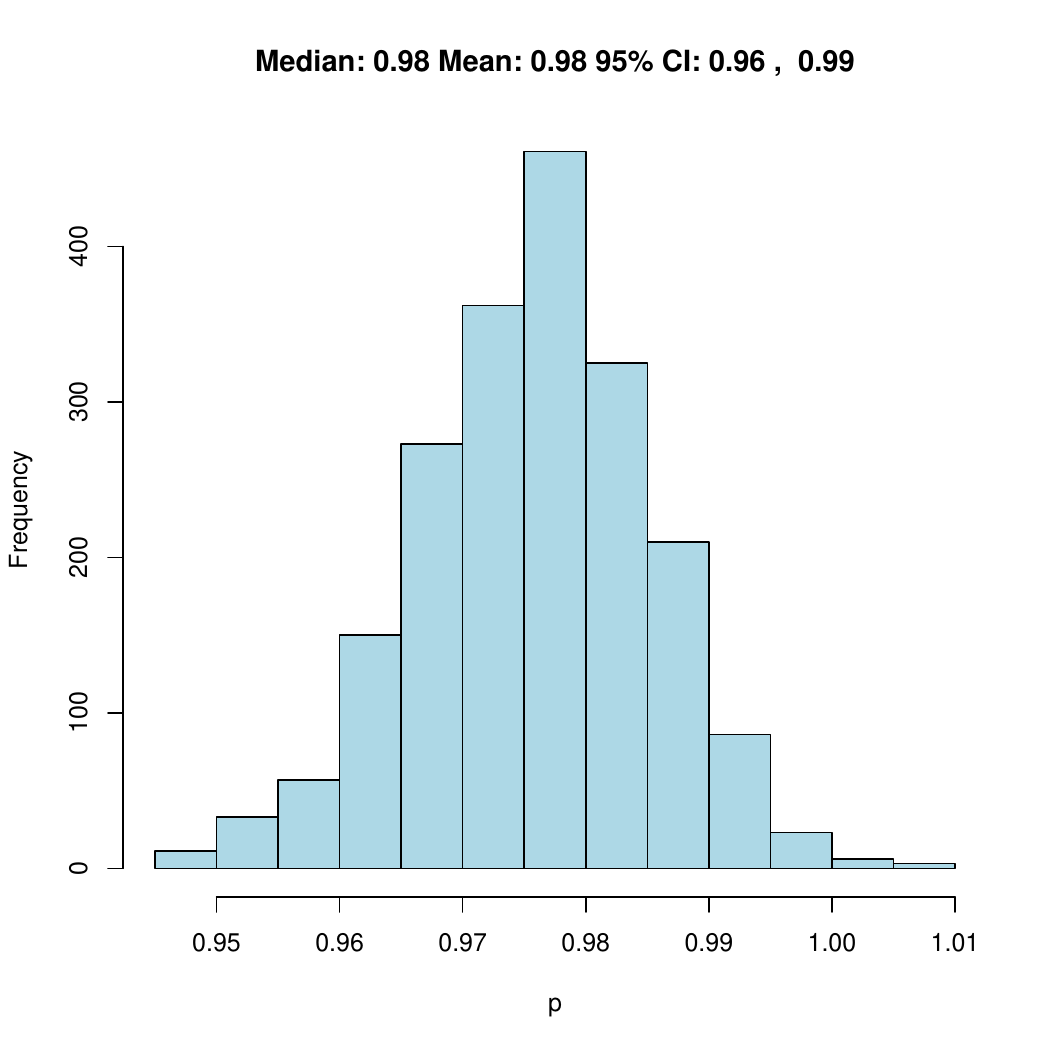}\\
	\includegraphics[width=0.2\textwidth]{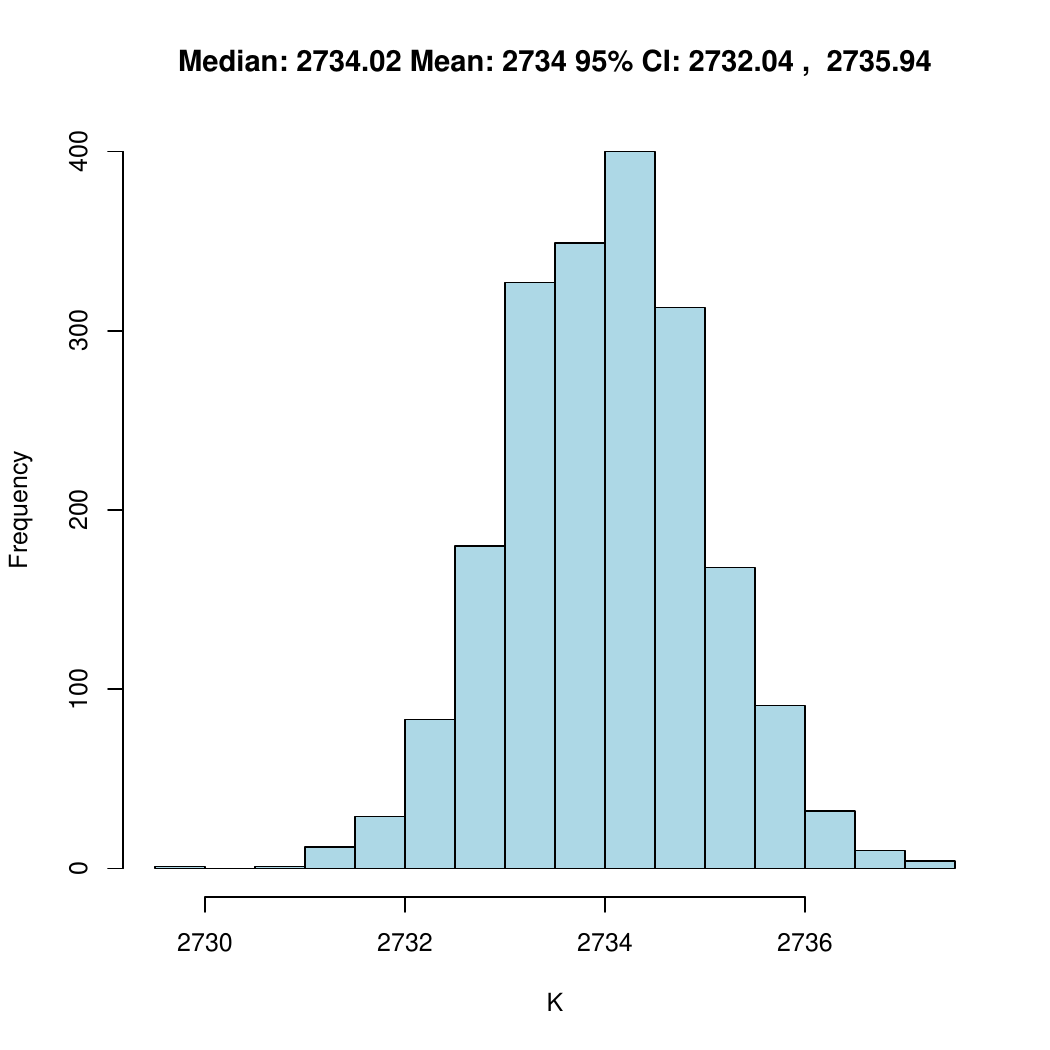}
	\caption{Histograms for GLM data with the $r$, $k$, and $p$ parameters.}
	\label{fig:parameters_Lung_injury}
\end{figure}

\begin{figure}[H]
	\centering
	\includegraphics[width=0.2\textwidth]{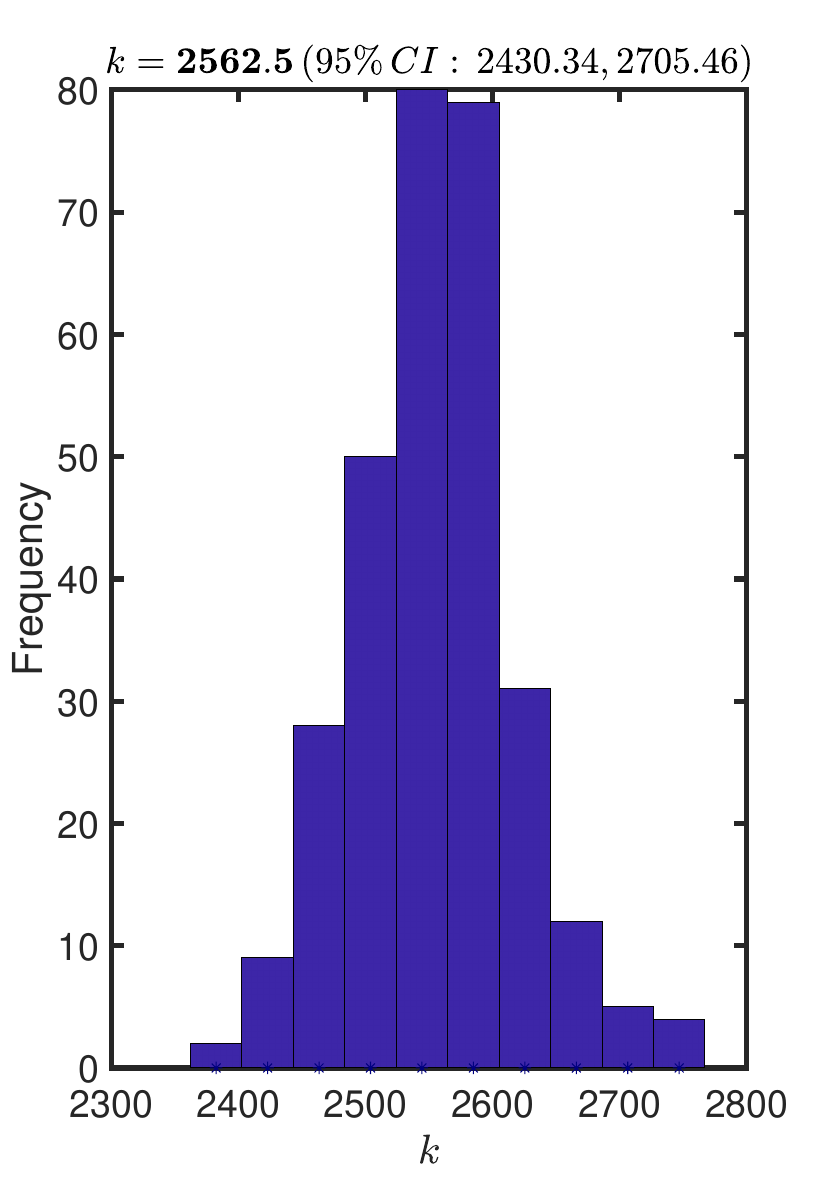}
	\includegraphics[width=0.2\textwidth]{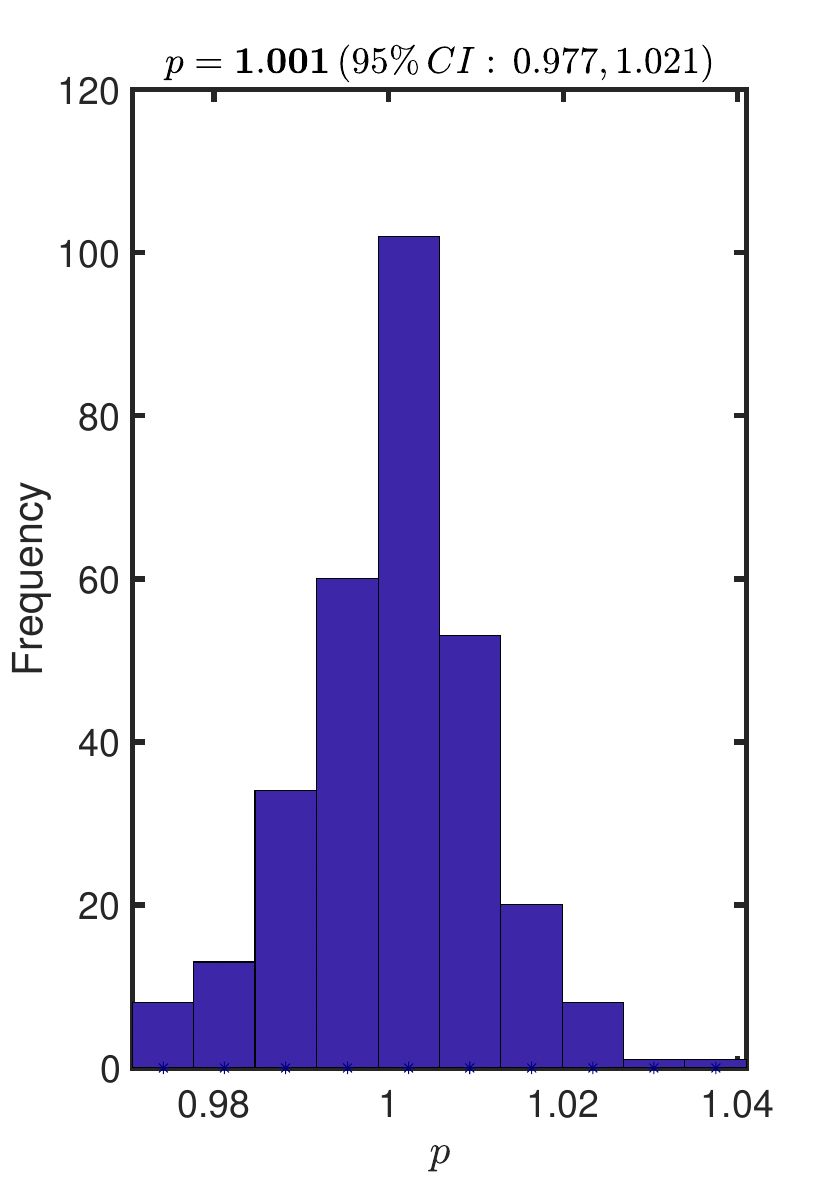}
	\includegraphics[width=0.2\textwidth]{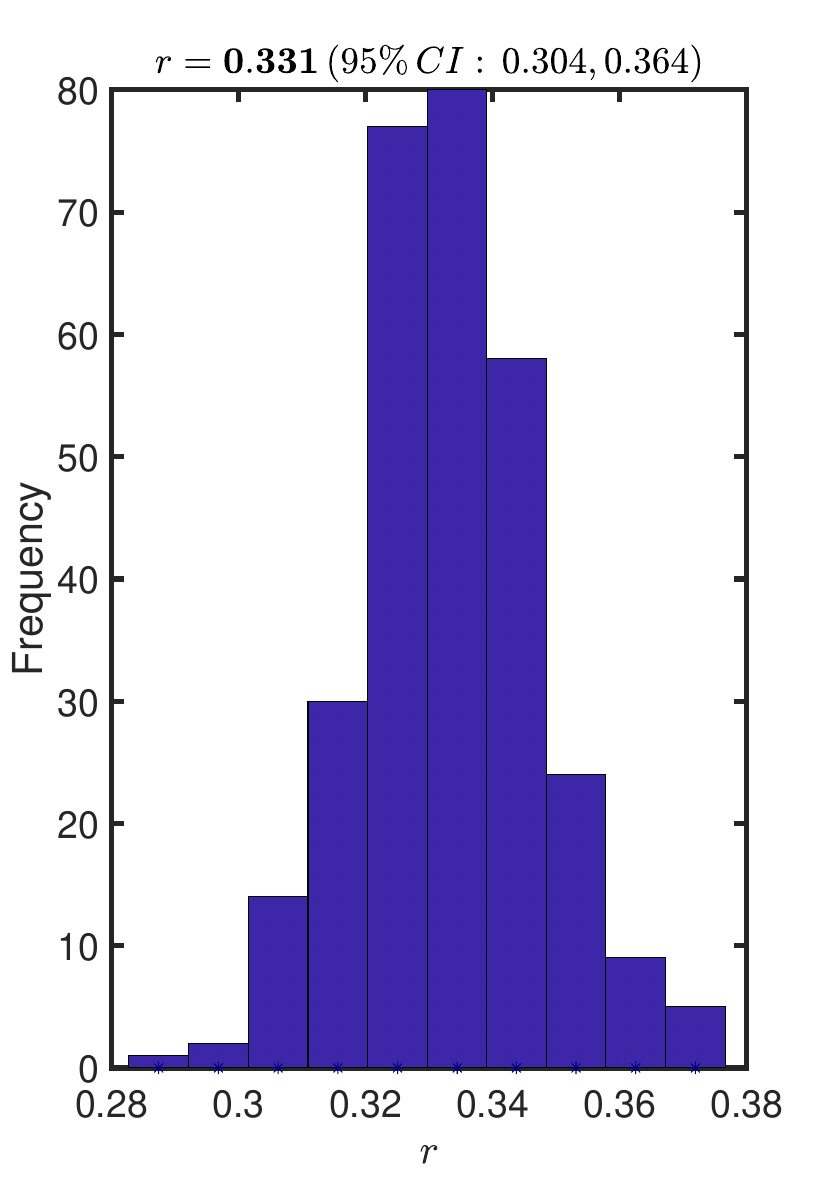}
	\caption{Histograms $r$, $p$, and $k$ parameters.}
	\label{fig:parameters_Lung_injury_1}
\end{figure}

The histogram summaries (Figures~\ref{fig:parameters_Lung_injury}--\ref{fig:parameters_Lung_injury_1}) present the posterior, or sampling, distributions of the GLM parameters $(r,k,p)$ under BFF and QDF. Each panel reports the mean, median, and 95\% confidence interval (CI), enabling a direct comparison of parameter uncertainty between the two estimation frameworks. Overall, both methods yield relatively concentrated distributions, suggesting that the GLM parameters are well identified; for example, $p \approx 1.0 \pm 0.02$.

%%%%%%%%%%%%%%%%%%%%%%%%%%%%%%%%%%%%%%%%%%%%%%%%%

\subsection{GLM model for mpox epidemic in the USA}
\begin{figure}[H]
\centering
\includegraphics[width=0.8\linewidth]{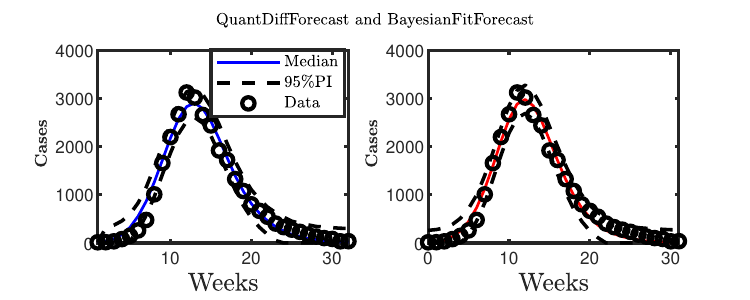}\vspace{0.5cm} 
\caption{Fitting visualization for the GLM model,  with US mpox data, using the QDF and BFF approaches. \textbf{Left:} QDF, \textbf{Right:} BFF.}
\label{fig:glm-mpox-fit}
\end{figure}

For the 2022 U.S.\ mpox epidemic, both methods reproduce the overall rise--fall pattern, indicating that the GLM captures the main outbreak trajectory under either inference framework (Figure~\ref{fig:glm-mpox-fit}). Parameter estimates are broadly consistent across methods, with modest differences in the uncertainty widths (Table~\ref{tab:glm-mpox-params}). For example, $p$ is tightly concentrated around $0.8$--0.85 for both methods, while $r$ differs more in its central estimate (BFF 1.95 vs.\ QDF 2.25), reflecting method-specific trade-offs in fitting the growth and saturation phases.

Calibration-period metrics favor QDF on point-error measures (MAE: 9.45 vs.\ 11.739; MSE:~150.47 vs.\ 399.43; WIS: 5.8 vs.\ 7.175), while both methods achieves coverage near the nominal~95\% level (97.8\% for BFF; 100\% for QDF), with QDF’s 100\% indicating slightly overconservative intervals; Table~\ref{tab:glm-mpox-metrics}). The Bayesian fit converged cleanly ($\hat{R}=1$ for all GLM parameters; Table~\ref{tab:glm-mpox-rhat}), supporting the stability of the reported BFF intervals.

\begin{table}[H]
\centering
\caption{Parameter estimates for the GLM model with US mpox data, using the QDF and BFF approaches.}\vspace{0.2cm} 
\begin{tabular}{lllll}
\hline
& r & p & K  \\ \hline
BFF & 1.95 (1.76, 2.15)  & 0.84 (0.82, 0.85)    & 29,041.01 (27,784.85, 30,164.68)     \\ 
QDF   & 2.25 (2.01, 2.7)  & 0.81 (0.80, 0.83) & 30,363.6 (29,168.42, 31,838.14)    \\ \hline
\end{tabular}
\label{tab:glm-mpox-params}
\end{table}

\begin{figure}[H]
\centering   \includegraphics[width=0.8\textwidth]{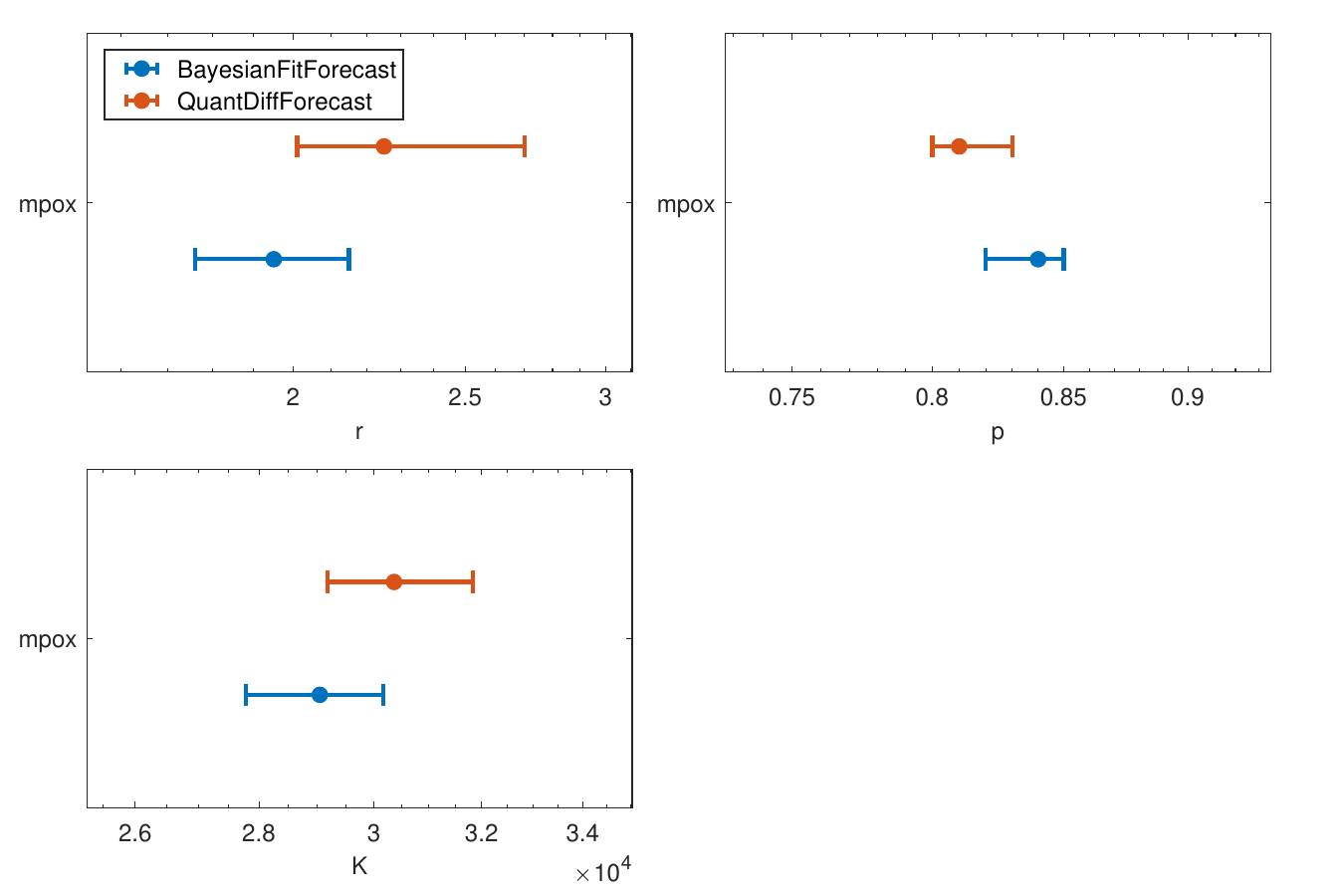}\vspace{0.5cm} 
\caption{Parameter comparison for the GLM model,  with U.S. mpox data, using QDF and BFF approaches. We use a log scale for the x-axis in the $K$ panel.}
\label{fig:glm-mpox-paramcmp}
\end{figure}

\begin{figure}[H]
% \centering
\centering
\includegraphics[width=0.75\textwidth]{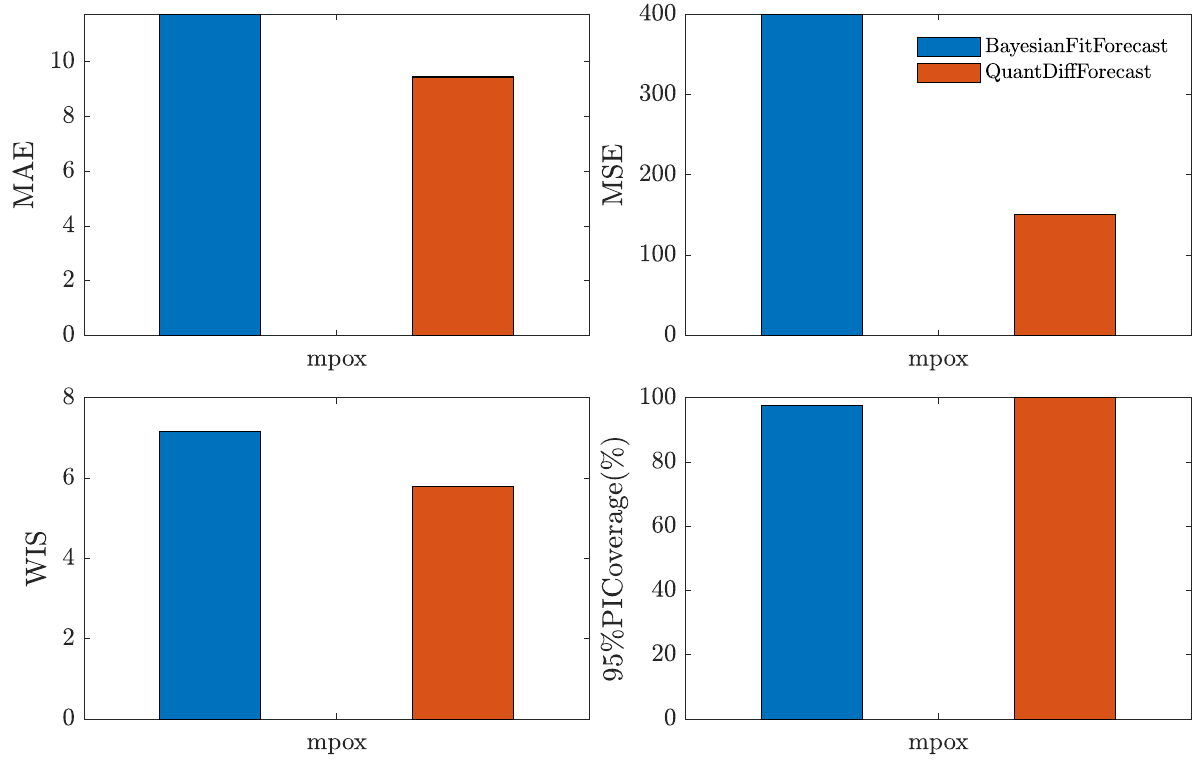}
\caption{Performance metrics comparison for the GLM model with US mpox data, using the QDF and BFF approaches.}
\label{fig:glm-mpox-perf}
\end{figure}

The cross-method parameter comparison (Figure~\ref{fig:glm-mpox-paramcmp}) and metric summary (Figure~\ref{fig:glm-mpox-perf}) reinforce these trends: parameter uncertainty is generally tight under both methods, and QDF achieves lower MAE, MSE, WIS while maintaining near-nominal PI coverage.

\begin{table}[H]
\centering
\caption{Performance metrics for the GLM model with US mpox data, using the QDF and BFF approaches.}\vspace{0.2cm} \setlength{\tabcolsep}{7mm} 
\begin{tabular}{lllll}
\hline
& MAE & MSE & WIS &95$\% $PI \\ \hline
\textbf{BFF} & 11.739  & 399.43    & 7.175  & 97.8   \\ 
\textbf{QDF}   & 9.45  & 150.47 & 5.8  & 100 \\ \hline
\end{tabular}

\label{tab:glm-mpox-metrics}
\end{table}
\noindent
\noindent
\begin{table}[H]
\centering
\caption{Convergence diagnostics for the GLM model with US mpox data, using the BFF~approache.}\setlength{\tabcolsep}{15mm}\vspace{0.2cm}  
\begin{tabular}{lllll}
\hline
& $r$ & $p$ & $k$  \\ \hline
$\textbf{$\hat{R} $}$ & 1     & 1    & 1        \\ \hline
\end{tabular}

\label{tab:glm-mpox-rhat}
\end{table}

The parameter bounds used for applying the BFF and QDF for this model are provided in Tables~\ref{tab:parameter_ranges_GLM_mpox_BFF} and~\ref{tab:parameter_ranges_GLM_mpox_QDF}.

\label{sec:glm_mpox_hists}

% ---------- GLM (mpox): parameter bounds tables ----------
\label{sec:glm_mpox_params_bff}

\begin{table}[H]
\centering
\caption{Parameter ranges for the GLM on mpox data under BFF with normal error.}\vspace{0.2cm} 
\setlength{\tabcolsep}{6mm} 
\begin{tabular}{llll}
\hline
Parameter & Prior distribution & Lower bound (LB) & Upper bound (UB) \\
\hline
$r$ & Normal $(0.5, 0.5)$ T$[0,\ ]$ & 0.0 & NA \\

$p$ & Normal $(0.5, 0.5)$ T$[0,\ ]$ & 0.0 & NA \\

$k$ & Normal $(29,751, 10)$ T$[0,\ ]$ & 0.0 & 30,000 \\
\hline
\end{tabular}
\label{tab:parameter_ranges_GLM_mpox_BFF}
\end{table}

\begin{table}[H]
\centering
\caption{Parameter ranges for the GLM on mpox data under QDF with normal error.}\setlength{\tabcolsep}{7mm} \vspace{0.2cm} 
\begin{tabular}{lll}
\hline
Parameter & Lower bound (LB) & Upper bound (UB) \\
\hline
$r$     & 0.01            & 5.0              \\

$p$      & 0.0            & 2.0              \\

$k$     & 0.01            & 1,000,000.0                         \\

%\bottomrule
\hline
\end{tabular}

\label{tab:parameter_ranges_GLM_mpox_QDF}
\end{table}

\begin{figure}[H]
	\centering
	\includegraphics[width=0.3\textwidth]{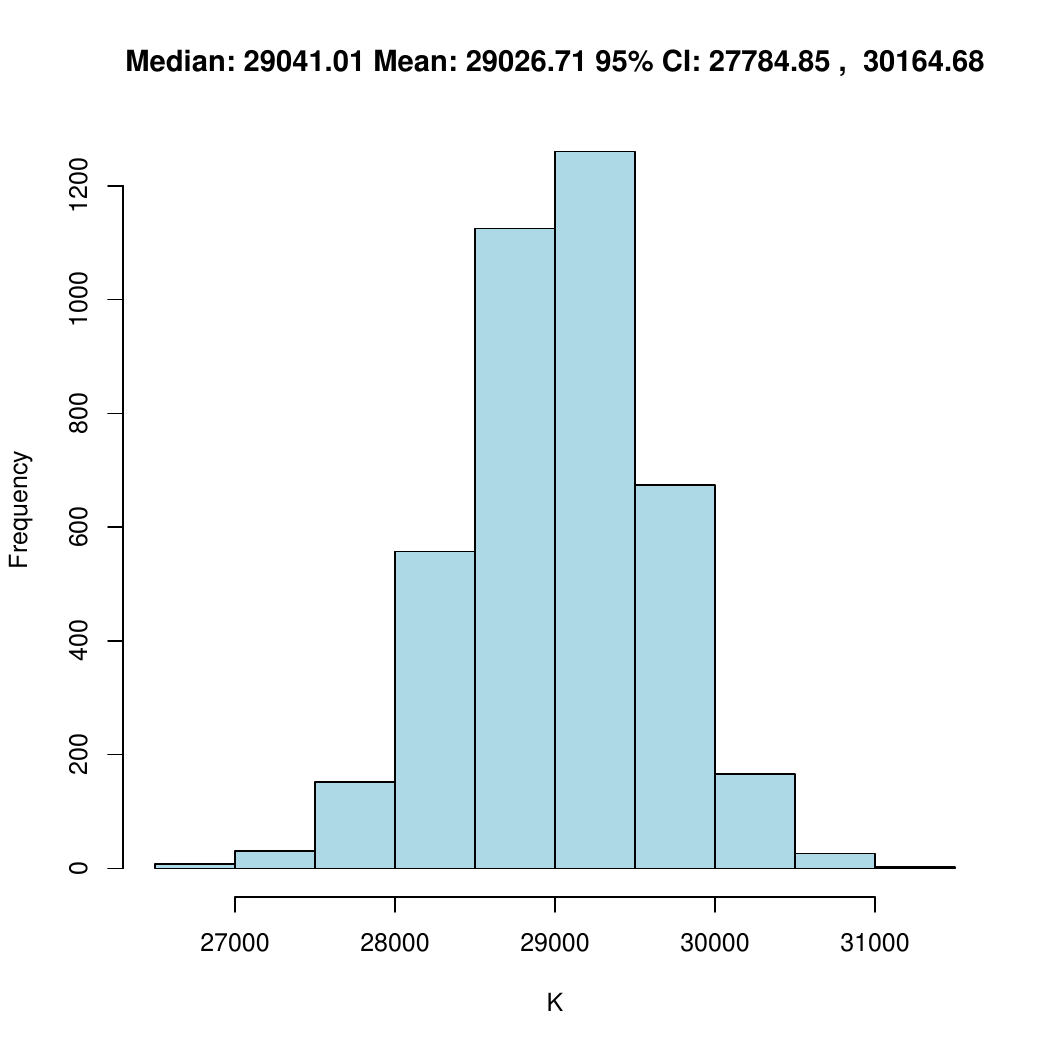}
	\includegraphics[width=0.3\textwidth]{K-histogram-GLM-mpox-normal-cal-32.pdf}\\
	\centering\includegraphics[width=0.3\textwidth]{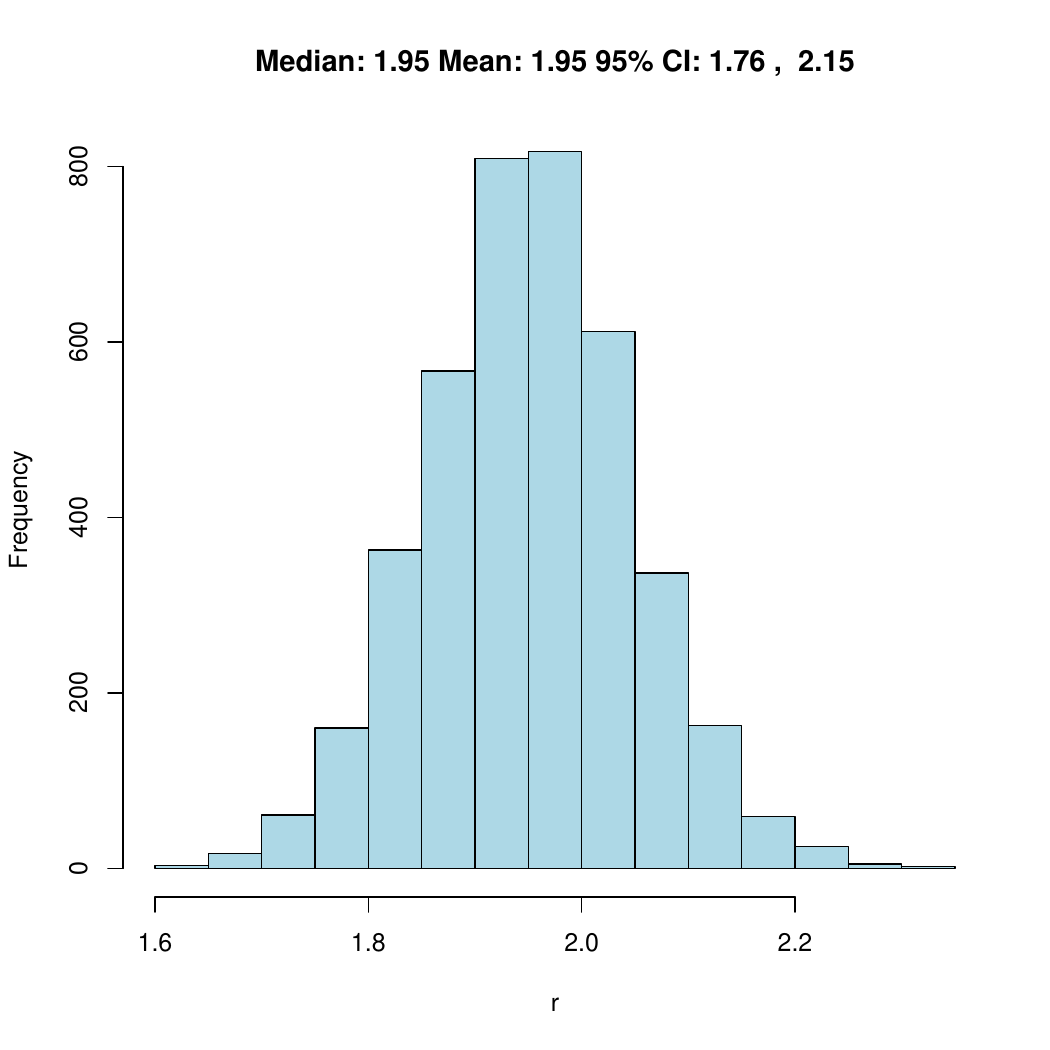}
	\caption{Histograms for GLM data with the $r$, $k$, and $p$ parameters.}
	\label{fig:parameters_mpox}
\end{figure}

\begin{figure}[H]
	\centering
	\includegraphics[width=0.2\textwidth]{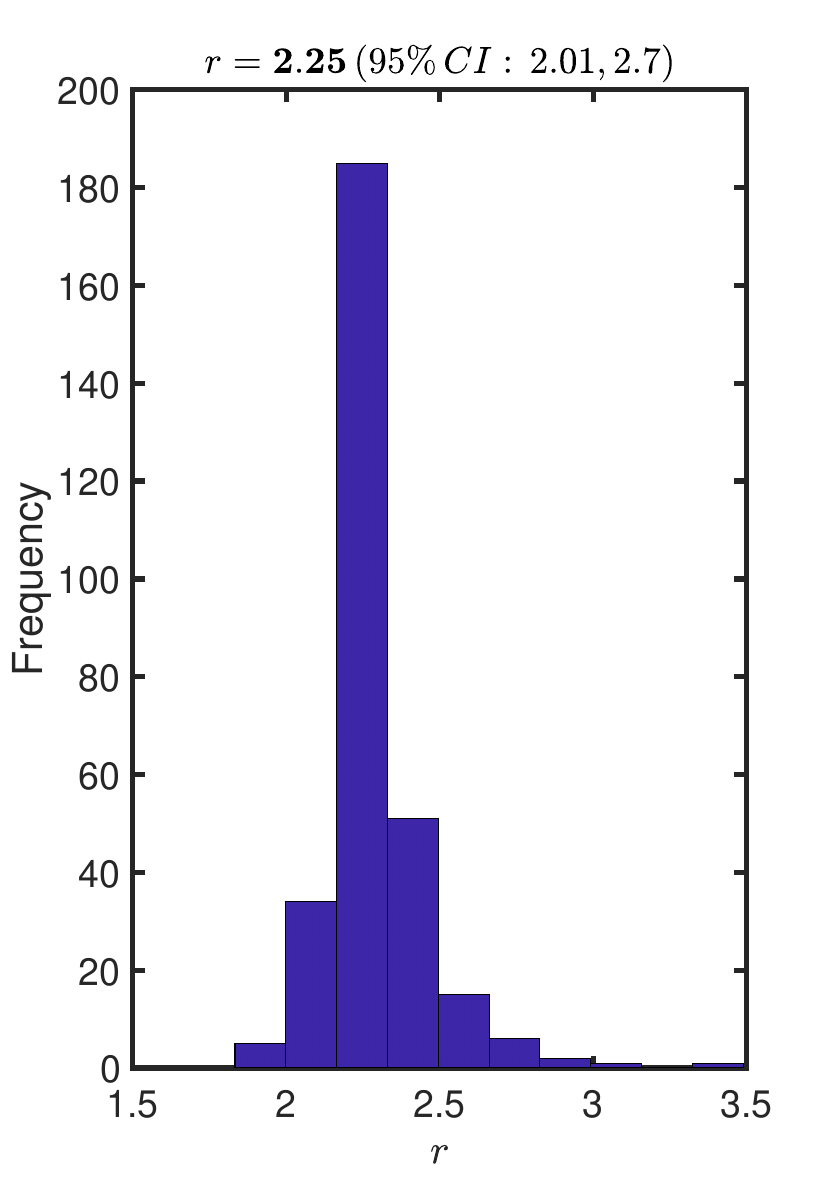}
	\includegraphics[width=0.2\textwidth]{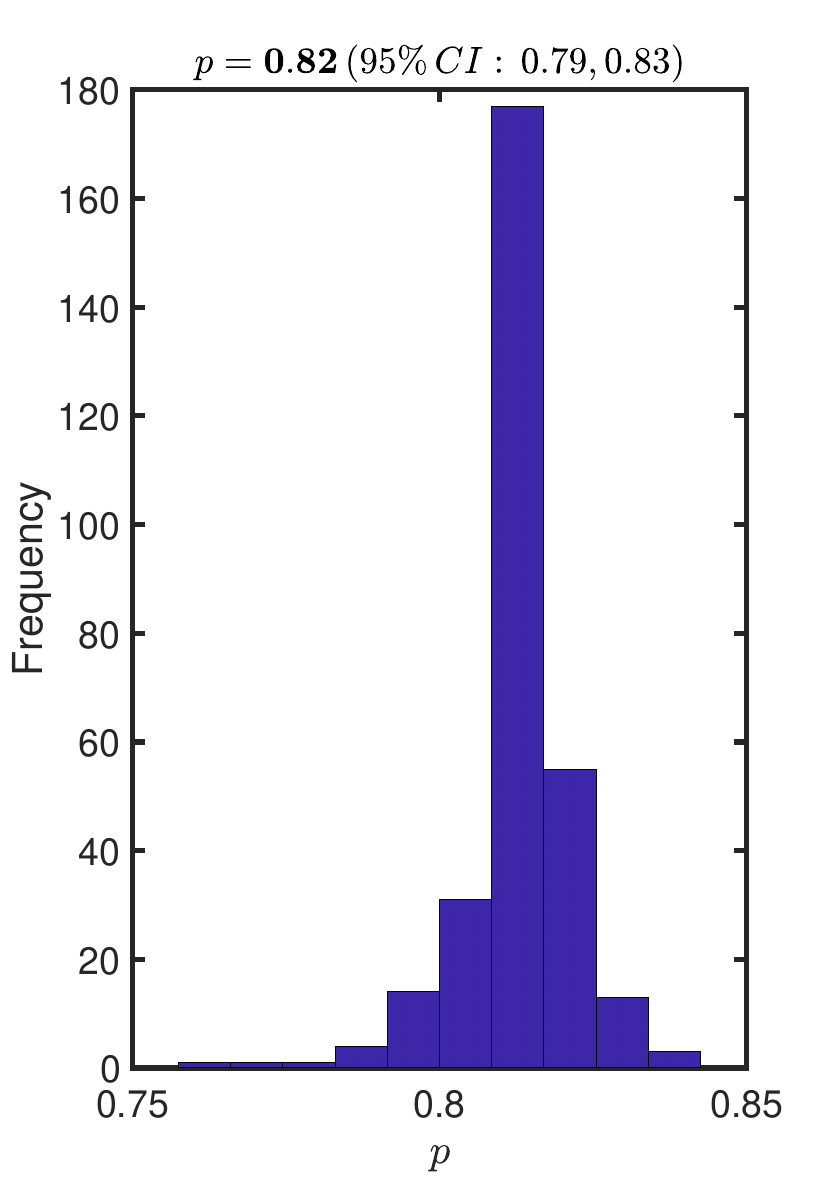}\\
	\includegraphics[width=0.2\textwidth]{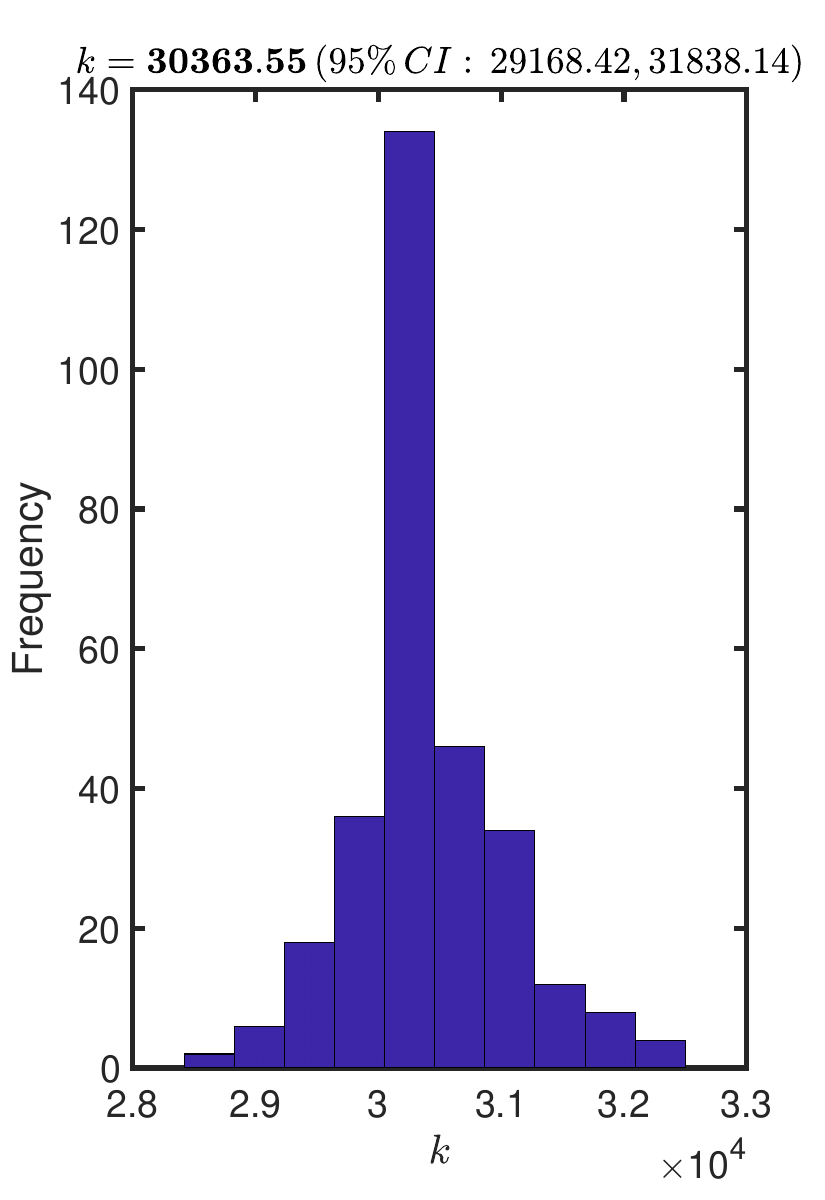}
	\caption{Histograms for GLM data with the $r$, $k$, and $p$ parameters.}
	\label{fig:parameters_mpox_1}
\end{figure}

The histogram summaries (Figures~\ref{fig:parameters_mpox}--\ref{fig:parameters_mpox_1}) show the posterior distributions under BFF and the sampling distributions under QDF for the GLM parameters $(r,k,p)$ fitted to the mpox incidence data. For each parameter, the panels report the mean, median, and 95\% confidence interval (CI), allowing direct side-by-side comparison of uncertainty between the two estimation frameworks. Overall, both methods produce tight, largely unimodal distributions, suggesting that the GLM parameters are well identified. In particular, the parameters such as $p$, $k$, and $r$ exhibit concentrated uncertainty summaries across the two approaches, though the exact figure labeling should be checked carefully before~submission.

%%%%%%%%%%%%%%%%%%%%%%%%%%%%%%%%%%%%%%%%%%%%%%%%%%

\subsection{SEIUR model of COVID-19 cases in Spain during the first wave}

\begin{figure}[H]
\centering   
\includegraphics[width=0.8\textwidth]{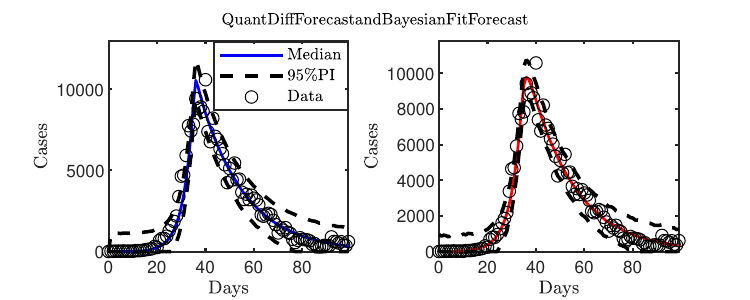}  \vspace{0.5cm} 
\caption{Fitting visualization for the SEIUR model with Spain’s first COVID-19 wave data, using the QDF and BFF approaches. \textbf{Left:} QDF, and \textbf{Right:} BFF.}
\label{fig:model_4_SEIR}
\end{figure}

For Spain’s first COVID-19 wave, both methods track the rise and decline in reported incidence (Figure~\ref{fig:model_4_SEIR}), but the inferred uncertainties and error metrics differ in ways that are consistent with the SEIUR model’s limited identifiability under incidence-only observations (Section~\ref{sec:SI}). 

Compared with QDF, BFF yields lower errors (MAE 294.57 vs.\ 352.30; MSE 210,754 vs.\ 311,054; WIS 211.30 vs.\ 250.46) with similar PI coverage (94\% vs.\ 92\%; Table~\ref{tab:seiur-metrics}), suggesting an advantage for Bayesian regularization in this partially observed, higher-dimensional setting.

The parameter summaries (Table~\ref{tab:parameter_estimates_bayesian}) differ most for $\rho$ and $q_1$, which are strongly coupled to unobserved compartments and are therefore weakly constrained by incidence alone. This helps explain why the two frameworks can yield different central estimates despite producing similar fitted~trajectories.

The convergence diagnostics indicate some remaining sampling difficulty for the Bayesian SEIUR fit (Table~\ref{tab:seiur-rhat}; several parameters have $\hat R>1.1$), which is expected in a coupled, weakly identified setting. Accordingly, we interpret SEIUR posterior summaries with appropriate caution.

\begin{table}[H]
\centering
\caption{Performance metrics for the SEIUR model,  with Spain’s first COVID-19 wave data, using the QDF and BFF approaches.}\vspace{0.2cm} \setlength{\tabcolsep}{7mm} 
\begin{tabular}{lllll}
\hline
& MAE & MSE & WIS &95$\% $PI \\ \hline
BFF & 294.57  &  210,754.1222   & 211.303  &  94  \\ 
QDF   & 352.2978  & 311,053.5723 & 250.4568  & 92 \\ \hline
\end{tabular}
\label{tab:seiur-metrics}
\end{table}

\begin{table}[H]
\centering
\caption{Parameter estimates for the SEIUR model with Spain’s first COVID-19 wave data, using the QDF and BFF approaches.}\vspace{0.2cm} 
\resizebox{\textwidth}{!}{%
\begin{tabular}{lllllll}
\hline
& \textbf{$\beta_0$} & \textbf{$\beta_1$} & \textbf{$q_1$} & \textbf{$\rho$} & \textbf{$\kappa$} & 
\textbf{$\gamma_1$}  \\ \hline
BFF
&   2.18 (1.6, 3.09)
&   1.69 (1.14, 2.24)
&   0.88 (0.64, 1.41)
&   0.6 (0.28, 0.92)
&   2.39 (1.23, 3.73)
&   1.78 (1.22, 2.59)
\\

QDF
& 1.93 (1.18, 1.96)  
& 1.35 (0.54, 1.38)  
& 1.99 (1.13, 2.00)  
& 0.96 (0.72, 1.00)  
& 1.62 (0.61, 1.75)  
& 1.44 (0.65, 1.47)  
\\ \hline
\end{tabular}%
}

\label{tab:parameter_estimates_bayesian}
\end{table}

\begin{table}[H]
\centering
\caption{Convergence diagnostics for the SEIUR model with Spain’s first COVID-19 wave data, using the BFF approach.}\setlength{\tabcolsep}{7mm} \vspace{0.2cm} 
\begin{tabular}{lllllllll}
\hline
& \textbf{$\beta_0$} & \textbf{$\beta_1$} & \textbf{$q_1$} & \textbf{$\rho$} & \textbf{$\kappa$} & 
\textbf{$\gamma_1$}    \\ \hline
$\textbf{$\hat{R}$}$ & 1.18     & 1.22    & 1.17  & 1 & 1.07 & 1.21      \\ \hline
\end{tabular}	
\label{tab:seiur-rhat}
\end{table}

\begin{figure}[H]
\centering
\makebox[\textwidth][c]{%
\includegraphics[width=0.8\textwidth]{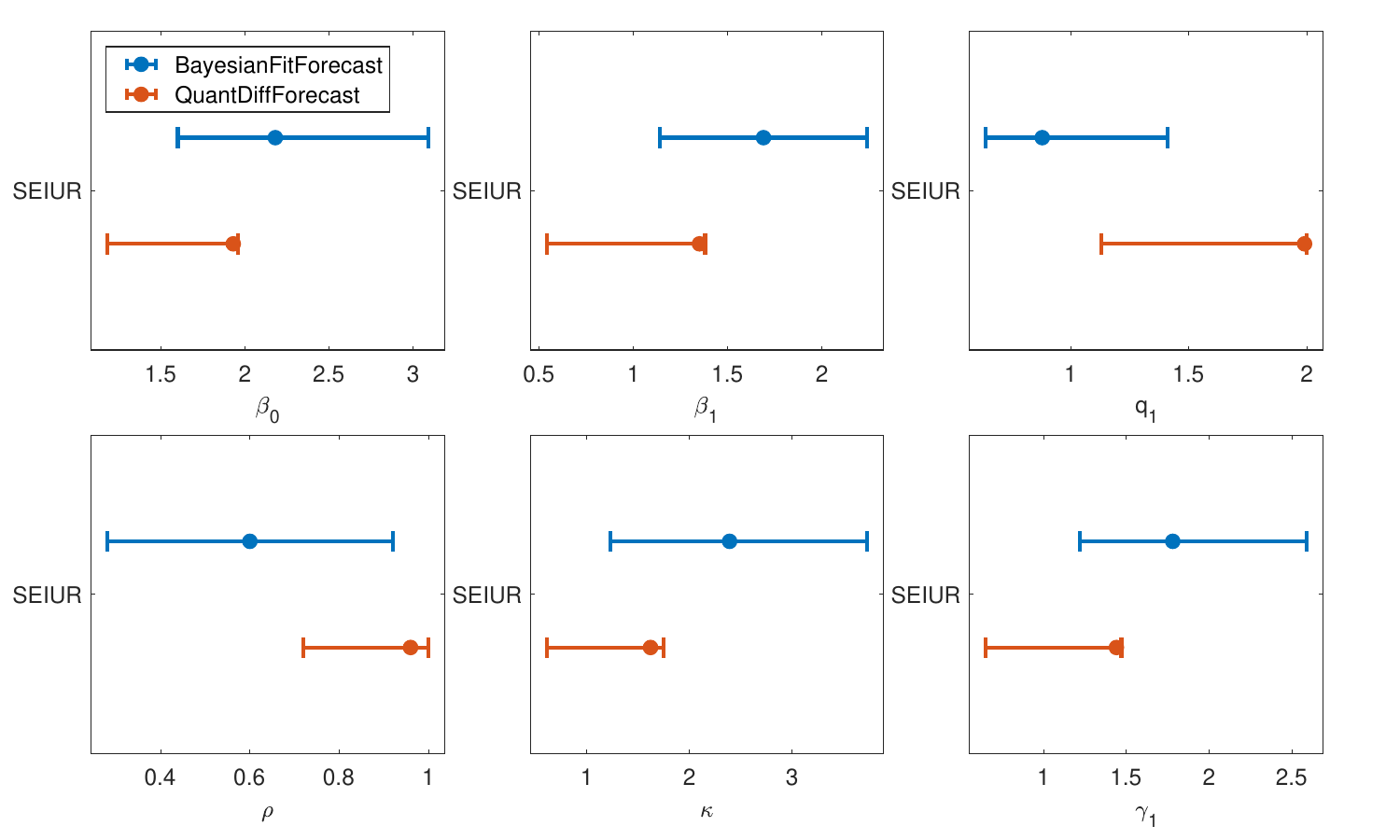}}\vspace{0.5cm} 
\caption{Parameter comparison for the SEIUR model,  with Spain’s first COVID-19 wave data, using BFF approach.}
\label{fig:seiur-paramcmp}
\end{figure}

The parameter comparison (Figure~\ref{fig:seiur-paramcmp}) highlights broader uncertainty for several rates under BFF, consistent with posterior uncertainty propagation under identifiability limitations.

\begin{figure}[H]
\centering
\includegraphics[width=0.75\textwidth]{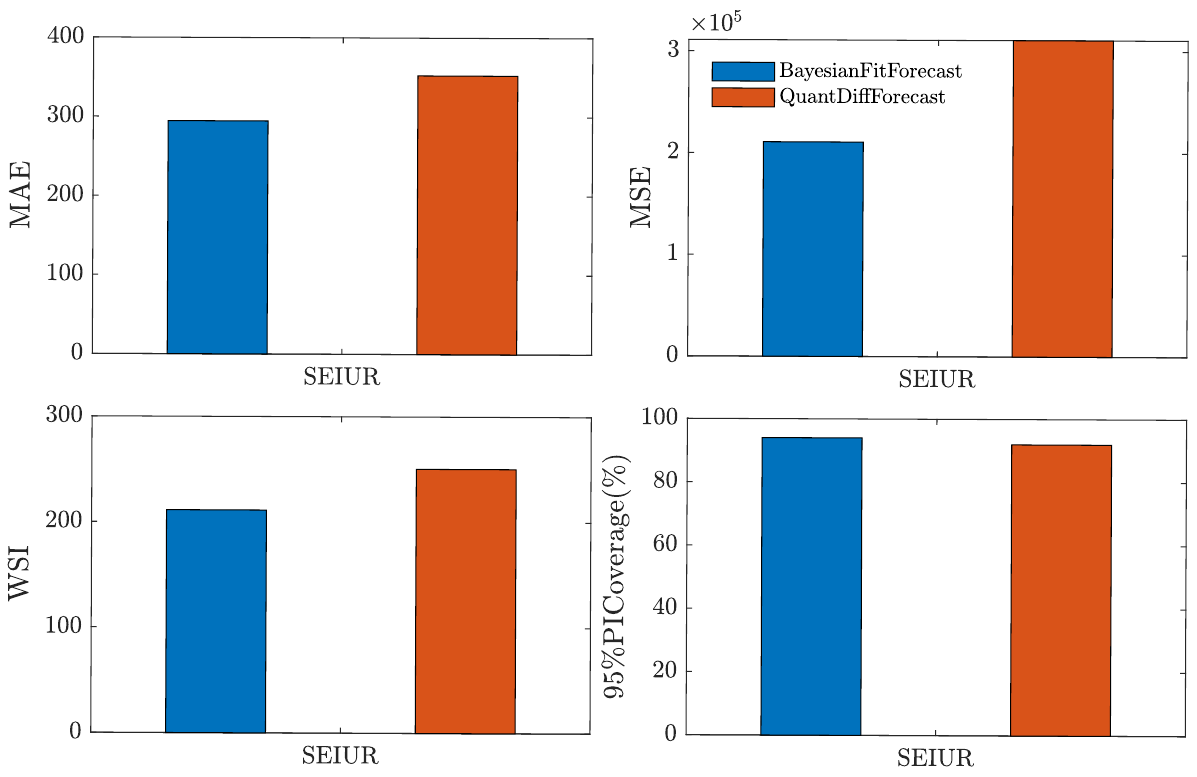} 
\caption{Performance metrics comparison for the SEIUR model,  with Spain’s first COVID-19 wave data, using BFF approach.}
\label{fig:seiur-perf}
\end{figure}

The metric summary (Figure~\ref{fig:seiur-perf}) reinforces the table-based ranking: BFF achieves lower MAE, MSE, WIS while PI coverage remains close to nominal for both methods.

The parameter bounds used for applying the BFF and QDF for this model are provided in Tables~\ref{tab:parameter_ranges_1} and~\ref{tab:parameter_ranges_2}.

\begin{table}[H]
\centering
\caption{Parameter ranges for the SEIUR model with Spain’s first COVID-19 wave data, using the BFF approach.}\vspace{0.2cm} 
\setlength{\tabcolsep}{7mm} 
\begin{tabular}{llll}
\hline
Parameter & Prior distribution & Lower bound (LB) & Upper bound (UB) \\
\hline
$\beta_0$ & Normal $(0.5, 0.5)$ T$[0,\ ]$ & 0.0 & NA \\

$\beta_1$ & Normal $(0.5, 0.5)$ T$[0,\ ]$ & 0.0 & NA \\

$q1$ & Normal $(0.5, 0.5)$ T$[0,\ ]$ & 0.0 & NA \\

$\rho$ & Normal $(0.5, 0.5)$ T$[0,\ ]$ & 0.0 & NA \\

$\kappa$ & Normal $(0.5, 0.5)$ T$[0,\ ]$ & 0.0 & NA \\

$\gamma_1$ & Normal $(0.5, 0.5)$ T$[0,\ ]$ & 0.0 & NA \\
\hline
\end{tabular}
\label{tab:parameter_ranges_1}
\end{table}

\begin{table}[H]
\centering
\caption{Parameter ranges for the SEIUR model with Spain’s first COVID-19 wave data, using the QDF approach.}\setlength{\tabcolsep}{7mm} \vspace{0.2cm} 

\begin{tabular}{llllll}
%\toprule
\hline
Parameter & Lower bound (LB) & Upper bound (UB) \\
\hline
$\beta_0$     & 0.001            & 2.0              \\

$\beta_1$      & 0.001            & 2.0              \\

$q1$     & 0.001            & 2.0                         \\

$\rho$     & 0.001            & 1.0                         \\

$\kappa$     & 0.001            & 2.0                         \\

$\gamma_1$     & 0.001            & 4.0                         \\
\hline
\end{tabular}

\label{tab:parameter_ranges_2}
\end{table}

\begin{figure}[H]
	\centering
	\includegraphics[width=0.20\textwidth]{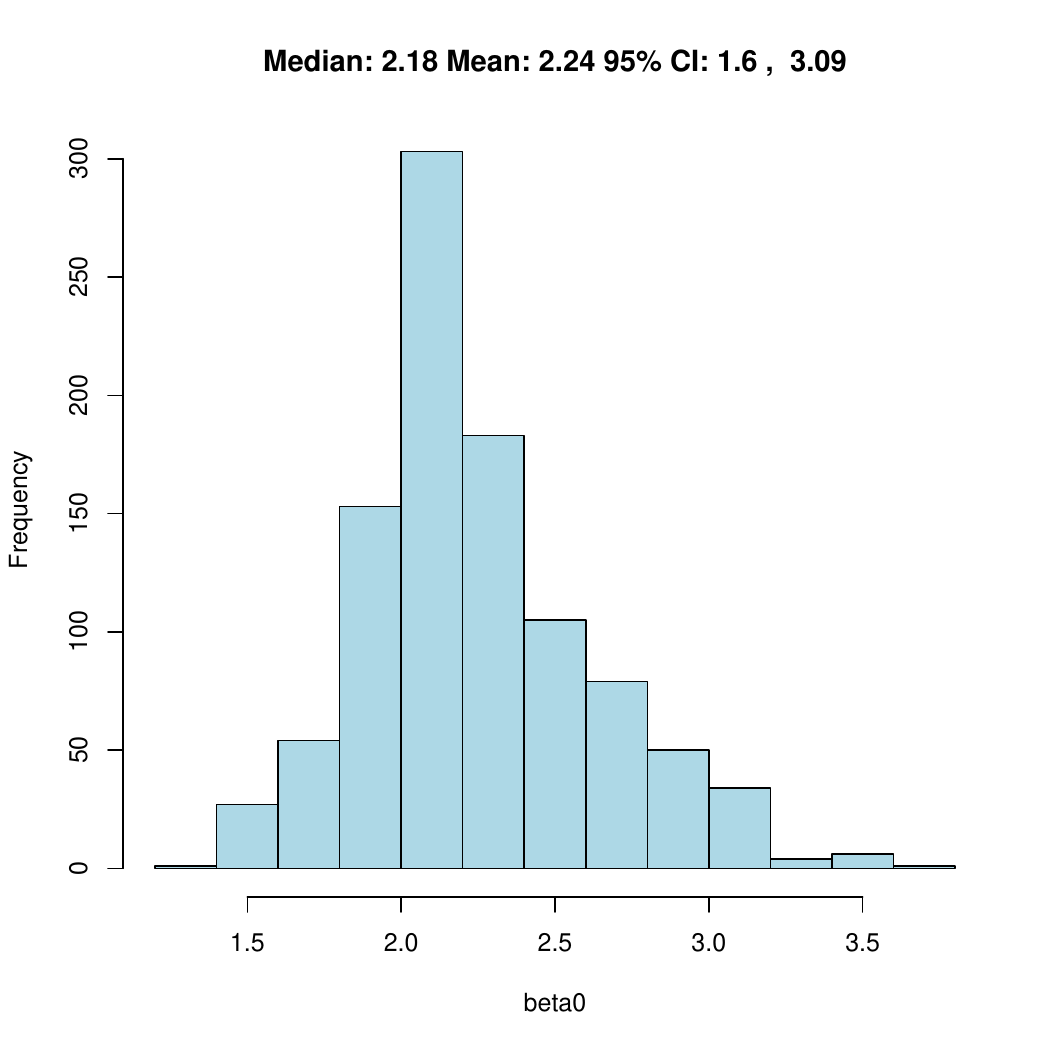}
	\includegraphics[width=0.20\textwidth]{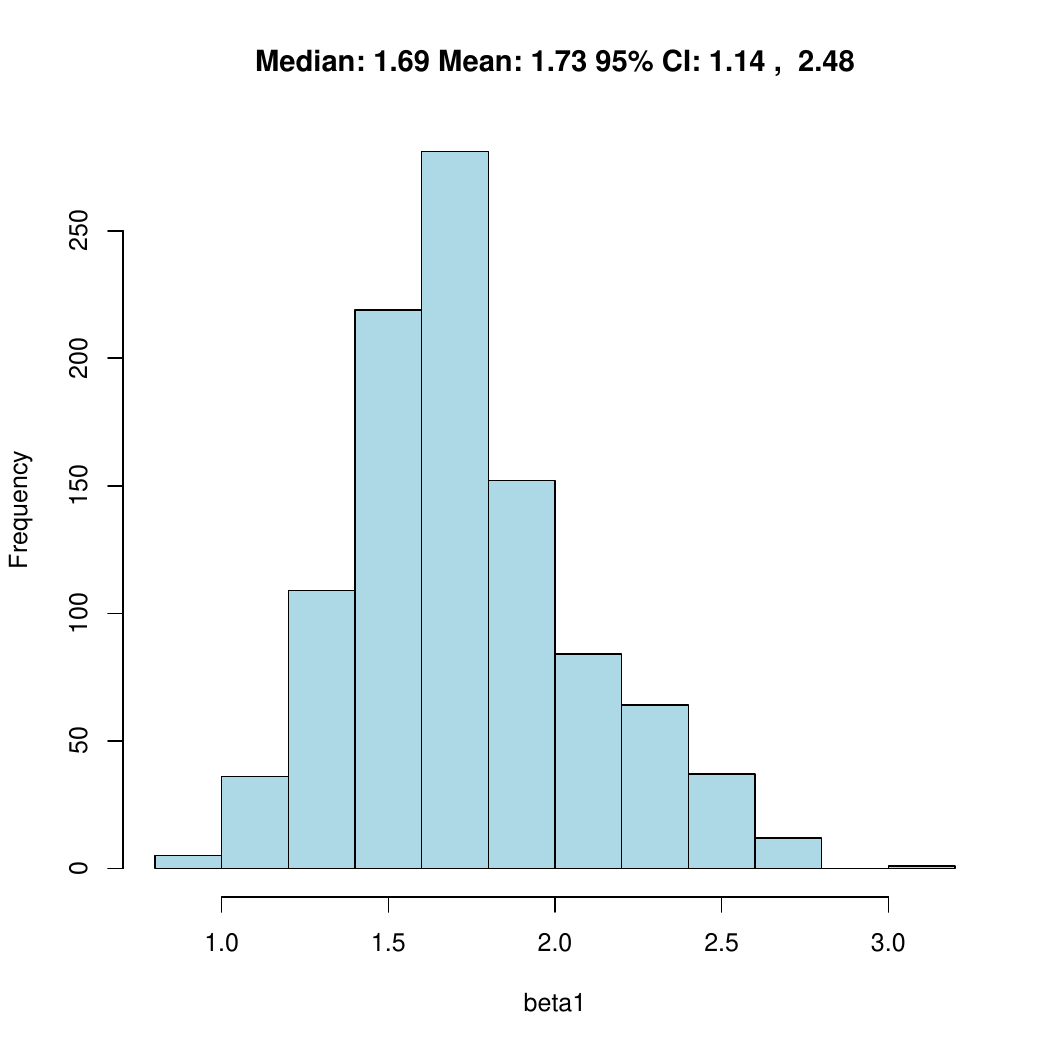}
	\includegraphics[width=0.20\textwidth]{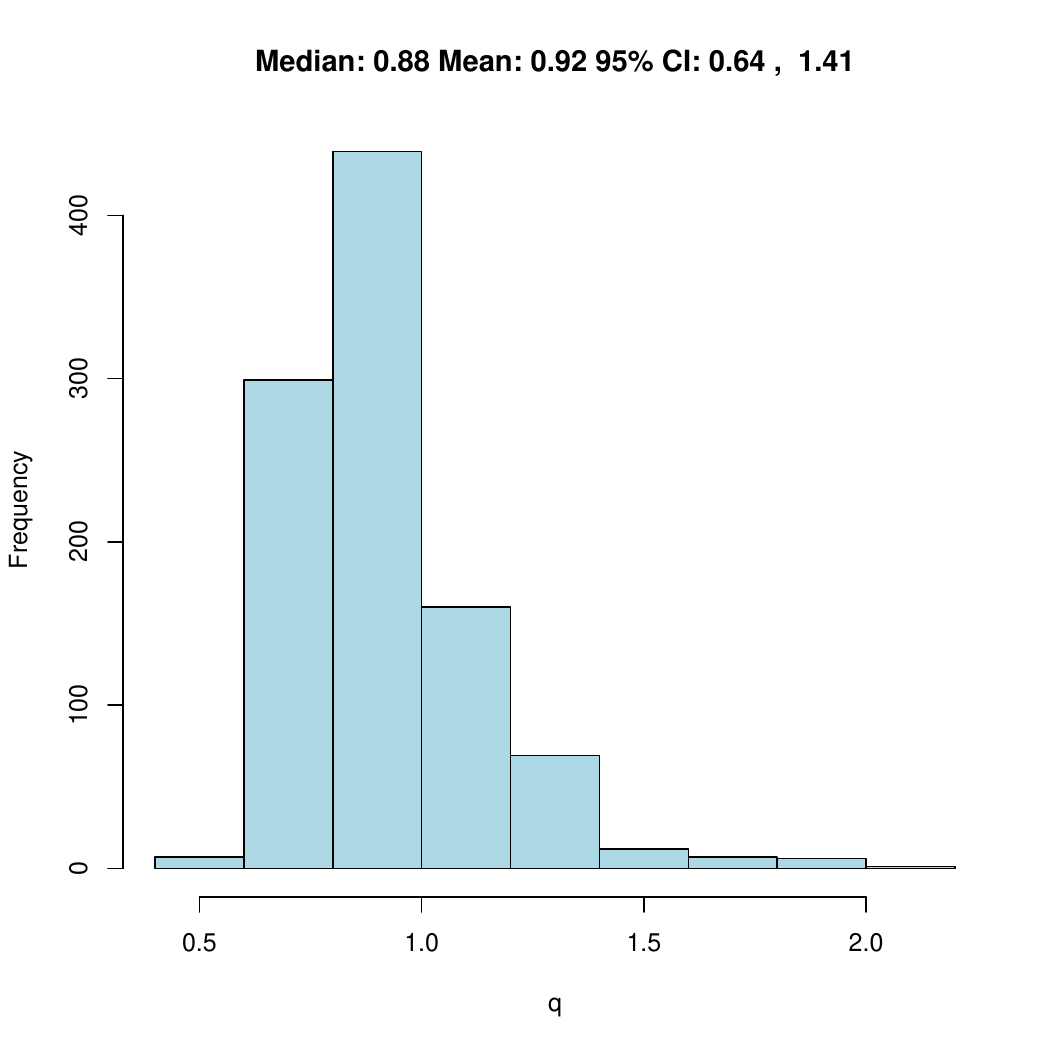}\\
	\includegraphics[width=0.20\textwidth]{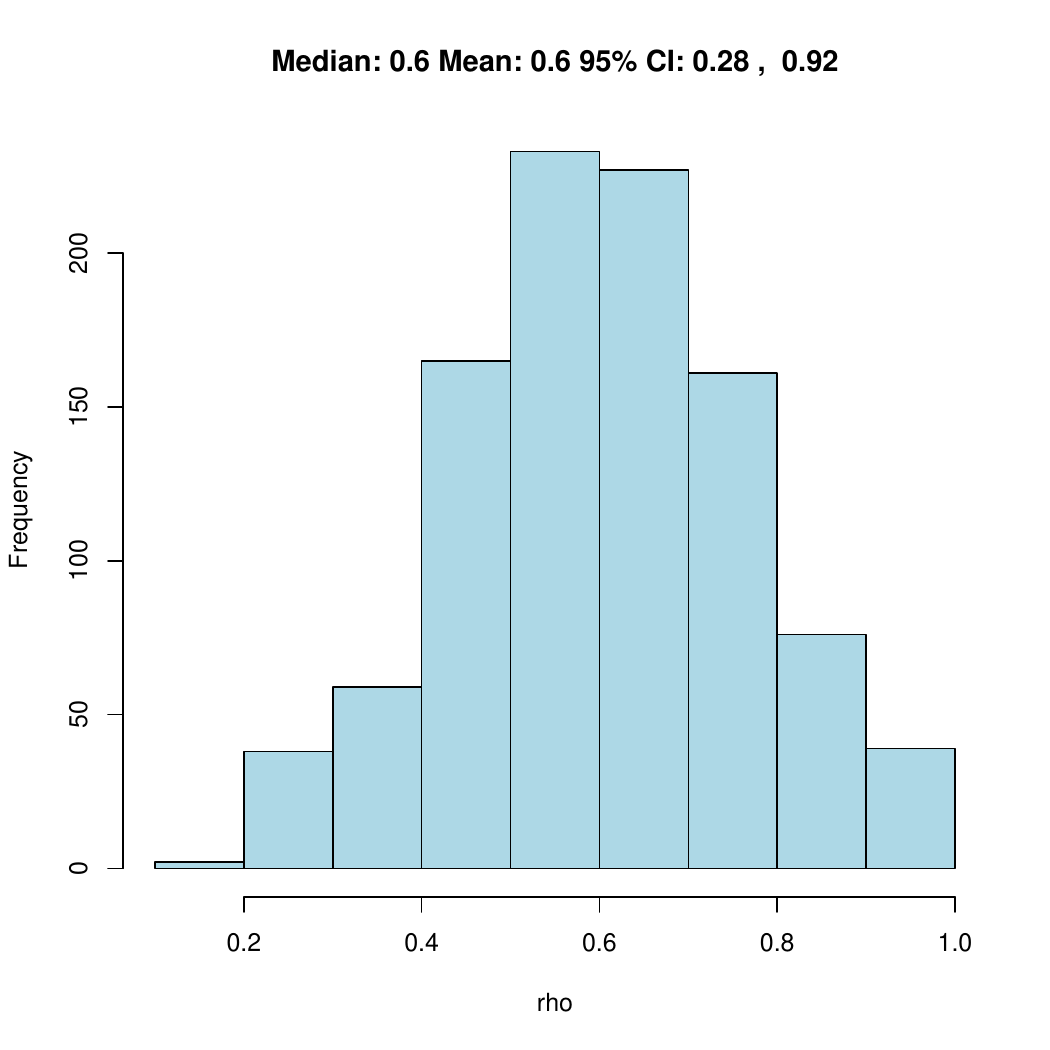}               
	\includegraphics[width=0.20\textwidth]{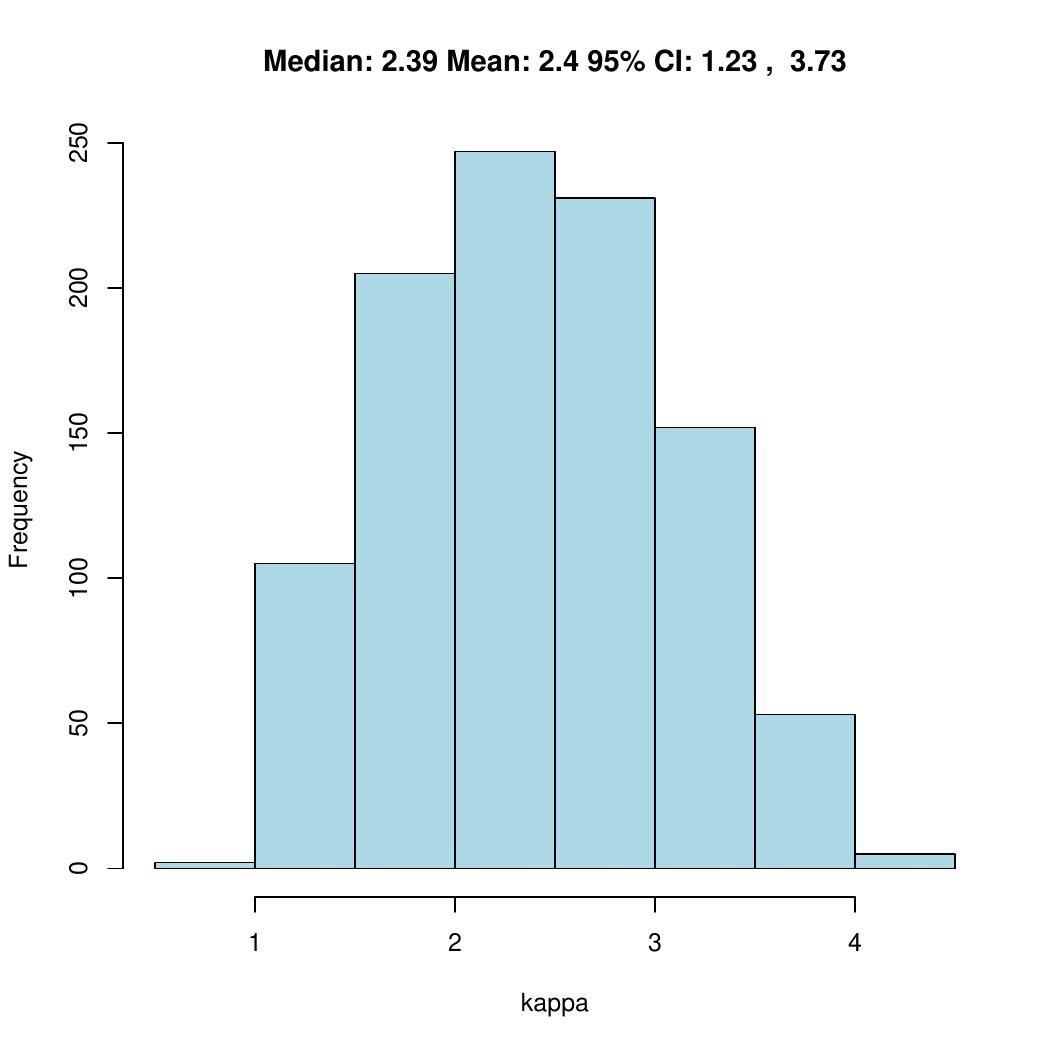}
	\includegraphics[width=0.20\textwidth]{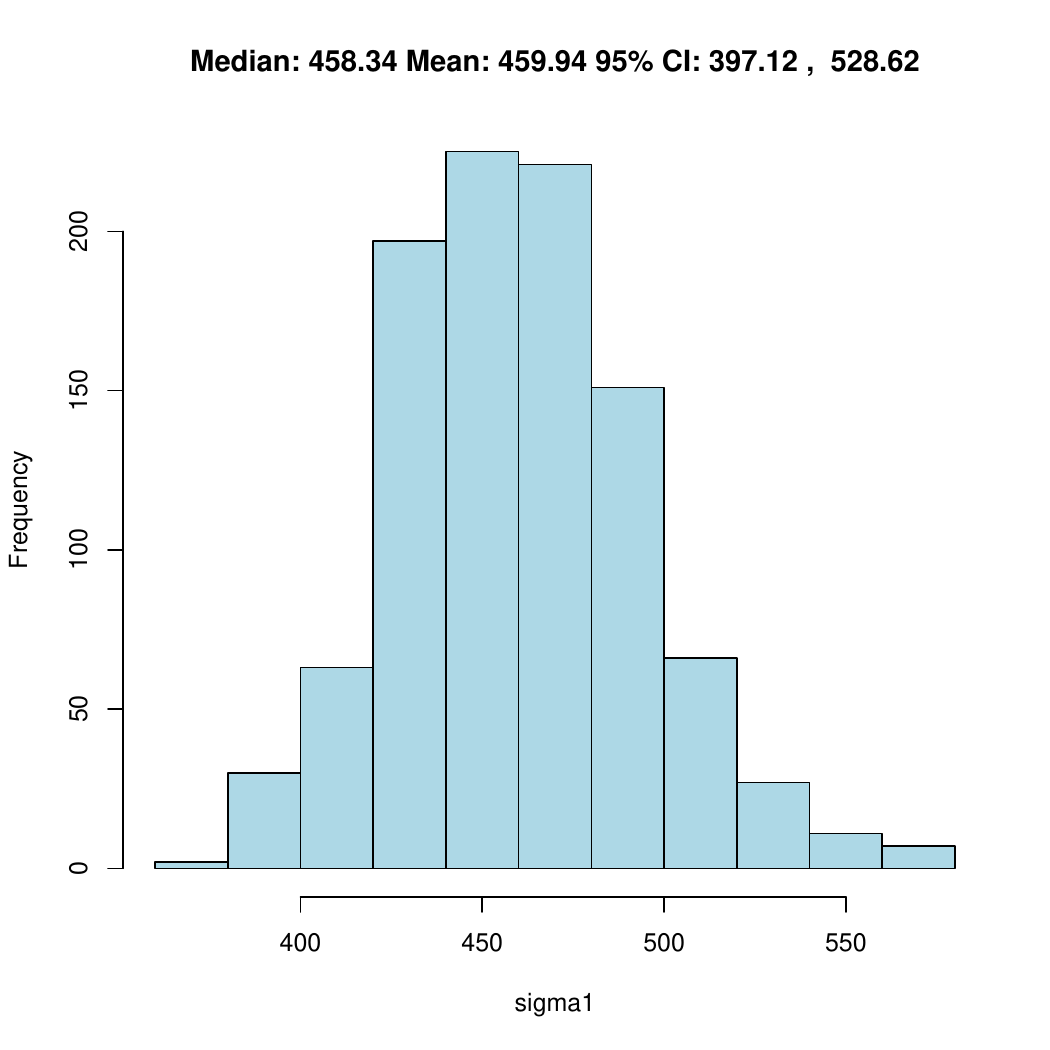}
	\caption{Histograms $\beta_0$, $\beta_1$, $q$, $rho$, $kappa$, and $sigma_1$ parameters.}
	\label{fig:parameters_SEIUR}
\end{figure}

\begin{figure}[H]
	\centering
	\includegraphics[width=0.253\textwidth]{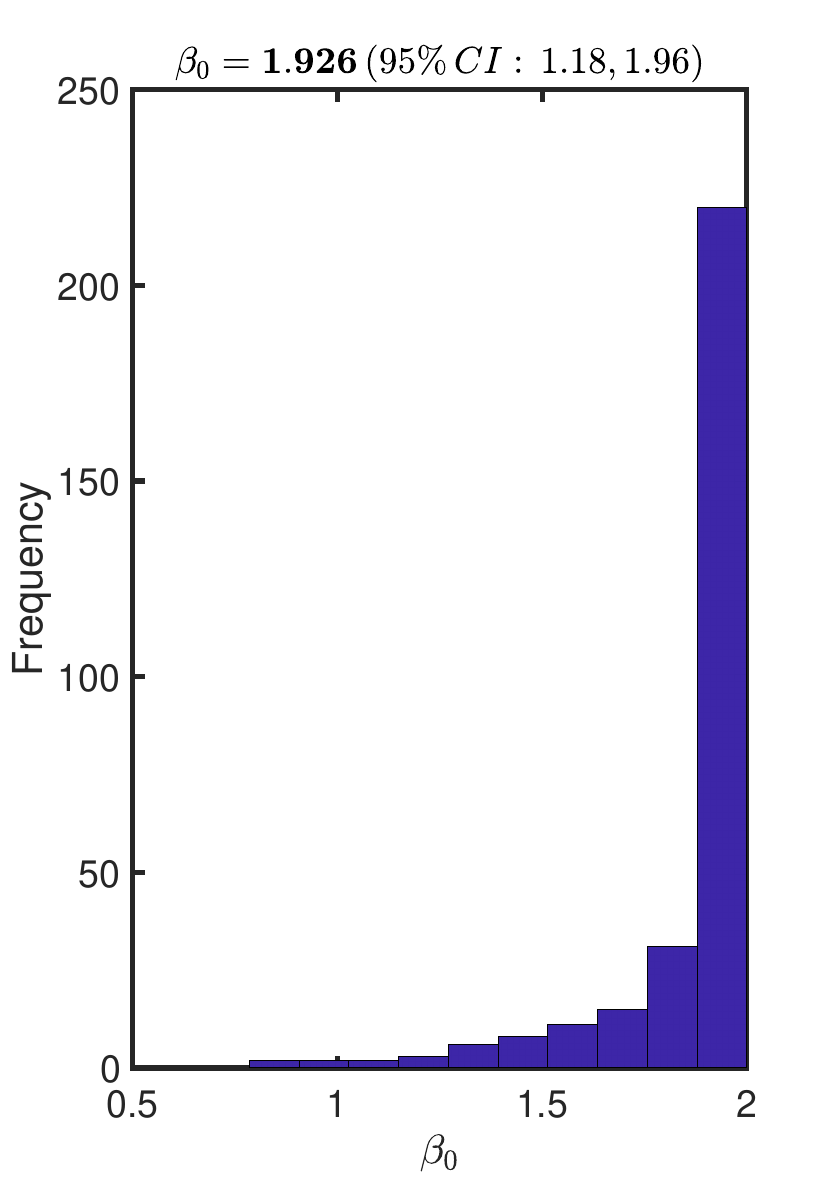}
	\includegraphics[width=0.253\textwidth]{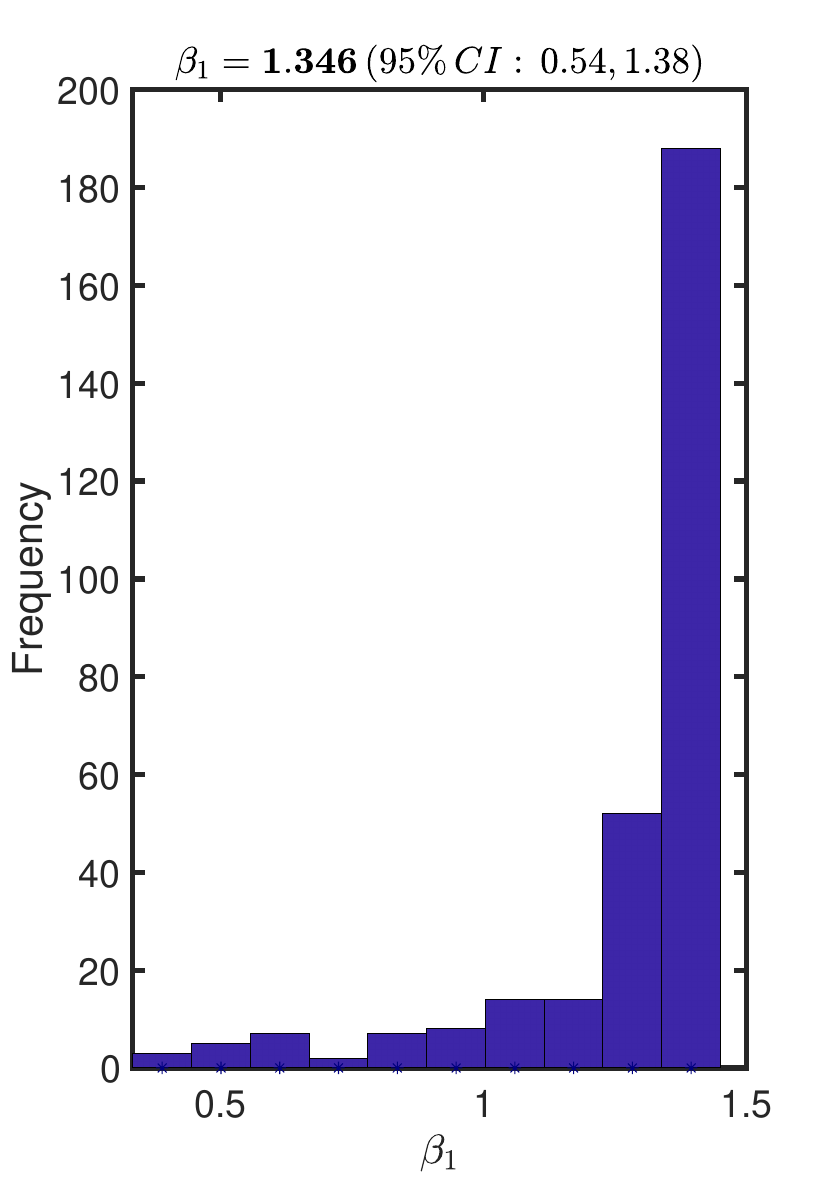}
	\includegraphics[width=0.253\textwidth]{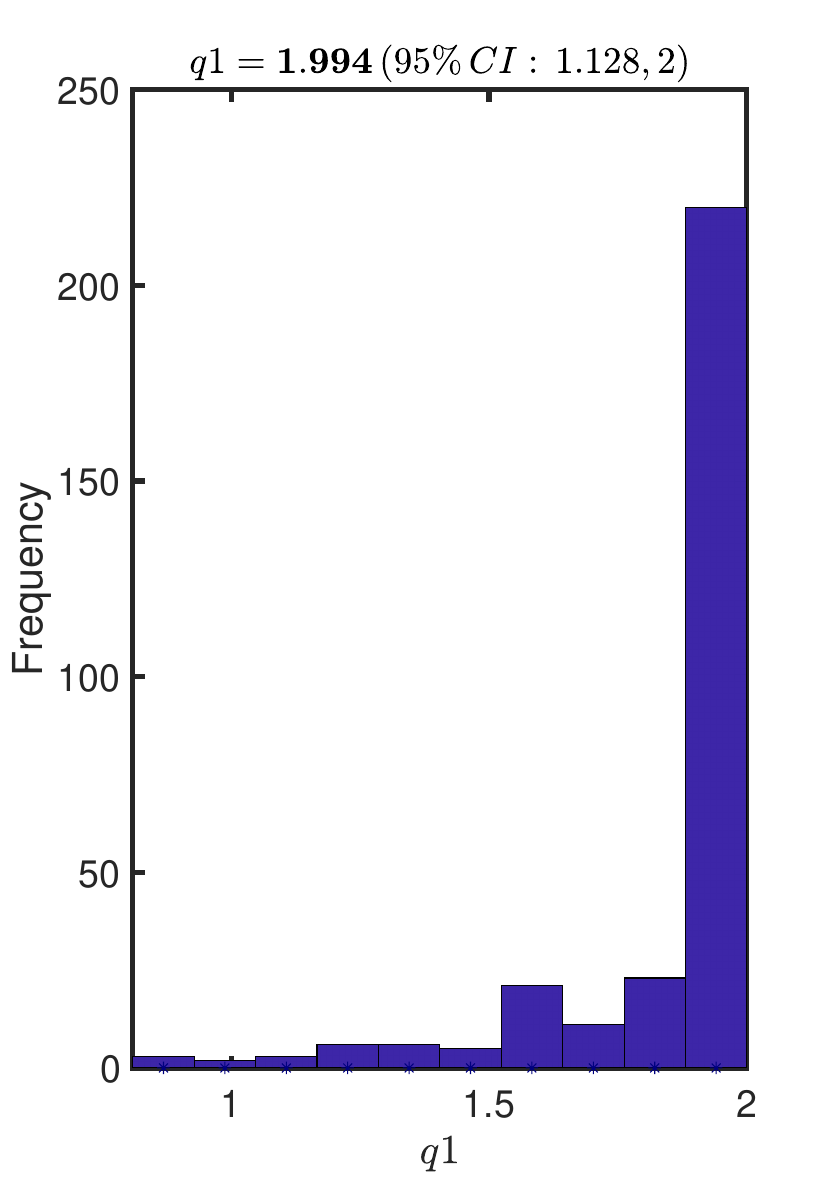}\\
	\includegraphics[width=0.253\textwidth]{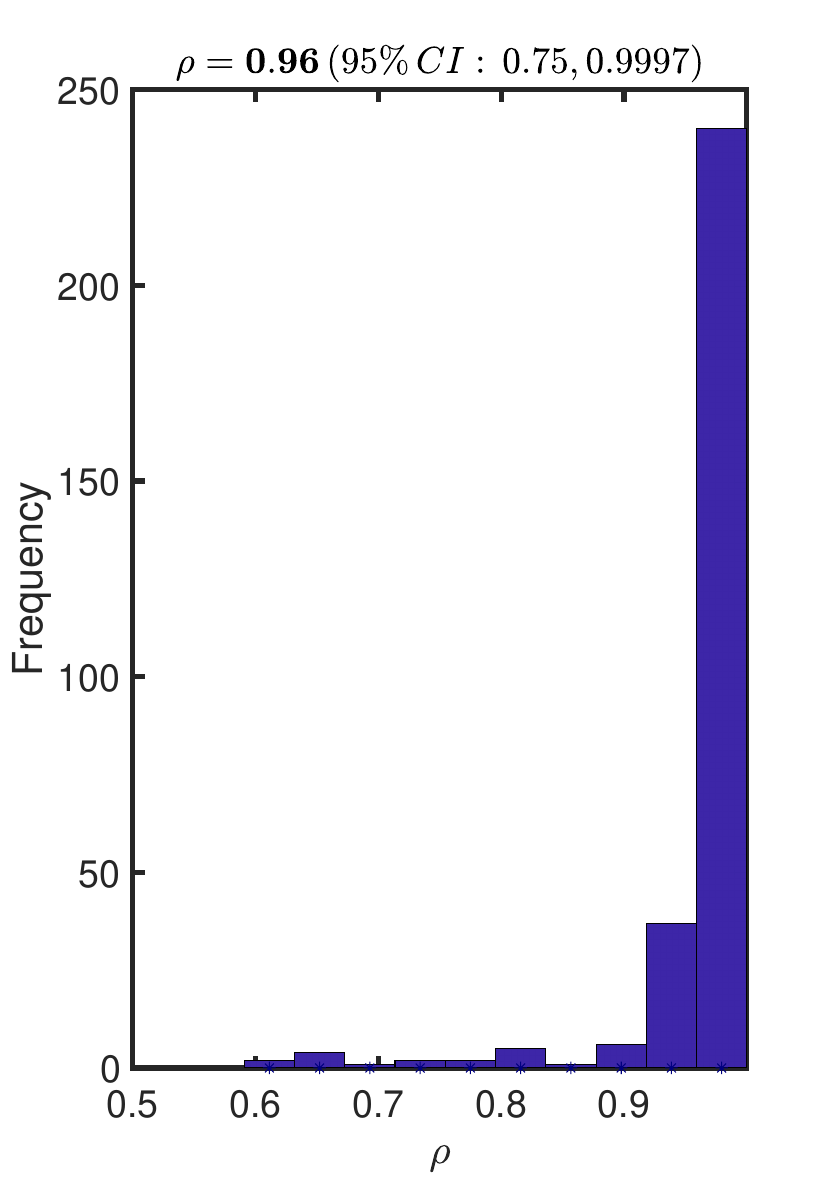}               
	\includegraphics[width=0.253\textwidth]{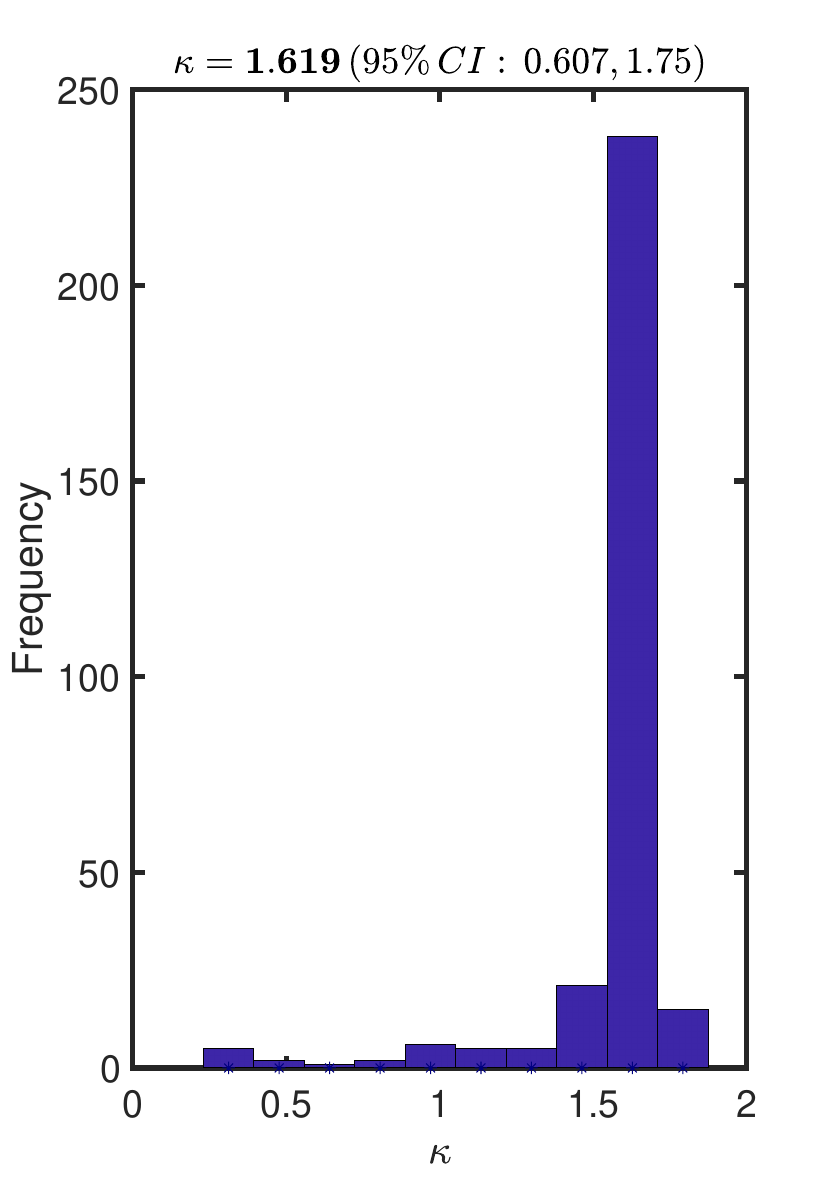}
	\includegraphics[width=0.253\textwidth]{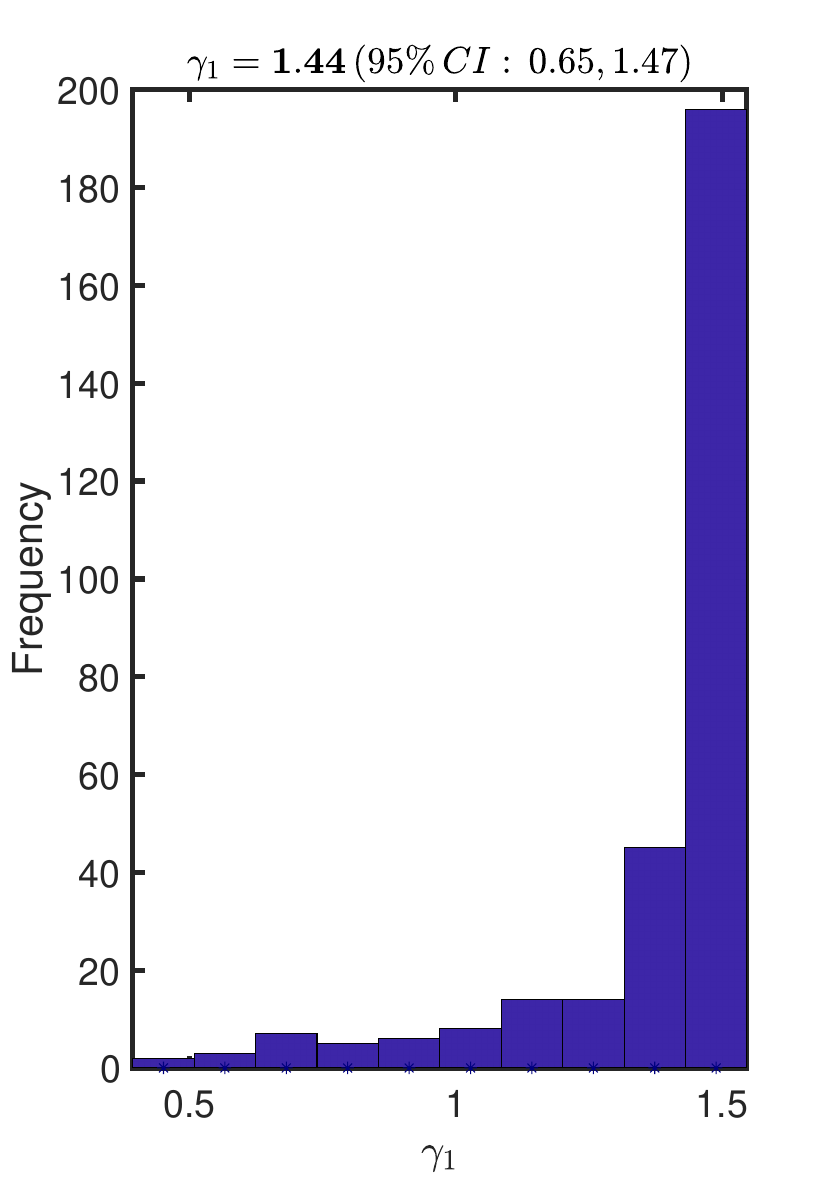}
	\caption{Histograms for $\beta_0$, $\beta_1$, $q$, $\rho$, $\kappa$, and $\gamma_1$ parameters.}
	\label{fig:parameters_SEIUR_1}
\end{figure}

The histogram summaries (Figures~\ref{fig:parameters_SEIUR}--\ref{fig:parameters_SEIUR_1}) show the posterior distributions under BFF and the sampling distributions under QDF for the SEIR model parameters. For each parameter, the panels report the mean, median, and 95\% confidence interval (CI), allowing side-by-side comparison of uncertainty between the two estimation frameworks. The BFF histograms are comparatively broad for several rate parameters, reflecting partial identifiability from the observed data, whereas the QDF histograms tend to be narrower and more clearly unimodal, indicating greater stability in the corresponding estimates. Together, these results emphasize the role of parameter coupling and limited data information in shaping uncertainty, and they suggest that additional data or reparameterization could improve identifiability.

%%%%%%%%%%%%%%%%%%%%%%%%%%%%%%%%%%%%%%%%%%%%%%%%%%

\section{Discussion}

This study compares Bayesian and frequentist inference across ecological, clinical, and epidemiological models, with a focus on how data availability, model structure, and identifiability shape the inference performance. Across all examples, we find that neither framework universally outperforms the other. Instead, performance depends critically on the degree of structural identifiability, the presence of latent states, and whether the primary goal is accurate point prediction or reliable uncertainty quantification. Bayesian methods, implemented via BFF, consistently provide stronger uncertainty estimates and diagnostic tools, while frequentist methods, implemented via QDF, often achieve lower point prediction errors with substantially lower computational cost when identifiability is strong.

For the LV model, inference performance varied markedly across observation scenarios. When both prey and predator populations were observed, both methods produced stable and plausible parameter estimates. QDF yielded slightly lower point errors for predator dynamics, while BFF achieved better uncertainty calibration for prey. When only a single population was observed, QDF continued to provide sharper point predictions, whereas BFF maintained superior PI coverage. Bayesian convergence diagnostics indicated stable posterior sampling even under partial observation, confirming the reliability of Bayesian inference in these settings.

Differences in parameter estimates across observation schemes reflect changes in practical identifiability. With full observation of both species, the interaction and turnover parameters are more strongly informed by the data, reducing confounding but also exposing differences between inference philosophies. In this regime, Bayesian and optimization-based approaches may favor different trade-offs when fitting coupled dynamics, leading to larger discrepancies in some parameters. Under partial observation, the LV system becomes more weakly identifiable, as multiple parameter combinations can reproduce similar marginal dynamics for the observed series while uncertainty is absorbed by the unobserved state. In this case, both methods are effectively constrained toward similar parameter regimes through regularization, bounds, or the prior structure, which can make several parameter estimates appear closer despite increased uncertainty.

These empirical findings closely follow the predictions of the structural identifiability analysis. When both the prey and predator are observed, all LV parameters are structurally identifiable, whereas observing only one species leads to nonidentifiability unless the initial conditions are known. This limitation affects both the Bayesian and frequentist approaches equally, underscoring that observability places fundamental limits on inference quality regardless of the  statistical framework.

For the GLM applied to lung injury and mpox outbreaks, QDF consistently achieved better point forecast accuracy while maintaining similar or better uncertainty calibration. In the mpox case, QDF produced lower MAE, MSE, and WIS, while achieving PI coverage at or slightly above the nominal~95\% level, indicating conservative uncertainty estimates rather than improved predictive accuracy. Bayesian inference showed excellent convergence for all parameters, confirming well-sampled posteriors. Similar patterns were observed for lung injury. These results are explained by the strong structural identifiability of the GLM: The single-equation structure and direct observation of incident cases ensure that all parameters are identifiable, even when the initial conditions are unknown. In this setting, optimization-based inference provides accurate estimates efficiently, while Bayesian inference adds computational overhead without clear gains in predictive~accuracy.

A contrasting pattern emerges for the SEIUR model applied to COVID-19 data from Spain. Here, BFF outperformed QDF across all metrics, including point accuracy and uncertainty calibration. The SEIUR model includes latent compartments and exhibits limited structural identifiability when only reported cases are observed and the initial conditions are unknown. Structural identifiability analysis shows that most parameters and state variables are unidentifiable in this setting, with only a small subset remaining identifiable. Under these conditions, Bayesian inference benefits from prior information, which regularizes the ill-posed inverse problem and stabilizes both parameter estimates and forecasts. In contrast, deterministic optimization must search a high-dimensional and weakly constrained parameter space, leading to less reliable results.

Taken together, these results show that structural identifiability sets hard limits on what can be learned from data. No statistical method can recover parameters that are structurally unidentifiable, and identifiability analysis should therefore precede model fitting. When the identifiability is strong and the data are rich, frequentist methods are well suited for minimizing point forecast error and computational cost. When the identifiability is weak, the data are sparse or noisy, or the models include latent states, Bayesian methods provide more reliable uncertainty quantification and stable inference.

Several methodological choices also influence these results. The use of different software environments reflects methodological requirements rather than implementation convenience: BFF relies on Stan-based Bayesian inference supported in R, while QDF depends on deterministic optimization and repeated numerical ODE solutions, for which MATLAB provides highly optimized solvers. Both approaches use identical model structures, data inputs, and evaluation metrics, ensuring that performance differences arise from the inference methodology rather than the computational~platform.

In the practical identifiability analysis, the initial conditions were treated as known and fixed to align with the structural identifiability setting and to isolate differences arising from the inference methodology. Allowing the initial conditions to be estimated would introduce additional confounding, particularly in partially observed systems, and would be expected to increase the uncertainty and variability in parameter estimates. Both QDF and BFF support joint estimation of the initial conditions, and examining the impact of this added uncertainty remains an important direction for future work.

This study intentionally focuses on low-dimensional models to allow transparent interpretation and the application of differential algebraic structural identifiability methods. While this enables a clean methodological comparison, it also imposes limitations. We assumed normal error models, analyzed the systems independently, and considered idealized identifiability conditions. For larger and more realistic models, practical identifiability tools based on Fisher information or sensitivity analysis may be necessary, especially under noisy or incomplete observations. Similarly, while Hamiltonian Monte Carlo performs well for moderate parameter dimensions, higher-dimensional models may require alternative sampling strategies such as adaptive Metropolis or Delayed Rejection Adaptive Metropolis (DRAM).

\subsection*{Practical implications}

Our results provide practical guidance for selecting between Bayesian and frequentist inference~frameworks:

\begin{itemize}
	\item \textbf{Use frequentist methods (QDF)} when: (1) the data are abundant and of high quality, (2) the models are structurally identifiable, (3) computational speed is critical, (4) the primary objective is minimizing point forecast error, and (5) the model's complexity is low to moderate. The GLM applications exemplify these conditions.
	
	\item \textbf{Use Bayesian methods (BFF)} when: (1) the data are sparse, noisy, or partially observed, (2) the models contain latent states or unobserved compartments, (3) the structural identifiability is limited; (4) comprehensive uncertainty quantification is essential; (5) prior information is available to constrain inference; and (6) diagnostic assurance via convergence metrics is important. The SEIUR application exemplifies these conditions.
	
	\item \textbf{Improve both methods} by: (1) increasing observational coverage (e.g., measuring both the prey and the predator rather than one species), (2) obtaining accurate estimates of the initial conditions, (3) incorporating auxiliary data sources to break the parameters' nonuniqueness, and (4) using an SI-guided experimental design to ensure parameters of interest are identifiable.
\end{itemize}

When practical identifiability is limited or interval calibration is critical, BFF is advantageous because posterior sampling naturally captures the parameter and predictive uncertainty while providing convergence assurance (e.g., GLM $\hat{R} = 1$; LV $\hat{R} \approx 1.01$). When the priority is minimizing point forecast error and the data are informative, QDF can be preferable (e.g., LV for prey only; GLM for mpox), though PIs may be under calibrated relative to BFF (e.g., the LV for prey only coverage was 90.48\% vs. BFF’s higher coverage, which reflects more conservative uncertainty rather than improved predictive accuracy). Across all systems, adding informative observations improves both methods (LV for predator + prey vs.\ single-series), and SI analysis clarifies when further accuracy gains are impossible without additional data or constraints.

Future work could explore hybrid inference approaches that combine the computational efficiency of optimization with Bayesian uncertainty quantification, as well as extensions to higher-dimensional, spatial, stochastic, or agent-based models. Structural identifiability analysis could also be integrated more directly into experimental design, guiding data collection before inference is attempted. Incorporating machine learning approaches such as neural ordinary differential equations or physics-informed neural networks may provide complementary advantages when combined with identifiability-aware modeling.

\section*{Use of AI tools declaration}
The authors declare they have not used artificial intelligence (AI) tools in the creation of this article.

%\section*{Acknowledgments}

%All sources of funding of the study must be disclosed

\section*{Conflict of interest}

The authors declare there is no conflict of interest.


\begin{thebibliography}{999}
	
\bibitem{modeling_overview}
 M. J. Keeling, P. Rohani,
 \emph{Modeling Infectious Diseases in Humans and Animals},
Princeton University Press, Princeton, NJ, 2008.
\doilink{https://doi.org/10.1515/9781400841035}

\bibitem{mohammed2024trophiccascade}
M. Mohammed, M. A. Y. Mohammed, A. Alsammani, M. Bakheet, C. Hui, P. Landi,
 Coexistence via trophic cascade in plant–herbivore–carnivore systems under intense predation pressure, preprint, arXiv:2408.04862.


\bibitem{alsammani2025cholera}
A. Alsammani, G. A. M. O. Farah, M. A. Y. Mohammed, M. Yavuz,
 Cholera transmission dynamics with sanitation control measures, preprint, arXiv:2505.08873.


\bibitem{dixon2022comparison}
S. Dixon, R. Keshavamurthy, D. H. Farber, A. Stevens, K. T. Pazdernik, L. E. Charles,
A comparison of infectious disease forecasting methods across locations, diseases, and time,
\emph{Pathogens}, \textbf{11} (2022), 185.
\doilink{https://doi.org/10.3390/pathogens11020185}

\bibitem{cheng2023real}
C. Cheng, W. M. Jiang, B. Fan, Y. C. Cheng, Y. T. Hsu, H. Y. Wu, et al.,
Real-time forecasting of COVID-19 spread according to protective behavior and vaccination: Autoregressive integrated moving average models,
\emph{BMC Public Health}, \textbf{23} (2023), 1500.
\doilink{https://doi.org/10.1186/s12889-023-16419-8}

\bibitem{lutz2019applying}
C. S. Lutz, M. P. Huynh, M. Schroeder, S. Anyatonwu, F. S. Dahlgren, G. Danyluk, et al.,
Applying infectious disease forecasting to public health: A path forward using influenza forecasting examples,
\emph{BMC Public Health}, \textbf{19} (2019), 1--12.
\doilink{https://doi.org/10.1186/s12889-019-7966-8}

\bibitem{shearer2020infectious}
F. M. Shearer, R. Moss, J. McVernon, J. V. Ross, J. M. McCaw,
Infectious disease pandemic planning and response: Incorporating decision analysis,
\emph{PLoS Med.}, \textbf{17} (2020), e1003018.
\doilink{https://doi.org/10.1371/journal.pmed.1003018}

\bibitem{chowell2022ensemble}
G. Chowell, S. Dahal, A. Tariq, K. Roosa, J. M. Hyman, R. Luo,
An ensemble n-sub-epidemic modeling framework for short-term forecasting epidemic trajectories: Application to the COVID-19 pandemic in the USA,
 \emph{PLoS Comput. Biol.}, \textbf{18} (2022), e1010602.
\doilink{https://doi.org/10.1371/journal.pcbi.1010602}

\bibitem{chowell2022sub}
G. Chowell, R. Rothenberg, K. Roosa, A. Tariq, J. M. Hyman, R. Luo,
Sub-epidemic model forecasts during the first wave of the COVID-19 pandemic in the USA and European hotspots,
in \emph{Mathematics of Public Health: Proceedings of the Seminar on the Mathematical Modelling of COVID-19}, Springer, (2022), 85--137.
\doilink{https://doi.org/10.1007/978-3-030-85053-1\_5}

\bibitem{karami2016comparative}
H. Karami, R. Luo, P. Sanaei, G. Chowell,
Comparative study of Bayesian and frequentist methods for epidemic forecasting: Insights from simulated and historical data,
\emph{Stat. Methods Med. Res.}, \textbf{35} (2026), 21--39.
\doilink{https://doi.org/10.1177/09622802251387451}

\bibitem{reich2019collaborative}
N. G. Reich, L. C. Brooks, S. J. Fox, S. Kandula, C. J. McGowan, E. Moore, et al.,
A collaborative multiyear, multimodel assessment of seasonal influenza forecasting in the United States, \emph{PNAS}, \textbf{116} (2019), 3146--3154.
\doilink{https://doi.org/10.1073/pnas.1812594116}

\bibitem{mcgowan2019collaborative}
C. J. McGowan, M. Biggerstaff, M. Johansson, K. M. Apfeldorf, M. Ben-Nun, L. Brooks, et al.,
Collaborative efforts to forecast seasonal influenza in the United States, 2015--2016,
 \emph{Sci. Rep.}, \textbf{9} (2019), 683.
\doilink{https://doi.org/10.1038/s41598-018-36361-9}

\bibitem{biggerstaff2014estimates}
 M. Biggerstaff, S. Cauchemez, C. Reed, M. Gambhir, L. Finelli,
 Estimates of the reproduction number for seasonal, pandemic, and zoonotic influenza: A systematic review of the literature,
\emph{BMC Infect. Dis.}, \textbf{14} (2014), 1--20.
\doilink{https://doi.org/10.1186/1471-2334-14-480}

\bibitem{chowell2017perspectives}
 G. Chowell, C. Viboud, L. Simonsen, S. Merler, A. Vespignani,
 Perspectives on model forecasts of the 2014--2015 Ebola epidemic in West Africa: Lessons and the way forward,
 \emph{BMC Med.}, \textbf{15} (2017), 1--8.
\doilink{https://doi.org/10.1186/s12916-017-0811-y}

\bibitem{funk2019assessing}
S. Funk, A. Camacho, A. J. Kucharski, R. Lowe, R. M. Eggo, W. J. Edmunds,
Assessing the performance of real-time epidemic forecasts: A case study of Ebola in the Western Area region of Sierra Leone, 2014--15,
 \emph{PLoS Comput. Biol.}, \textbf{15} (2019), e1006785.
\doilink{https://doi.org/10.1371/journal.pcbi.1006785}

\bibitem{meltzer2014estimating}
M. I. Meltzer, C. Y. Atkins, S. Santibanez, B. Knust, B. W. Petersen, E. D. Ervin, et al.,
Estimating the future number of cases in the Ebola epidemic---Liberia and Sierra Leone, 2014--2015,
\emph{MMWR Surveillance Summ.}, \textbf{63} (2014), 1--14.


\bibitem{chretien2015mathematical}
 J. P. Chretien, S. Riley, D. B. George,
Mathematical modeling of the West Africa Ebola epidemic,
 \emph{eLife}, \textbf{4} (2015), e09186.
\doilink{https://doi.org/10.7554/eLife.09186}

\bibitem{chowell2014catastrophic}
 G. Chowell, L. Simonsen, C. Viboud, Y. Kuang,
Is West Africa approaching a catastrophic phase or is the 2014 Ebola epidemic slowing down? Different models yield different answers for Liberia,
\emph{PLOS Curr. Outbreaks}, \textbf{2014} (2014).
\doilink{https://doi.org/10.1371/currents.outbreaks.b4690859d91684da963dc40e00f3da81}


\bibitem{roosa2020multimodel}
 K. Roosa, A. Tariq, P. Yan, J. M. Hyman, G. Chowell,
Multimodel forecasts of the ongoing Ebola epidemic in the Democratic Republic of Congo, March--October 2019,
 \emph{J. R. Soc. Interface}, \textbf{17} (2020), 20200447.
\doilink{https://doi.org/10.1098/rsif.2020.0447}

\bibitem{bleichrodt2023mpox_rt}
A. Bleichrodt, S. Dahal, K. Maloney, L. Casanova, R. Luo, G. Chowell,
Real-time forecasting the trajectory of monkeypox outbreaks at the national and global levels, July--October 2022,
 \emph{BMC Med.}, \textbf{21} (2023).
\doilink{https://doi.org/10.1186/s12916-022-02725-2}

\bibitem{bleichrodt2023mpox_ensemble}
 A. Bleichrodt, R. Luo, A. Kirpich, G. Chowell,
 Retrospective evaluation of short-term forecast performance of ensemble sub-epidemic frameworks and other time-series models: The 2022--2023 mpox outbreak across multiple geographical scales, July 14th, 2022, through February 26th, 2023, medRxiv, 2023.
\doilink{https://doi.org/10.1101/2023.05.15.23289989}

\bibitem{charniga2024nowcasting}
 K. Charniga, Z. J. Madewell, N. B. Masters, J. Asher, Y. Nakazawa, I. H. Spicknall,
Nowcasting and forecasting the 2022 US mpox outbreak: Support for public health decision making and lessons learned,
\emph{Epidemics}, \textbf{47} (2024), 100755.
\doilink{https://doi.org/10.1016/j.epidem.2024.100755}

\bibitem{chowell2024growthpredict}
 G. Chowell, A. Bleichrodt, S. Dahal, A. Tariq, K. Roosa, J. M. Hyman, et al.,
GrowthPredict: A toolbox and tutorial-based primer for fitting and forecasting growth trajectories using phenomenological growth models,
 \emph{Sci. Rep.}, \textbf{14} (2024), 1630.
\doilink{https://doi.org/10.1038/s41598-024-51852-8}

\bibitem{maclulich1937hare}
D. A. MacLulich,
 \emph{Fluctuations in the Numbers of the Varying Hare (Lepus americanus)},
University of Toronto Studies, Biological Series, 1937.
\doilink{https://doi.org/10.3138/9781487583064}

\bibitem{hyndman2021forecasting}
 R. J. Hyndman, G. Athanasopoulos,
 \emph{Forecasting: Principles and Practice},
3rd ed., OTexts, Melbourne, Australia, 2021.


\bibitem{chowell2024quantdiffforecast}
 G. Chowell, A. Bleichrodt, R. Luo,
Parameter estimation and forecasting with quantified uncertainty for ordinary differential equation models using QuantDiffForecast: A MATLAB toolbox and tutorial,
\emph{Stat. Med.}, \textbf{43} (2024), 1--23.
\doilink{https://doi.org/10.1002/sim.10036}

\bibitem{karami2025bayesianfitforecast}
H. Karami, A. Bleichrodt, R. Luo, G. Chowell,
 BayesianFitForecast: A user-friendly R toolbox for parameter estimation and forecasting with ordinary differential equations,
 \emph{BMC Med. Inf. Decis. Making}, \textbf{25} (2025), 1--40.
\doilink{https://doi.org/10.1186/s12911-025-03208-z}

\bibitem{raue2009profilelikelihood}
 A. Raue, C. Kreutz, T. Maiwald, J. Bachmann, M. Schilling, U. Klingmüller, et al.,
Structural and practical identifiability analysis of partially observed dynamical models by exploiting the profile likelihood,
 \emph{Bioinformatics}, \textbf{25} (2009), 1923--1929.
\doilink{https://doi.org/10.1093/bioinformatics/btp358}

\bibitem{chis2011structural}
O. T. Chis, J. R. Banga, E. Balsa-Canto,
Structural identifiability of systems biology models: A critical comparison of method,
\emph{PLoS One}, \textbf{6} (2011), e27755.
\doilink{https://doi.org/10.1371/journal.pone.0027755}

\bibitem{atanasov2025honeybee}
 A. Atanasov, S. Georgiev, L. Vulkov,
 Parameter identification analysis of food and population dynamics in honey bee colonies,
in \emph{Advanced Computing in Industrial Mathematics}, Springer, Cham, (2025), 29--41.
\doilink{https://doi.org/10.1007/978-3-031-76782-1\_3}

\bibitem{alsammani2025vaccination}
A. Alsammani, C. N. Ngonghala, M. Martcheva,
 Impact of vaccination behavior on COVID-19 dynamics and economic outcomes,
 \emph{Math. Biosci. Eng.}, \textbf{22} (2025), 2300--2338.
\doilink{https://doi.org/10.3934/mbe.2025084}

\bibitem{gneiting2008probabilistic}
T. Gneiting, Probabilistic forecasting,
\emph{J. R. Stat. Soc. Ser. A: Stat. Soc.}, \textbf{171} (2008), 319--321.
\doilink{https://doi.org/10.1111/j.1467-985X.2007.00522.x}

\bibitem{mwambi2011forceofinfection}
 H. Mwambi, S. Ramroop, L. J. White, E. A. Okiro, D. J. Nokes, Z. Shkedy, et al., A frequentist approach to estimating the force of infection for a respiratory disease using repeated measurement data from a birth cohort,
\emph{Stat. Methods Med. Res.}, \textbf{20} (2011), 551--570.
\doilink{https://doi.org/10.1177/0962280210385749}

\bibitem{chowell2017primer}
 G. Chowell,
Fitting dynamic models to epidemic outbreaks with quantified uncertainty: A primer for parameter uncertainty, identifiability, and forecasts,
\emph{Infect. Dis. Modell.}, \textbf{2} (2017), 379--398.
\doilink{https://doi.org/10.1016/j.idm.2017.08.001}

\bibitem{chowell2020ensembles}
G. Chowell, R. Luo, K. Sun, K. Roosa, A. Tariq, C. Viboud,
Real-time forecasting of epidemic trajectories using computational dynamic ensembles,
 \emph{Epidemics}, \textbf{30} (2020), 100379.
\doilink{https://doi.org/10.1016/j.epidem.2019.100379}

\bibitem{banks2014uncertainty}
H. T. Banks, S. Hu, W. C. Thompson,
 \emph{Modeling and Inverse Problems in the Presence of Uncertainty},
 CRC Press, Boca Raton, FL, 2014.



\bibitem{pruitt2024likelihood}
C. D. Pruitt, A. E. Lovell, C. Hebborn, F. M. Nunes,
The role of the likelihood for elastic scattering uncertainty quantification, preprint, arXiv:2403.00753.


%%%%%%%%%%%
\bibitem{transtrum2012optimal}
 M. K. Transtrum, P. Qiu,
Optimal experiment selection for parameter estimation in biological differential equation models,
\emph{BMC Bioinf.}, \textbf{13} (2012), 1--12.
\doilink{https://doi.org/10.1186/1471-2105-13-181}

\bibitem{huang2024nonlinear}
 H. H. Huang, Q. He,
Nonlinear regression analysis, preprint, arXiv:2402.05342.



\bibitem{bates1988nonlinear}
D. M. Bates, D. G. Watts,
\emph{Nonlinear Regression Analysis and Its Applications},
 Wiley, New York, 1988.
\doilink{https://doi.org/10.1002/9780470316757}

\bibitem{cao2012penalized}
J. Cao, J. Z. Huang, H. Wu,
Penalized nonlinear least squares estimation of time-varying parameters in ordinary differential equations,
 \emph{J. Comput. Graphical Stat.}, \textbf{21} (2012), 42--56.
\doilink{https://doi.org/10.1198/jcgs.2011.10021}

\bibitem{ramsay2017dynamic}
J. Ramsay, G. Hooker,
 \emph{Dynamic Data Analysis: Modeling Data with Differential Equations},
Springer, New York, NY, 2017.
\doilink{https://doi.org/10.1007/978-1-4939-7190-9}

\bibitem{seber2003nonlinear}
 G. A. F. Seber, C. J. Wild,
 \emph{Nonlinear Regression},
John Wiley \& Sons, Hoboken, NJ, 2003.
\doilink{https://doi.org/10.1002/0471725315}


\bibitem{nocedal2006optimization}
J. Nocedal, S. J. Wright,
\emph{Numerical Optimization},
 2nd ed., Springer, New York, 2006.
\doilink{https://doi.org/10.1007/978-0-387-40065-5}

\bibitem{greenland2009bias}
 S. Greenland,
Bayesian perspectives for epidemiologic research: III. Bias analysis via missing-data methods,
 \emph{Int. J. Epidemiol.}, \textbf{38} (2009), 1662--1673.
\doilink{https://doi.org/10.1093/ije/dyp278}

\bibitem{mckinley2014bayesian}
T. J. McKinley, J. V. Ross, R. Deardon, A. R. Cook,
Simulation-based Bayesian inference for epidemic models,
\emph{Comput. Stat. Data Anal.}, \textbf{71} (2014) 434--447.
\doilink{https://doi.org/10.1016/j.csda.2012.12.012}

\bibitem{kypraios2017abc}
 T. Kypraios, P. Neal, D. Prangle,
 A tutorial introduction to Bayesian inference for stochastic epidemic models using approximate Bayesian computation,
 \emph{Math. Biosci.}, \textbf{287} (2017), 42--53.
\doilink{https://doi.org/10.1016/j.mbs.2016.07.001}

\bibitem{girolami2008diffeq}
 M. Girolami,
 Bayesian inference for differential equations,
 \emph{Theor. Comput. Sci.}, \textbf{408} (2008), 4--16.
\doilink{https://doi.org/10.1016/j.tcs.2008.07.005}

\bibitem{grinsztajn2021workflow}
 L. Grinsztajn, E. Semenova, C. C. Margossian, J. Riou,
 Bayesian workflow for disease transmission modeling in Stan,
\emph{Stat. Med.}, \textbf{40} (2021), 6209--6234.
\doilink{https://doi.org/10.1002/sim.9164}

\bibitem{bouman2024timevarying}
J. A. Bouman, A. Hauser, S. L. Grimm, M. Wohlfender, S. Bhatt, E. Semenova, et al., Bayesian workflow for time-varying transmission in stratified compartmental infectious disease transmission models,
 \emph{PLoS Comput. Biol.}, \textbf{20} (2024), e1011575.
\doilink{https://doi.org/10.1101/2023.10.09.23296742}

\bibitem{gelman2020workflow}
A. Gelman, A. Vehtari, D. Simpson, C. C. Margossian, B. Carpenter, Y. Yao, et al., Bayesian workflow, preprint, arXiv:2011.01808.

\bibitem{belasso2023bayesian}
C. J. Belasso, Z. Cai, G. Bezgin, T. Pascoal, J. Stevenson, N. Rahmouni, et al., Bayesian workflow for the investigation of hierarchical classification models from tau-PET and structural MRI data across the Alzheimer’s disease spectrum,
\emph{Front. Aging Neurosci.}, \textbf{15} (2023), 1225816.
\doilink{https://doi.org/10.3389/fnagi.2023.1225816}

\bibitem{martin2020computingbayes}
G. M. Martin, D. T. Frazier, C. P. Robert,
 Computing Bayes: Bayesian computation from 1763 to the 21st century, preprint,
arXiv:2004.06425.


\bibitem{dunson2001advantages}
 D. B. Dunson,
Commentary: Practical advantages of Bayesian analysis of epidemiologic data,
 \emph{Am. J. Epidemiol.}, \textbf{153} (2001), 1222--1226.
\doilink{https://doi.org/10.1093/aje/153.12.1222}

\bibitem{harel2018imputation}
O. Harel, E. M. Mitchell, N. J. Perkins, S. R. Cole, E. J. T. Tchetgen, B. Sun, et al.,
 Multiple imputation for incomplete data in epidemiologic studies,
 \emph{Am. J. Epidemiol.}, \textbf{187} (2018), 576--584.
\doilink{https://doi.org/10.1093/aje/kwx349}

\bibitem{annis2017stan}
J. Annis, B. J. Miller, T. J. Palmeri,
Bayesian inference with Stan: A tutorial on adding custom distributions,
\emph{Behav. Res. Methods}, \textbf{49} (2017), 863--886.
\doilink{https://doi.org/10.3758/s13428-016-0746-9}

\bibitem{kelter2020survival}
 R. Kelter,
Bayesian survival analysis in Stan for improved measuring of uncertainty in parameter estimates,
\emph{Meas.: Interdiscip. Res. Perspect.}, \textbf{18} (2020),  101--109.
\doilink{https://doi.org/10.1080/15366367.2019.1689761}

\bibitem{gelman1996pk}
 A. Gelman, F. Bois, J. Jiang,
Physiological pharmacokinetic analysis using population modeling and informative prior distributions,
 \emph{J. Am. Stat. Assoc.}, \textbf{91} (1996), 1400--1412.
\doilink{https://doi.org/10.1080/01621459.1996.10476708}

\bibitem{huang2006hiv}
 Y. Huang, D. Liu, H. Wu,
Hierarchical Bayesian methods for estimation of parameters in a longitudinal HIV dynamic system,
\emph{Biometrics}, \textbf{62} (2006), 413--423.
\doilink{https://doi.org/10.1111/j.1541-0420.2005.00447.x}

\bibitem{huang2020bayesianode}
 H. Huang, A. Handel, X. Song,
A Bayesian approach to estimate parameters of ordinary differential equation,
\emph{Comput. Stat.}, \textbf{35} (2020), 1481--1499.
\doilink{https://doi.org/10.1007/s00180-020-00962-8}

\bibitem{gelman2013bayesiandata}
 A. Gelman, J. B. Carlin, H. S. Stern, D. B. Dunson, A. Vehtari, D. B. Rubin,
\emph{Bayesian Data Analysis},
Chapman \& Hall/CRC, 3rd edition, 2013.
\doilink{https://doi.org/10.1201/b16018}

\bibitem{vehtari2021rhat}
 A. Vehtari, A. Gelman, D. Simpson, B. Carpenter, P. C. Bürkner,
Rank-normalization, folding, and localization: An improved $\hat{R}$ for assessing convergence of MCMC,
\emph{Bayesian Anal.}, \textbf{16} (2021), 667--718.
\doilink{https://doi.org/10.1214/20-BA1221}

\bibitem{gelman1992rubin}
 A. Gelman, D. B. Rubin,
Inference from iterative simulation using multiple sequences,
 \emph{Stat. Sci.}, \textbf{7} (1992), 457--472.
\doilink{https://doi.org/10.1214/ss/1177011136}

\bibitem{monnahan2017hmc}
 C. C. Monnahan, J. T. Thorson, T. A. Branch,
Faster estimation of Bayesian models in ecology using Hamiltonian Monte Carlo,
\emph{Methods Ecol. Evol.}, \textbf{8} (2017), 339--348.
\doilink{https://doi.org/10.1111/2041-210X.12681}

\bibitem{burkner2017brms}
 P. C. Bürkner,
brms: An R package for Bayesian multilevel models using Stan,
\emph{J. Stat. Software}, \textbf{80} (2017), 1--28.
\doilink{https://doi.org/10.18637/jss.v080.i01}

\bibitem{gneiting2014probabilistic}
 T. Gneiting, M. Katzfuss, Probabilistic forecasting,
\emph{Annu. Rev. Stat. Appl.}, \textbf{1} (2014), 125--151.
\doilink{https://doi.org/10.1146/annurev-statistics-062713-085831}

\bibitem{richards1959growth}
F. J. Richards,
A flexible growth function for empirical use,
\emph{J. Exp. Bot.}, \textbf{10} (1959), 290--301.
\doilink{https://doi.org/10.1093/jxb/10.2.290}

\bibitem{curran2022hivmpx}
 K. G. Curran, K. Eberly, A. O. Russell, R. E. Snyder, E. Phillips, E. Tang, et al., HIV and sexually transmitted infections among persons with monkeypox --- eight U.S. jurisdictions, May--July 2022,
 \emph{Morb. Mortal. Wkly. Rep.},  \textbf{71} (2022), 1141--1147.
\doilink{https://doi.org/10.15585/mmwr.mm7136a1}

\bibitem{arenas2020covidmodel}
 A. Arenas, W. Cota, J. Gómez-Gardeñes, S. Gómez, C. Granell, J. T. Matamalas, et al.,
A mathematical model for the spatio-temporal epidemic spreading of COVID-19,
 medRxiv, 2020.
\doilink{https://doi.org/10.1101/2020.03.21.20040022}

\bibitem{walter1997identification}
E. Walter, L. Pronzato,
\emph{Identification of Parametric Models from Experimental Data},
Springer, Berlin, 1997.

\bibitem{stigter2023identifiability}
 J. D. Stigter,
Computing parameter identifiability and other structural properties for natural resource models,
 \emph{Nat. Resour. Model.}, \textbf{37} (2023), e12382.
\doilink{https://doi.org/10.1111/nrm.12382}

\bibitem{dankwa2022structural}
 E. A. Dankwa, A. F. Brouwer, C. A. Donnelly,
 Structural identifiability of compartmental models for infectious disease transmission is influenced by data type,
 \emph{Epidemics}, \textbf{41} (2022), 100643.
\doilink{https://doi.org/10.1016/j.epidem.2022.100643}

\bibitem{bracher2021evaluating}
 J. Bracher, E. L. Ray, T. Gneiting, N. G. Reich,
Evaluating epidemic forecasts in an interval format,
 \emph{PLoS Comput. Biol.}, \textbf{17} (2021), e1008618.
\doilink{https://doi.org/10.1371/journal.pcbi.1008618}

\bibitem{gneiting2007scoring}
T. Gneiting, A. E. Raftery,
 Strictly proper scoring rules, prediction, and estimation,
 \emph{J. Am. Stat. Assoc.}, \textbf{102} (2007), 359--378.
\doilink{https://doi.org/10.1198/016214506000001437}

\bibitem{brooks1998convergence}
 S. P. Brooks, A. Gelman,
General methods for monitoring convergence of iterative simulations,
\emph{J. Comput. Graphical Stat.}, \textbf{7} (1998), 434--455.
\doilink{https://doi.org/10.1080/10618600.1998.10474787}

\bibitem{carpenter2018lotkavolterra}
 B. Carpenter, Predator--prey population dynamics: The Lotka--Volterra model in Stan, 2018.

\bibitem{elton1924periodic}
 C. Elton, M. Nicholson,
 Periodic fluctuations in the numbers of animals: Their causes and effects,
 \emph{J. Exp. Biol.}, \textbf{2} (1924), 119--163.
\doilink{https://doi.org/10.1242/jeb.2.1.119}

\bibitem{krishnasamy2020evali}
 V. P. Krishnasamy, B. D. Hallowell, J. Y. Ko, A. Board, K. P. Hartnett, P. P. Salvatore, et al.,
Characteristics of a nationwide outbreak of e-cigarette, or vaping, product use--associated lung injury --- United States, August 2019--January 2020,
 \emph{Morb. Mortal. Wkly. Rep.}, \textbf{69} (2020), 90--94.
\doilink{https://doi.org/10.15585/mmwr.mm6903e2}

\bibitem{lozier2019evaliupdate}
 M. J. Lozier, B. Wallace, K. Anderson, S. Ellington, C. M. Jones, D. Rose, et al.,
Update: Demographic, product, and substance-use characteristics of hospitalized patients in a nationwide outbreak of e-cigarette, or vaping, product use--associated lung injury --- United States, March 31--December 3, 2019,
\emph{Morb. Mortal. Wkly. Rep.}, \textbf{68} (2019), 1142--1148.
\doilink{https://doi.org/10.15585/mmwr.mm6849e1}

\bibitem{cdc2022mpox}
 Centers for Disease Control and Prevention, 2022 U.S. mpox outbreak global map. Available from: {https://www.cdc.gov/poxvirus/mpox/response/2022/index.html}.


\bibitem{chowell2016zika}
G. Chowell, D. Hincapie-Palacio, J. Ospina, B. Pell, A. Tariq, S. Dahal, et al., Using phenomenological models to characterize transmissibility and forecast patterns and final burden of Zika epidemics,
 \emph{PLOS Curr. Outbreaks}, \textbf{2016} (2016).
\doilink{https://doi.org/10.1371/currents.outbreaks.f14b2217c902f453d9320a43a35b9583}


\bibitem{pell2018ebola}
B. Pell, Y. Kuang, C. Viboud, G. Chowell,
 Using phenomenological models for forecasting the 2015 Ebola challenge,
 \emph{Epidemics}, \textbf{22} (2018), 62--70.
\doilink{https://doi.org/10.1016/j.epidem.2016.11.002}

\bibitem{vanDeSchoot2021}
 R. van de Schoot, S. Depaoli, R. King, B. Kramer, K. Märtens, M. G. Tadesse, et al., Bayesian statistics and modelling,
 \emph{Nat. Rev. Methods Primers}, \textbf{1} (2021).
\doilink{https://doi.org/10.1038/s43586-020-00001-2}

\bibitem{grinsztajn2021bayesian}
 L. Grinsztajn, E. Semenova, A. Margossian, E. B. Fox,
Bayesian workflow for disease transmission modelling,
 \emph{Infect. Dis. Modell.}, \textbf{6} (2021), 409--423.
\doilink{https://doi.org/10.1016/j.idm.2021.02.001}

\bibitem{giampiccolo2024hnode}
 S. Giampiccolo, F. Reali, A. Fochesato, G. Iacca, L. Marchetti,
Robust parameter estimation and identifiability analysis with hybrid neural ordinary differential equations in computational biology,
\emph{npj Syst. Biol. Appl.}, \textbf{10} (2024), 139.
\doilink{https://doi.org/10.1101/2024.06.04.597372}

\end{thebibliography}
\end{document}